\documentclass[12pt]{article}
\usepackage{amsfonts,amsmath,amsxtra,dsfont,verbatim}
\usepackage{theorem}
\usepackage{graphics}
\usepackage{epsfig}
{ \theorembodyfont{\normalfont}
 }

{ \theorembodyfont{\normalfont}
 }

\newcommand{\newsection}[1]{
\refstepcounter{section} %\setcounter{equation}{0}
\setcounter{subsection}{0} \addcontentsline{toc}{section}{\protect
\numberline{\arabic{section}}{{\rm #1}}} \vglue .0cm \pagebreak[3]
\noindent{\large \bf  \thesection. #1}\nopagebreak[4]\par\vskip .3cm}
\newcommand{\newsubsection}[1]{
\refstepcounter{subsection}
\addcontentsline{toc}{subsection}{\protect
\numberline{\arabic{section}.\arabic{subsection}}{ #1}} \vglue .0cm
\pagebreak[3] \noindent{\it \thesubsection. #1}\nopagebreak[4]\par\vskip .3cm}

\def \ov {\over}
\def \s{\sigma}

\def \ha {{1 \over 2}}

\def \a {\alpha}

\def\s{\sigma}

\def\ov{\over}
\def\la{\label}

\def\l{\lambda}
\def\eps{\epsilon}

\def \adss{$AdS_5\!\times\!S^5$\ }

\newcommand{\ootimes}{\mbox{\large $\otimes$}}
\newcommand{\bea}{\begin{eqnarray}}
\newcommand{\eea}{\end{eqnarray}}
\newcommand{\be}{\begin{equation}}
\newcommand{\ee}{\end{equation}}
\newcommand{\nn}{\nonumber}
\newcommand{\ds}{\displaystyle}
\newcommand{\com}{\!,\,}

\newcommand{\indups}[1]{_{\mathrm{\scriptscriptstyle #1}}}

\newcommand{\alg}[1]{\mathfrak{#1}}

\newcommand{\comm}[2]{[#1,#2]}

\newcommand{\superN}{\mathcal{N}}

\newcommand{\ham}{\mathcal{H}}
\newcommand{\gym}{g\indups{YM}}
\newcommand{\Op}{\mathcal{O}}
\newcommand{\diag}{\mathop{\mathrm{diag}}}

\newcommand{\fld}[1]{\mathcal{#1}}

\newcommand{\Pib}{\mathbf{\Pi}}
\newcommand{\Pb}{\mathbf{P}}
\newcommand{\tb}{\mathbf{t}}
\newcommand{\Tb}{\mathbf{T}}
\newcommand{\xb}{\mathbf{x}}

\newcommand{\phib}{\mathbf{\phi}}

\newcommand{\onebb}{\mathds{1}}
\newcommand{\Cbb}{\mathds{C}}
\newcommand{\Rbb}{\mathds{R}}
\newcommand{\Zbb}{\mathds{Z}}
\newcommand{\Ac}{\mathcal{A}}

\newcommand{\Hc}{\mathcal{H}}

\newcommand{\Kc}{\mathcal{K}}
\newcommand{\Lc}{\mathcal{L}}

\newcommand{\Nc}{\mathcal{N}}
\newcommand{\Oc}{\mathcal{O}}

\newcommand{\Rc}{\mathcal{R}}
\newcommand{\Sc}{\mathcal{S}}

\newcommand{\Uc}{\mathcal{U}}
\newcommand{\Wc}{\mathcal{W}}
\newcommand{\e}{\mathrm{e}}
\newcommand{\f}{\mathrm{f}}
\newcommand{\E}{\mathrm{E}}
\newcommand{\gf}{\mathfrak{g}}
\newcommand{\hf}{\mathfrak{h}}

\newcommand{\Rf}{\mathfrak{R}}
\newcommand{\mt}{\tilde{m}}
\newcommand{\nt}{\tilde{n}}

\newcommand{\avg}[1]{\left\langle #1 \right\rangle}
\newcommand{\cg}{\Cbb[\Gamma]}
\newcommand{\pr}{\partial}
\newcommand{\half}{\frac{1}{2}}

\newcommand{\clebsch}[3]{C^{#1}_{#2\!;\, #3}}
\newcommand{\invclebsch}[3]{\overline{C^{#1}_{#2\!;\, #3}}}
\newcommand{\K}[2]{\Kc^{#1}_{#2}}
\newcommand{\Kinv}[2]{\overline{\Kc^{#1}_{#2}}}
\newcommand{\rep}[3][]{\rrho^{#1}_{#2}(#3)}
\newcommand{\repconj}[3][]{\overline{\rrho^{#1}_{#2}}(#3)}

\newcommand{\xq}{\xb}

\newcommand{\phiq}{\phib}

\newcommand{\act}{\mathop{\triangleright}}
\newcommand{\ad}{\mathop{\mathrm{ad}}}

\newcommand{\End}{\mathop{\mathrm{End}}}
\newcommand{\gl}{\mathfrak{gl}}

\newcommand{\Hom}{\mathop{\mathrm{Hom}}}
\newcommand{\ord}[1]{\left\vert #1 \right\vert}
\newcommand{\rk}{\mathop{\mathrm{rk}}}
\newcommand{\sgn}{\mathop{\mathrm{sgn}}}
\newcommand{\su}{\mathfrak{su}}
\newcommand{\sdet}{\mathop{\mathrm{sdet}}}
\newcommand{\SU}{\mathrm{SU}}
\newcommand{\SO}{\mathrm{SO}}
\newcommand{\STr}{\mathop{\mathrm{STr}}{}}
\newcommand{\Tr}{\mathop{\mathrm{Tr}}{}}

\newcommand{\reg}{\mathrm{reg}}
\newcommand{\triv}{\mathrm{triv}}

\newcommand{\ie}{\emph{i.e.}}
\newcommand{\eg}{\emph{e.g.}}

\newcommand{\GG}{\Gamma}

\newcommand{\PG}{{\mathrm{P}}_{{}_\GG} }

\newcommand{\ggamma}{\mbox{\large\bf $\gamma$}}

\newcommand{\ba}{\begin{eqnarray}}
\newcommand{\ea}{\end{eqnarray}}

\newcommand{\ket}[1]{\left\vert #1 \right\rangle}

\def \la{\label}

\newcommand{\overgamma}{{\raisebox{1pt}{\small $1$}\over \raisebox{-2.5pt}{\small $|\Gamma|$}}}

 \newcommand{\cover}{{{}^{{\cal N}=4}}}
  \newcommand{\Rff}{{\mbox{\small $\Rf$}}}
 \newcommand{\ooplus}{ \raisebox{-5pt}{\Large ${\bigoplus}\atop{\mbox{ \scriptsize \!\!\!\!\! $\lambda$}}$}\, }
\newcommand{\oootimes}{ \raisebox{-5pt}{\Large ${\bigotimes}\atop{\mbox{ \scriptsize $\lambda$}}$}\, }
\newcommand{\omplus}{ \raisebox{-6pt}{\Large ${\bigoplus}\atop{\raisebox{1pt}{ \scriptsize \!\!\!\!\!\! $\mu$}}$}\, }

\newcommand{\ttt}{\mbox{ \large\!\!\!\! t\,}}
\newcommand{\II}{{\mbox{\tiny $I$}}}
\newcommand{\JJ}{{\mbox{\tiny $J$}}}
\newcommand{\rrho}{\mbox{\large $\, \rho$}}
\newcommand{\rrhobar}{\; \overline{\! \! \rrho}}
\newcommand{\aA}{{{}^{{}_A}}}
\newcommand{\aAe}{{{}^{{}_{A_1}}}}
\newcommand{\aAt}{{{}^{{}_{A_2}}}}
\newcommand{\aAL}{{{}^{{}_{A_L}}}}
\newcommand{\bB}{{{}^{{}_B}}}
\newcommand{\bBe}{{{}^{{}_{B_1}}}}
\newcommand{\bBt}{{{}^{{}_{B_2}}}}
\newcommand{\bBL}{{{}^{{}_{B_L}}}}

\newcommand{\aAb}{{{{}_A}}}
\newcommand{\aAeb}{{{{}_{A_1}}}}
\newcommand{\aAtb}{{{{}_{A_2}}}}
\newcommand{\aALb}{{{{}_{A_L}}}}

\newcommand{\dddots}{\; \dot{}\; \dot{}\; \dot{}\; }
\newcommand{\ihalf}{{i\over 2}}
\newcommand{\Nsuper}{\Nc\!=\!}
\newcommand{\da}{\downarrow}
\newcommand{\ua}{\uparrow}

\begin{document}

\thispagestyle{empty}
\vspace*{1cm}
\begin{flushright}
PUPT-2249\\
ITEP-TH-16/07\\
arXiv:0711.1697\\[20mm]
\end{flushright}
\begin{center}
{\Large \scshape Bethe Ansatz Equations\\[5mm] For General Orbifolds of $\Nsuper 4$ SYM} \\[15mm]
Alexander Solovyov\,${}^{a,\,b}$\\[10mm]
${}^a$ \emph{Physics Department,}\\
\emph{Princeton University, Princeton, NJ 08544}\\[5mm]
${}^b$ \emph{Bogolyubov Institute for Theoretical Physics,}\\
\emph{Kiev 03680, Ukraine}\\[30mm]
\textbf{Abstract}
\end{center}
{\small We consider the Bethe Ansatz Equations for orbifolds of $\Nsuper 4$ SYM w.r.t.\ an arbitrary discrete group.
Techniques used for the Abelian orbifolds can be extended to the generic non-Abelian case with minor modifications.
We show how to make a transition between the different notations in the quiver gauge theory.}

%\newpage
%\tableofcontents

\newpage
\newsection{Introduction}

For a long time in high energy physics there exists a strong interest in the web of dualities
between gauge theory and closed strings. The relationship essentially started with
the inception of string theory as a dual model
of hadronic interactions (for a recent review see, \eg, \cite{Br}).
From the point of view of QCD, the modern
theory of strong interactions, hadronic strings would be interpreted as color electric flux tubes
between quarks. A concrete version of gauge/string correspondence was proposed by 't~Hooft
in~\cite{'tHooft} (see also~\cite{Wil},\cite{KS},\cite{Po}) in the form of the $1/N$ -expansion,
the central idea of which is that the Feynman graphs
of a large $N$ gauge theory naturally organize themselves as triangulations of a string surface.
The rank of the gauge group $N$ is related to the string coupling via $g_S = 1/N$, and counts the
number of handles of the surface spanned by the non-planar graphs.
The gauge theory/closed string duality is expected to be a limit of a more general open/closed
string correspondence, which should hold at the world sheet level. Open string diagrams are
equivalent to closed string world sheets with holes. The idea behind the open/closed string
correspondence is that the holes can be replaced by closed string vertex operators, and absorbed
into an adjustment of the sigma model that  governs the motion of the closed string.
From the perspective of the low energy effective field theory, this relation between open and
closed strings gives rise to the famous duality between gauge theory and gravity, the central
example of which is the celebrated AdS/CFT correspondence~\cite{Ma},\cite{GKP},\cite{Wit},\cite{AdSCFT}.
The key physical insight that spurned this development was the discovery of the D-branes~\cite{Pol}, followed
by understanding of the geometrical nature of the non-Abelian Chan-Paton factors in terms of
stacks of coincident branes~\cite{Wit2}. On the other side, in terms of the Matrix Theory proposal~\cite{BFSS} non-Abelian gauge degrees of freedom are just a part of a more general theory; and thus they naturally incorporate into the web of dualities.

However, there are some difficulties in studying the AdS/CFT conjecture. One of them is the fact that the weak coupling on the gravity side (closed strings) corresponds to the strong coupling regime on the gauge theory side (open strings); and this prevents one from performing simple perturbative checks.
It was major breakthrough when it was realized that some integrable structures were present in the scalar subsector of $\Nsuper 4$ SYM~\cite{MZ}, and this result was extended to the complete set of operators in~\cite{BKS},\cite{Be},\cite{BS}.
At the same time there was investigated the integrability of the closed string motion in~\cite{GKP2} and the following works. This opened the new opportunities of understanding the AdS/CFT duality beyond perturbation theory.

Another idea commonly used in string theory since~\cite{DHVW} is
that of the orbifold space. An orbifold is a quotient of some
manifold w.r.t.\ a discrete group. The procedure of orbifoldization
was expected to be useful in particular for model building. A strong
motivation for this is the usage of quotient spaces for
(super)symmetry breaking. Another way to use orbifold construction
which was used recently is to embed some models into quiver gauge
theories.

Even though there are some works studying these dualities for some special orbifolds or some special limits~\cite{AS-J},\cite{KPRT},\cite{BdBHIO},\cite{WaW},\cite{Ide},\cite{DRGOS},\cite{SSJ},\cite{BR},\cite{AFGS}; they mainly deal with the Abelian orbifolds and the corresponding quiver gauge theories.
The main goal of this paper is to extend some of these studies to the generic orbifolds with an arbitrary non-Abelian orbifold group.
Organization of the paper is as follows.
In the second Section we summarize the results regarding the closed string motion on orbifolds. We discuss the subtleties specific to the general non-Abelian orbifolds.
In the third Section we introduce the orbifold gauge theory which is the low-energy limit of the corresponding open string theory.
We introduce the two different descriptions, the one using the twist fields (and most closely resembling the original unorbifolded theory) as well as the one using the quiver diagram.
Then we develop the transition formulae between them.
In the fourth Section we introduce the Feynman rules for the quiver gauge theory and study the field theory dynamics. This leads to the representation of the matrix of anomalous dimensions as a spin chain Hamiltonian. The Hamiltonian locally coincides with that of the unorbifolded theory.
In the fifth Section we review the Bethe Ansatz Equations (BAE) and generalize them to the generic orbifold theories. The key ingredients of the construction are essentially the same as those for the Abelian orbifolds. The idea is that one can diagonalize the twist field in each given twisted sector, and then the setup reduces to the Abelian case modulo some subtleties.
In the sixth Section we study some applications of the BAE. We find the solutions in the long spin chain limit and compare them with the closed string energies. We also consider particular quivers (both Abelian and non-Abelian) and show how the eigenvectors of the matrix of anomalous dimensions are rewritten in terms of the quiver notation.
Appendix A introduces the basic group theory notations and conventions.
Appendix B contains the calculations related to the conversion between the two descriptions in the orbifold gauge theory as well as the construction of observables.\\

\bigskip
\newsection{Orbifold String Theory}
\label{sec:closed_string}

Besides exploring the integrability of the orbifold gauge theories,
an important motivation for our study is to test the correspondence between
large $N$ gauge theory and closed string theory. The closed string dual to the orbifold gauge theories
follows from the
AdS/CFT dictionary. The stack of $N$ D3-branes, located on the fixed point of the orbifold space $\Cbb^3\!/\GG$,
induce via their gravitational backreaction a near-horizon geometry that is given by
\be
AdS_5\times S^5\!/\GG.
\ee
The AdS/CFT correspondence states that the planar diagrams of the orbifold gauge theory span
the worldsheet of closed strings propagating on this near-horizon geometry.

The orbifold group $\Gamma$ can be an arbitrary finite subgroup of $SO(6)$, the
isometry group of the sphere $S^5$.  In general, the finite group does not commute
with supersymmetry, and the resulting orbifold string theory is therefore non-supersymmetric.
It can be shown, however, that all such non-supersymmetric orbifolds of \adss
are unstable, due to the
presence of localized tachyonic modes. For this reason we will restrict ourselves to supersymmetric
orbifolds, for which $\Gamma$ defines a finite subgroup of $SU(3)$.
Let us parameterize $S^5$ as a sphere of radius $R$ inside $\Cbb^3$, with coordinates
$(Z_1,Z_2,Z_3)$:
\be
\sum_\II \overline{Z}_\II Z_\II = R^2\, .
\ee
 $SU(3)$ naturally acts on $\Cbb^3$ and on the $S^5$.  In the special case
that the finite group $\GG$ fits inside an $SU(2)$ subgroup of $SU(3)$, the orbifold theory is ${\cal N}\!=\! 2$ supersymmetric.\footnote{
Finite subgroups of $SU(2)$ have a well-known classification: they organize into an ADE series.
The A-type subgroups are Abelian, while the D-type and exceptional type subgroups are non-Abelian.
Under the McKay correspondence these correspond to the cyclic
groups, the double covers of the dihedral groups, and the double
covers of the  rotational symmetry groups of the tetrahedron,
cube/octahedron, and dodecahedron/octahedron, respectively.
The finite
subgroups of $SU(3)$ are less familiar, but have a similar classification.
Finite subgroups of $SU(3)$ other than $SU(2)$ and direct products of Abelian
phase groups fall into 2 series:  analogues of dihedral subgroups, denoted by $\Delta(3n^2)$ with $n$ a
positive integer, and  $\Delta(6n^2 )$  with $n$ a positive even integer, and analogues of exceptional subgroups,
denoted by $\Sigma(60)$, $\Sigma(168)$, $\Sigma(360k)$, $\Sigma(36k)$, $\Sigma(72k)$, $\Sigma(216k)$ with
$k$ = 1, 3. The number in braces is the order of the group. As an example, the discrete
$SU(3)$ subgroup $\Delta(3n^2)$ has $3n^2$ elements, generated by
the three $\Zbb^3$ transformations (here $\omega = e^{\frac{2\pi i}{n}}$)
\ba
\label{gggg}
g_1 \,: \qquad
(Z_1, Z_2 , Z_3) & {\longrightarrow} & (\, \omega \, Z_1\, , \, \omega^2 \, Z_2 , Z_3 ) \,,
\nonumber\\[.5mm]
g_2 \,: \qquad
(Z_1, Z_2 , Z_3) & {\longrightarrow} & (\, Z_1, \omega \, Z_2\, , \omega^2 \, Z_3 ) \,,
\\[.5mm]
g_3 \, : \qquad (Z_1,Z_2,Z_3)
 & \longrightarrow & \;  (\; Z_1 ,\; Z_2 \; ,\; Z_3\; )\nonumber \,.
\ea
}%end footnote

\begin{figure}[htb]
\begin{center}
\includegraphics[width=2in,height=2in]{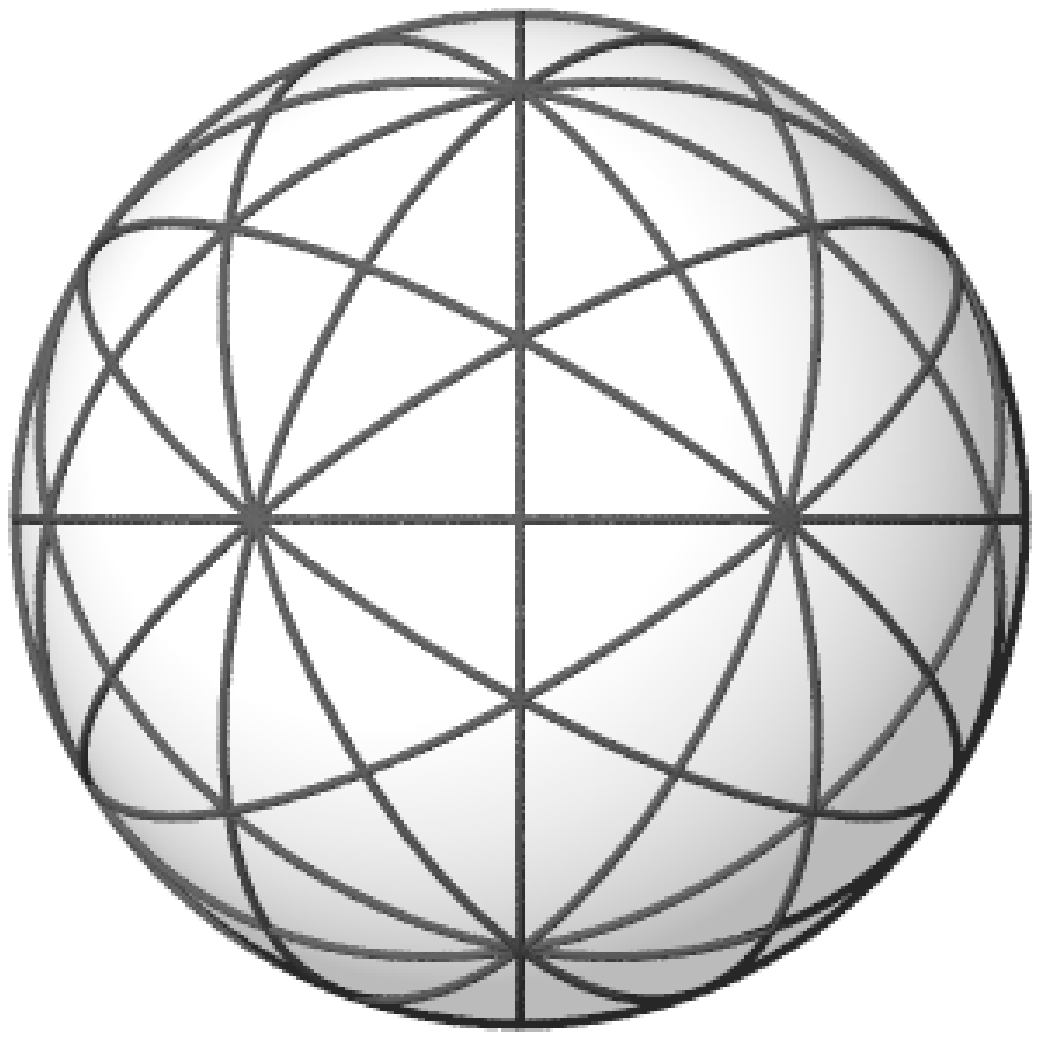}
\qquad\qquad\qquad\qquad
\includegraphics[width=2in,height=2in]{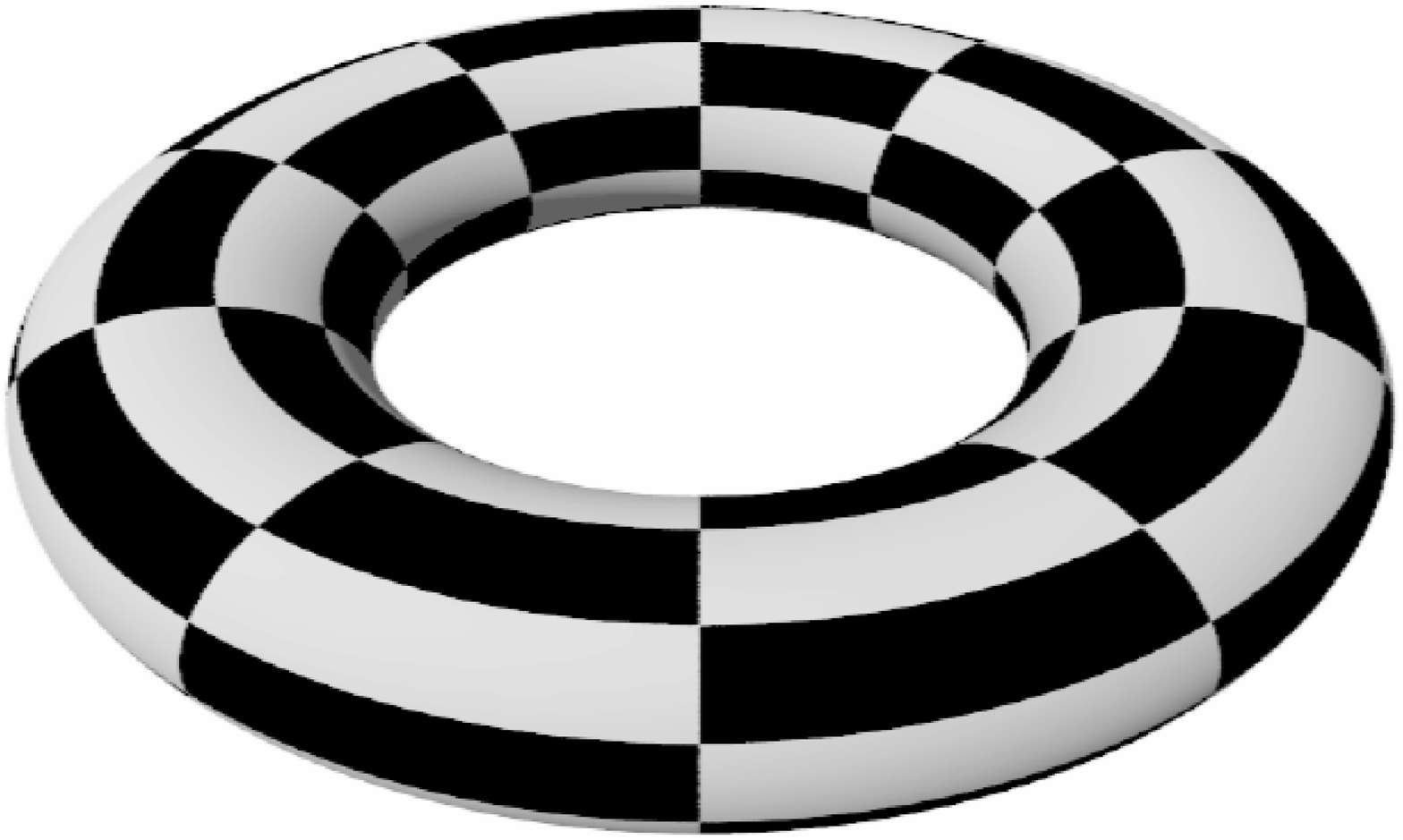}
\end{center}
\caption{\small Finite group transformations acting on $S^5$ may have fixed points or act freely. In the former
case the group action looks similar to the isometries that act on the sphere on the left.
In the free case, the group element can be viewed as a combination of commuting Abelian isometries,
analogous to the isometries that act on the torus on the right.
For supersymmetric orbifolds, transformations with fixed points are contained inside an $SU(2)$ subgroup. (The above pattern on the sphere has icosahedral symmetry, which is one of the exceptional subgroups of $SU(2)$.) }
\end{figure}

In this section we will  summarize the semiclassical treatment of closed
strings moving on $AdS_5\!\times\!S^5\!/\GG$.  We will mostly focus on string configurations in the twisted
sectors, since  the properties of untwisted states simply follow from the parent theory on $AdS_5\!\times\!S^5$.
Twisted sector strings connect two different points on $S^5$ that are related  via some element $g\in \GG$.
Since the finite group still acts on the twist $g$ by conjugation, twisted sectors are labeled by conjugacy
classes in~$\GG$.

To characterize the twisted string states, we note that the $S^5$ metric allows for three commuting Abelian
isometries.  In general, these are broken by the orbifold group. For a given twist element $g$, however, we can
orient things such that $g$ acts by a combination of the three isometries, and thus preserves all three of them.
So to specify a given twisted sector, we are free to assume that the twist $g$ acts
via a diagonal matrix on the $Z_\II$. If $g$ is an element of order $S$ inside $\GG$,
$g^S= 1$, we can write
\be
\label{ab}
g\; : \ \ ( Z_1,\ Z_2,\ Z_3) \; \to \; ( \omega^{s_1} Z_1,\ \omega^{s_2} Z_2,\ \omega^{s_3} Z_3) \,,
\ee
\be
\omega = e^{2\pi i/S} \, ,  \qquad \qquad \quad \sum_\II^{} s_\II = 0 \, . \label{orb}
\ee
We see that,  in this given twisted sector,
the string is free to move along three circle directions, and one can define corresponding conserved angular momenta
$J_\II$, with ${I}=1,2,3$.
We further observe that in general, the group element $g$ acts freely on $S^5$. The corresponding
twisted sector strings thus have a minimal length. However, when one of the three integers
$s_\II$, say $s_1$, vanishes  --- so when $g$ in fact fits inside $SU(2)$ ---  the action of $g$ on $S^5$ has an obvious fixed point at $(Z_1,Z_2,Z_3) = (R,0,0)$. \\

\medskip
\newsubsection{Classical strings on $S^5\!/\GG$: Semiclassical Treatment}

We now summarize some relevant results on the classical motion of strings along the $S^5$ having in mind the future comparison with the gauge theory side. The more general calculations can be found in~\cite{ART} and references therein; in particular, \cite{FT}, \cite{AFRT} and \cite{AS}.
We restrict ourselves to the strings moving in $S^5$ directions only and trivially embedded into $AdS^5$.
This motion is governed by the sigma model action (restricted to the bosonic string coordinates)
\begin{equation}
 S \sim %\sqrt{2\l} % { \sqrt{\lambda} \over 2\pi }
 \int\! d\tau d\sigma \left({\ha}\partial_a \ttt \partial^a \ttt - \partial_a \overline Z_\II \partial^a Z_\II
+ \Lambda (\overline Z_\II Z_\II -R^2)\right) \, .
\end{equation}
Here $\ttt$ denotes the $AdS$ time coordinate, and $\Lambda$ is a Lagrange multiplier field.
This action is obtained as a reduction of the string worldsheet action in the conformal gauge;
thus the equations of motion derived from this action
must be supplemented by the corresponding Virasoro constraints:
\bea
\dot{\,\ttt}\!{}^2 &=& \dot{Z}_\II \dot{\bar{Z}}_\II + Z^\prime_\II \bar{Z}^\prime_\II \,,
\\
0 &=& \dot{Z}_\II \bar{Z}_\II^\prime \,.
\eea

We want to solve for the motion of the string in
the twisted sector defined by the twist element (\ref{ab}).
As it was emphasized in~\cite{Ide}, the closedness requirement then allows for the fractional winding numbers,
\be
\label{per}
Z^{}_\II(\sigma + 2\pi) = \e^{2\pi i \mt_\II} Z^{}_\II(\sigma)\,, \qquad \mt_\II=m_\II+\frac{s_\II}{S} \,.
\ee
The $S^5$ metric has three commuting isometries. In general, these are broken by the
orbifold group. However, from the explicit form of the twist element given in (\ref{ab}), we see that
{\it  in this given twisted sector},
the string is free to move along three circle directions. It is therefore natural to choose the following {\it Ansatz,}
\be
\ttt = \kappa \tau\,, \qquad \qquad Z_\II =z^{}_\II (\sigma)\ e^{i\omega^{}_\II \tau}\, .
\ee
Inserting this Ansatz, the Lagrangian governing the dependence $z_\II(\s)$ becomes
\be
\label{L}
\Lc= \half \sum_{\II} (\bar z'_\II  z'_\II -  \omega_\II ^2  \bar z^{}_\II
z^{}_\II ) - \half
\Lambda(\sum_\II \bar z^{}_\II z^{}_\II-R^2) \, .
\ee
The Virasoro constraints simplify to
\ba \la{cve}
\kappa^2\!\! & =& \! \! \sum_{\II}( \bar z'_\II z'_\II      + \omega_\II^2 \bar z^{}_\II z^{}_\II )\,,    \\
0 \!\! & =& \! \! \sum_\II \omega_\II (\bar z{}'_\II z^{}_\II - z'_\II \bar z^{}_\II) \,.
\ea
Note that both Lagrangian and Virasoro constraints have a $U(1)^3$-invariance w.r.t.\ the multiplication by a phase,
\be
z_\II\to \e^{i\alpha_\II}z_\II\,, \qquad \qquad \bar{z}_\II\to \e^{-i\alpha_\II}\bar{z}_\II \,.
\ee
This invariance leads to the three integrals of motion,
\be\la{ank}
\ell_\II = \ihalf\, (\bar z{}'_\II z^{}_\II - z'_\II \bar z^{}_\II)\,.
\ee
This allows us to eliminate the angular variables. Then denoting $r_\II^2 = z_\II \bar{z}_\II$ and
substituting this back into the action we get the following effective  Lagrangian:
\be
\label{Laa}
\Lc = \frac{1}{2}\sum_{\II}
\Bigl(  r'^2_\II   -   w_\II^2  r^2_\II + { \ell_\II^2 \ov r_\II^2} \Bigr)
-  \frac{1}{2} \Lambda\Bigl(\sum^3_{\II}  r^2_\II-1\Bigr) \,.
\ee
This system is called the \emph{Neumann-Rosochatius} (NR)
integrable system (e.g., \cite{NR});
and its detailed analysis in the context of the closed string dynamics is given in~\cite{ART}.
Here we restrict ourselves to the simplest example, the circular strings.
These solutions take the following forms:
\be
z_\II = a_\II\, \e^{i\mt_\II\sigma} \,, \qquad \Lambda = {\rm const.}
\ee
With this ansatz integrals $\ell_\II=\mt_\II a_\II^2$; while the dynamical equations yield
\be
\label{Lag}
w_\II^2 = - \Lambda - \mt_\II^2 \,, \qquad \sum_{\II=1}^3 a_\II^2 = 1 \,;
 \ee
 \be
\label{Vira}
\kappa^2 = \sum_{\II=1}^3 a_\II^2 (w_\II^2 + \mt_\II^2)\,, \qquad
\sum_{\II=1}^3 a_\II^2 w_\II \mt_\II=0 \,.
\ee

The space-time energy for the circular string configuration is
\bea \la{enn}
E = \sqrt{8\pi^2\l} \int_0^{2\pi}\!\frac{d\sigma}{2\pi} \dot{\,\ttt} = \sqrt{8\pi^2\lambda}\, \kappa \,;
\eea
while the spins are
\be
\label{spins}
J_\II=\sqrt{8\pi^2\lambda}\, w_\II
\int_0^{2\pi}\! \frac{d\sigma}{2\pi}\,
 r_\II^2(\sigma) = \sqrt{8\pi^2\lambda}\, w_\II a_\II^2\,.
\ee
One can define the total spin $L = \sum_{\II=1}^3 |J_\II|$.
The relations (\ref{Lag}) and (\ref{Vira}) can be rewritten in terms
of the energy and the spins. Eqs.~(\ref{Vira}) read
\be
\label{Vira2}
\frac{E^2}{8\pi^2\l} = -\Lambda \,,
\qquad \sum_{\II=1}^3 \mt_\II J_\II =0 \,;
\ee
while (\ref{Lag}) becomes
\be
\sum_{\II=1}^3 \frac{|J_\II|}{\sqrt{-\Lambda-\mt_\II^2}}=\sqrt{8\pi^2\l} \,.
\ee

We will consider the two spin solution with $J_3=0$.
Then in the large $L$ limit one can solve these equations and find the following expansion for the energy:
\be
E = L + \frac{4\pi^2\l}{L}\, |\mt_1|\, |\mt_2| \,.
\ee

Given that $J_1,J_2>0$, one must have $\mt_1 \mt_2<0$.
Recalling the definition of the fractional winding numbers $\mt_2$, $\mt_2$ one can write the string energy as
\be
\label{string_energy}
E = L + \frac{4\pi^2\l}{L}\, \Bigl(m-\frac{s_1}{S}\Bigr) \Bigl(m^\prime+\frac{s_2}{S}\Bigr) \,;
\ee
$m$ and $m^\prime$ being some positive integers.
In Section~\ref{sec:BAE_apps} we will see that this expression matches the one-loop anomalous dimensions for the corresponding $\su_2$ subsector formed by the two scalars in the field theory.\\

\bigskip
\newsection{Orbifold Gauge Theory}
\label{sec:gauge_theory}
We now turn to the study the class of quiver gauge theories obtained by taking
an arbitrary (Abelian or non-Abelian) orbifold of $\Nsuper 4$ supersymmetric $U(N)$ gauge theory.
Our motivation is to investigate to what extent the recently uncovered large $N$ integrability
of $\Nsuper 4$ SYM can be extended to this general class of orbifold  gauge theories. In
this section we will summarize some of their relevant properties. The relevant references are~\cite{KaS},\cite{GP},\cite{DM},\cite{LNV},\cite{BKV},\cite{BJ}.\\

\medskip
\newsubsection{Orbifold Projection}

It will be convenient to think of the quiver gauge theory as the low energy limit of the open string theory
on a stack of $N$ D3-branes located near an orbifold singularity.
Before taking the orbifold quotient, the transverse space of the D3-branes is $\Rbb^6 \simeq \Cbb^3.$
The low energy field theory on the D3-branes is $\Nsuper 4$ SYM, with
its field
content  (in $\Nsuper 1$ superfield notation):
a vector multiplet $\Ac$ and three chiral multiplets $\Phi^\II$, with {\small $I =1,2,3$}, that parameterize the
three complex transverse positions of the D3-branes along  $\Cbb^3$.

Let $\GG$ be some finite group  of order $|\GG|$,  that acts on $\Cbb^3$.
The orbifold space  is obtained by dividing out the action of the discrete group $\Gamma$.
The transverse space to the D3-branes thus becomes
$$
\Cbb^3\!/\Gamma\, .
$$
When viewed from the covering space, the stack of $N$ D3-branes in the orbifold space give rise to the total of $|\Gamma|N$ image D3-branes.  It is convenient to label the
image D3-branes by a pair of Chan-Paton indices $(i,h)$ with $i=1,\ldots, N$ and $h \in \GG$, so that the brane labeled by $(i,h)$ is the image of the $i$-th brane inside the D3-stack under the group element $h\in \Gamma$. The group $\GG$ thus acts on the Chan-Paton indices as
\be
\label{gaction}
g \ :
\ (i,h) \ \to \ (i , gh)\, .
\ee
When the $N$ coincident D-branes all approach the orbifold fixed point,  the image branes all coincide and the strings stretched between them have massless ground states.
The vector multiplet $\Ac$ has  a separate matrix entry for  each open string stretching between
two image branes, and thus defines an $| \GG| N\!\times\!|\Gamma| N$ matrix. Before
imposing invariance under the orbifold group $\GG$, the full collection of image branes
thus supports an $U(|\Gamma| N)$ gauge symmetry.  The orbifold projection, however,
selects only those fields that are invariant under the discrete group $\Gamma$. The discrete group
acts on the vector multiplet $\Ac$ only via the Chan-Paton indices and on the chiral multiplets $\Phi_\II$  via both the Chan-Paton and transverse indices.

Although the orbifold theories all have less supersymmetry,
their action is assumed to be identical to that of the  parent ${\cal N} \!= \!4$ theory,
which in ${\cal N}\! =\! 1$ superfield notation reads
\ba
\label{YM_action}
\Lc  =   \int\limits^{\mbox{${}$}}_{\mbox{${}$}} \!\! d^4 \theta \, \Tr\Bigl( {\cal W}^\alpha {\cal W}_\alpha  %\frac14\,{\cal F}_{\mu\nu} {\cal F}^{\mu\nu} +% \sum_I \Tr
+e^ { {\Ac}} \Phi^\dagger_\II e^{-{\cal A} } \Phi_\II \Bigr)
  %\covd_\mu\Phi_\II^{\dagger}\, \covd^\mu\Phi_\II - \frac12\, g^2\,
%\sum_{I\com J} \Tr{} \bigl(
\ + \  \int\! \! d^2 \theta\, \epsilon^{IJK} \Tr( \Phi_I [ \Phi_J,\Phi_K] ) \; + \; c.c.
\ea

\noindent
Here the trace $\Tr$ is over the adjoint representation of the full $U(|\Gamma| N)$
gauge group of the $\Nsuper 4$ theory.
The fields $(\Ac,\Phi)$ of the orbifold  theory, however, are special matrices that are
obtained by applying a linear
projection $\PG$ on the fields
of the parent theory
\be
\label{projection}
\Ac\,  = \, \PG \,  \Ac^\cover\, ,  \qquad \qquad \Phi = \, P_{{}_\GG} \, \Phi^\cover\, .
\ee
The projection operator $\PG$ takes the generic form\footnote{Here and in the following, the summation over $g$
runs over the whole finite group $\GG$, unless otherwise indicated.}
\be
\label{proj}
\PG ={\raisebox{-1pt}{\small $1$}\over \raisebox{0pt}{\small $|\Gamma|$}} \sum_{g}\; g\, ,
\ee
where $g$ acts in an appropriate representation on $(\Ac,\Phi)$, that we will specify momentarily.
This projection operator does not commute with the full ${\cal N}\! = \! 4$ superconformal invariance,
but in the special case that $\GG$ forms a subgroup of $SU(3)$, the orbifold quotient
preserves $\Nsuper 1$ superconformal invariance. More generally, one could consider
orbifolds with $\Gamma$ some subgroup of $SO(6)$.
However, it has been shown that for non-supersymmetric orbifolds, the quantum
theory has non-zero $\beta$-functions for certain double-trace operators and is therefore
not conformally invariant. The renormalized Hamiltonian of non-supersymmetric
orbifolds thus contains extra terms that do not descend from the $\Nsuper 4$ Hamiltonian~\cite{DKR}.
For this reason we will restrict ourselves to the supersymmetric subclass.

Let us specify the action of $g$ in  (\ref{proj}) on $\Ac$ and $\Phi^\II$.
First, we recall the definition of the regular  representation of the finite group $\GG$. The group algebra
${\Cbb }[\GG]$
is the $|\GG|$-dimensional
vector space of $\Cbb$-valued functions on $\GG$
\be
{\Cbb}[\GG] = \Bigl\{x=\sum_{h} x(h)\; h
\Bigr\}
\ee
with $x(h) \in \Cbb$.
The group $\GG$ acts on the group algebra via
$g :  x\ \to \,  \sum_{h} x(h)\,
 g h. $
From this,
 we obtain a natural representation of $\GG$
in terms of $|\GG|\!\times\!|\GG|$ matrices,
known as the regular representation $\ggamma_{\reg}$. It acts on the $|\Gamma|$ dimensional
vector space
\be
V_{\reg}\; \simeq \; \Cbb[\GG]\, .
\ee

The regular representation is directly relevant to our problem. As is evident from our definition of the Chan-Paton indices,
the vector multiplet ${\cal A}$ is naturally
identified as an element of
\be\label{ain}
\Ac \in V_{\reg}^{\, \oplus N} \, \ootimes \, \overline{V}_{\!\reg}^{\, \oplus N} \,.
\ee
The finite group $\Gamma$ acts on ${\cal A}$ both from the left and from the right, both actions being implemented via the regular representation. Geometrically, we can think of each as applying the group element $g$ to the D3-brane at respectively the begin- and end-point of the
open strings.
The orbifold symmetry transformation acts simultaneously on both ends, and thus acts on $\Ac$
via conjugation. (Note that $\overline{\ggamma}_\reg(g) = \ggamma(g^{-1})$.)
Similarly, the chiral multiplets $\Phi$ are elements of
\be\label{phin}
\Phi \in \Cbb^3 \otimes V_{\reg}^{\, \oplus N} \, \ootimes \, \overline{V}_{\! reg}^{\, \oplus N} .
\ee
To write their transformation rule, we must also account for the action
discrete group on $\Cbb^3$, which proceeds via the
3-dimensional defining representation, that we will denote by $\Rf$. With these definitions,
$\GG$  acts on the fundamental fields as
\be
\label{Action}
g: \; \; \Ac \ \ \to\ \  \ggamma_\reg(g) \, \Ac \; \overline\ggamma_\reg(g)\, ,
\ee
\be
\label{vproj}
\qquad g: \; \; \Phi^\II \ \ \to \ \
\Rff(g)^{{}^{{}_{\II}}}_{\JJ}
\, \ggamma_{\reg}(
g )\, \Phi^\JJ \, \overline\ggamma_{\reg}(g)\, .
\ee
Combined with (\ref{projection}) and (\ref{proj}), this specifies the projection from the $\Nsuper 4$
parent theory to the orbifold gauge theory. Written out explicitly in terms of the group valued Chan-Paton indices (\ref{gaction}), the orbifold projected fields take the form\footnote{Here we suppress the other $U(N)$ valued Chan-Paton
index.}
\ba
\label{orbitbasis1}
\Ac(g)\!\! & = & \!\!  \overgamma \sum_h \Ac_{h,\bar{g}\bar{h}}\, \\[2mm]
\Phi^\II(g)\!\! & = & \!\!  \overgamma \sum_h \Rff(h)^\II_\JJ \, \Phi^\JJ_{h,\bar g \bar h}
\label{orbitbasis2}
\ea
We see that after the projection, the  $\GG$ valued left and right Chan-Paton indices have collapsed to a single group valued index. The gauge and matter fields can thus be thought of as group algebra valued $N\!\times\!N$ matrices. We will refer to the above basis of orbifold projected fields as the {\it orbit basis} (as distinguished from the {\it quiver basis}, that will be introduced later).

Note that, since the orbifold projection (\ref{projection})
does not commute with $U(|\Gamma| N)$, the gauge symmetry gets broken to a subgroup.
This unbroken gauge group is identified as follows.
The regular representation $\ggamma_{\reg}$ decomposes into irreducible representations $\rrho_\lambda$ via
\be
\label{decomp}
\qquad \ggamma_{\reg}(g) =\ooplus \rrho_\lambda(g)^{\oplus N_\lambda} \,,
\qquad \qquad N_\lambda={\rm dim} \rrho_\lambda .
\ee
In words, each irreducible representation $\rrho_\lambda$ occurs $N_\lambda$
times in the decomposition of the regular representation.  In explicit matrix notation, we have
\be\label{greg} \ggamma_{\reg}=  \mbox{\small $
\left(\begin{array}{cccc}{\rho_1 \otimes {\onebb}_{N_1}} & {0} & ... & 0 \\
0 & \rho_2 \otimes {\onebb}_{N_2} & ... & 0 \\
\vdots& \vdots & \ddots & \vdots \\
0 & 0 & ... & \rho_r \otimes {\onebb }_{N_r}  \end{array} \right) $}
\ee
where $\rrho_\lambda \otimes \onebb_{N_\lambda}$ is the $N_\lambda \times N_\lambda$ matrix
\be
\rrho_\lambda \otimes \onebb_{N_\lambda} =  \mbox{\small $
\left( \begin{array}{cccc} \rho_\lambda & 0 & ... & 0 \\
0 & \rho_\lambda & ... & 0 \\
\vdots& \vdots & \ddots & \vdots \\
0 & 0 & ... & \rho_\lambda \end{array}\right) $}\, .
\ee
Here each $\rrho_\lambda$ denotes an $N_\lambda\! \times\! N_\lambda$ matrix.
By inspecting the explicit form (\ref{greg}) of $\ggamma_{\reg}$, it is not difficult to derive that the orbifold gauge group, defined as
the subgroup of $U(|\GG|N)$ transformations that commutes with $\ggamma(g)$ for all $g \in \GG$, takes the
product form
\be
\label{sub}
\oootimes U(NN_\lambda) \,.
\ee
Here the product runs over all representations of $\GG$ and each factor $U(NN_\lambda)$ is the subgroup that rearranges the $NN_\lambda$ copies of the representation space
$V_\lambda$ of $\rrho_\lambda$ --- it therefore  obviously commutes with $\GG$. Using Schur's lemma, one
proves that (\ref{sub}) indeed defines the maximal unbroken gauge group:
physical operators  need only be gauge invariant under this group.\footnote{In other words, the decomposition (\ref{decomp}) of the regular representation shows that the representation vector space $V_{\reg}$ decomposes  as
\be
\label{vdeco}
V_{\reg} =\; \raisebox{-5pt}{\Large ${\bigoplus}\atop{\mbox{\scriptsize $\lambda$}}$}\, V_\lambda\otimes \Cbb^{N_\lambda} \,,
\ee
 where $V_\lambda$ denotes the $N_\lambda$ dimensional representation space of $\rho_\lambda$.
Combining (\ref{vdeco}) with (\ref{ain}) and the invariance condition (\ref{A_inv_cond}), we find that
\be
\Ac \in \ooplus
\; ( V_\lambda^{ \, \oplus N_\lambda N} )
\, \ootimes \;   \, ( \bar{V}_{\! \lambda}^{\, \oplus N_\lambda N} ) \; = \; \ooplus
(V_\lambda \otimes \bar{V}_\lambda) \, \otimes\, ( \Cbb^{N N_\lambda} \! \otimes \bar{\Cbb}{}^{NN_\lambda}) .
\ee
This decomposition of $\Ac$ makes manifest that the commutant of $\GG$ within $U(|\GG|N)$ takes the form (\ref{sub}).}
 \\

\medskip
\newsubsection{The Untwisted Sector}

The untwisted sector states of the orbifold theory have a  simple relation to those of the parent theory: they
are obtained by applying the projection operator $\PG$ to (every
elementary field inside)
gauge invariant single-trace operators of the parent theory of the form
\be
\label{Spinoperator}
\Op=\Tr\bigl( \fld{W}_{\aAeb}\fld{W}_{\aAtb}\ldots\fld{W}_{\aALb}\bigr).
\ee
Here $\fld{W}_{\aAb}$
stands for a multiple covariant derivative of one of the fields of the theory, in $\Nsuper 1$ notation:
\be
\label{wact}
\fld{W}_{\aAb}\in \Bigl\{ \, \fld{D}^n\Phi_I,\,
\fld{D}^m \fld{W}_\alpha\, \Bigr\}
\ee
It should be emphasized, however, that not all single trace
expressions of the form (\ref{Spinoperator}) lead to non-vanishing operators of the orbifold theory:
the orbifold projection, combined with the cyclicity of the trace, implies that only the $\Gamma$-invariant
combinations survive. A complete basis of untwisted operators thus takes the
form\footnote{This obvious calculation is given in Appendix~\ref{app:untwisted_observables}.}
\be
\Oc_\Kc = \Kc^{\aAe \aAt \dddots \aAL} \Tr\bigl(
\fld{W}_{{}_{A_1}}\fld{W}_{{}_{A_2}}\ldots\fld{W}_{{}_{{A_L}}}\bigr),
\ee
where ${\cal K}^{\aAe \aAt \dddots \aAL}$ denotes any $\GG$-invariant tensor.

In recent years, there has been enormous
progress in understanding of the quantum properties of such
single trace operators in the large $N$ limit of $\Nsuper 4$ super Yang-Mills theory.
The central insight that has precipitated this breakthrough development, is that local operators of the form (\ref{Spinoperator})
can be represented as spin chain states
\be
|{\cal O} \rangle = |\mbox{\small $ A_1, A_2, \ldots,
A_L$} \rangle\, .
\ee
Under this identification, the gauge theory generator of conformal rescalings
maps to the spin chain Hamiltonian $\ham$. The energy eigenvalues $E_{\cal O}$ of $\ham$ are related to the
anomalous dimensions $\Delta_{\cal O}$ of local single trace operators ${\cal O}$  by
\be \label{eq:SpinChainEnergy}
\Delta_{\Op}= \lambda
\,E_{{\Op}},
\ee
with $\lambda = \frac{\gym^2 N}{8\pi^2}$ the 't~Hooft coupling.  Most remarkably, there is a fast growing
and convincing body of evidence that the spin chain system defined via this correspondence is
completely integrable, at least at leading order at large $N$ and for large length $L$ of the spin chain
(many operators $\fld{W}_{{}_A}$ in (\ref{Spinoperator})).
This realization has lead to a flurry of new gauge theory results and non-trivial
quantitative checks of the AdS/CFT correspondence.

In the following, we will investigate how these
results can be generalized to the full class of supersymmetric orbifolds of $\Nsuper 4$ super Yang-Mills theory.
Although these orbifolds all have less supersymmetry and  are less special than the parent theory, there are
several direct and indirect arguments to support the conjecture that these orbifold gauge theories
are indeed still integrable. Our goal is to extend the known results (\cite{BR} and related works) to the general  orbifold gauge theories.

It is a well known fact that in the large $N$ limit, the planar Feynman diagram expansion in the untwisted sector coincides with that of the unorbifolded theory modulo the projection onto the $\GG$-invariant states. As a consequence, the Hamiltonian $\Hc$  of the orbifold theory, when acting on untwisted states, is simply equal to the $\Nsuper 4$ Hamiltonian $\Hc^{\Nc=4}$ acting on the $\Gamma$-invariant subspace of the ${\cal N}\!= \! 4$ Hilbert space, $\Hc=\Pib^\GG \Hc^{\Nc=4}$. The original action being $\GG$-invariant, the $\Nc\!=\!4$ Hamiltonian commutes with the projector, $\comm{\Hc^{\Nc=4}}{\Pib^\GG}=0$; and thus the action of $\Hc$ is defined correctly. In particular, the conformal dimensions of the gauge invariant untwisted sector operators of the orbifold theory are identical to those of the parent theory.
This correspondence extends to arbitrary loop order.\\

\medskip
\newsubsection{The Twisted Sectors}

\begin{figure}[htb]
\begin{center}
\epsfig{figure=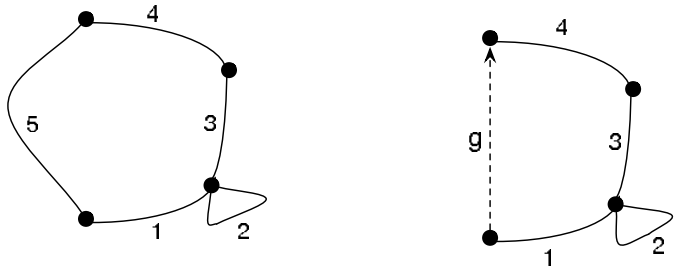,scale=0.55}
\end{center}
\caption{\small An untwisted state (left) and twisted state (right). Both are
concatenated arrays of open strings (lines) stretched between D3-branes (dots).
The end-point brane of the twisted state is the image under a finite group transformation
$g$ on its begin-point brane. }
\label{fig:twisted_state}
\end{figure}

So far, we have restricted our attention to the untwisted sector, the subclass of operators
that arise as a proper subsector of the parent ${\cal N}\!=\! 4$ theory.  In the open string language, the untwisted
operators can be thought of as arrays of concatenated open strings attached to several
image D3-branes, as indicated on the left in the Fig~\ref{fig:twisted_state}.
Each operator $\fld{W}_{A}$ corresponds to a ground state of one of the open strings. The gauge invariant
trace implies that the array is closed: it begins and
ends on the same D3-brane, and thus represents a proper closed string state in the unorbifolded
theory.  A twisted sector state, on the
other hand, corresponds to a concatenated array of open strings that ends on a different
D3-brane, related via a finite group transformation $g \in \GG$ to the D3-brane where it begins. This configuration looks like an open string on the covering space, but
represents a closed string on the orbifold space. Correspondingly,
it is associated with an operator that is not gauge invariant under the full $U(|\GG|N)$ symmetry of the
cover theory, but that {\it is} invariant under
the gauge group (\ref{sub}) of the orbifold theory.

In the gauge theory, the twisted states are
represented as single trace expressions
\be\label{OrbOperator}
\Op(g)=\Tr\bigl( \ggamma(g)\,
\fld{W}_{{}_{A_1}}\fld{W}_{{}_{A_1}}\ldots\fld{W}_{{}_{{A_L}}}\bigr),
\ee
where we introduced a twist operator $\ggamma(g)$, defined as follows. When $\ggamma(g)$ acts from the left
on a matrix-valued operator $\fld{W}_\aA$, it acts via the group action (\ref{gaction}) --- the regular representation
$\ggamma_\reg(g)$ --- on the left Chan-Paton
index. When $\ggamma(g)$ acts from the right, it acts via the complex conjugate
group action $\overline{\ggamma}_\reg(g)$ on the right Chan-Paton index.
The actions from the left and from the right
are not identical; instead, the operators $\fld{W}_\aA$ of the orbifold theory
satisfy a relation of the form
\be
\label{twistshift}
\ggamma(g)\, {\fld W}_{{}_A} =\;  \Rf(g)_{{}_A}^{\bB}\; {\fld W}_{{}_B}  \ggamma(g) \,,
\ee
where $\Rff(g)_\aA^\bB$ denotes a matrix representation of the finite group $\GG$, acting on the $\Cbb^3$
index of $\fld{W}_{{}_A}$.\footnote{Here $\Rff(g) = 1$ in case $\fld{W}_A$ has no $\Cbb^3$ index.
Note further that inserting multiple twist operators in the
trace does \emph{not} introduce a new class of operators, since by using the exchange relation (\ref{twistshift}) and
the property $\ggamma(g_1)\ggamma(g_2) = \ggamma(g_1g_2)$, one can always reduce any number of
twist operators to a single overall twist. This is as one would have expected from the string interpretation.}
% end footnote

As a consequence of the orbifold projection, some of the physical operators (\ref{OrbOperator}) vanish
identically.  Using equation \eqref{twistshift} to commute
$\ggamma(g)$ past all the fields in the operator shows
that a necessary condition for non-vanishing operators
is that the total single trace operator must be invariant under the simultaneous action of
$\Rff(g)_{{}_A}^{{}^{{}_B}}$ on all the spins. However, while necessary, this is not sufficient.
More generally, physical operators are of the form
\be
\label{twistbasis}
{\cal O}_{\cal K}(g) \; =\;  {\cal K}(g)^{\aAe \aAt \dddots \aAL} \Tr\bigl( \ggamma(g)\,
\fld{W}_{{}_{A_1}}\fld{W}_{{}_{A_1}}\ldots\fld{W}_{{}_{{A_L}}}\bigr),
\ee
where ${\cal K}(g)$ must be an invariant tensor under the complete stabilizer subgroup $S_g$ of $g$, defined
as the subgroup within $\GG$ of all elements that commute with $g$.\footnote{It can be the case that even for some $S_g$-invariant tensor $\Kc(g)$ the corresponding operator $\Oc_\Kc(g)$ vanishes identically due to some symmetry reasons --- for instance, this is the case in the $\Zbb_6$ quiver we consider in Section~\ref{sec:BAE_apps}.}
%end footnote
In the untwisted case, where $g$
is the identity element in $\GG$, the stabilizer subgroup is the whole group $\GG$ and indeed, as we saw before,
untwisted operators are in one-to-one correspondence with $\GG$-invariant tensors. For non-trivial twisted
sectors, it would be too strong a condition to impose that ${\cal K}(g)$ is invariant under the full group $\GG$,
since the group acts non-trivially on the twist element. Correspondingly, we must define
the $S_g$ invariant tensors ${\cal K}(g)$ to transform non-trivially under the full group action via
\be
\label{sinv}
{\cal K}(g)^{\aAe \aAt \dddots \aAL} \mbox{\small $ \Rff(h)_{{}_\aAe}^{\bBe}  \Rff(h)_{{}_\aAt}^{\bBt}
 \ldots \Rff(h)_{{}_\aAL}^{\bBL} $}\, =\;
{\cal K}(hgh^{-1})^{\bBe \bBt \dddots \bBL} \,
\ee
for all $h \in\GG$. We see that, in a given twisted sector, the $\GG$-invariance
is spontaneously broken to the stabilizer subgroup of the twist element $g$:
the property (\ref{sinv}) states the ${\cal K}(g)$ is
invariant under the simultaneous action of $\Rff(h)$ on all spins,
provided that $h$ commutes with $g$.

As one could have expected from the string dictionary~\cite{DVVV}, the twisted operators  (\ref{twistbasis}) depend only on
the conjugacy class of the twist element $g$:
\be
\label{conj}
\Op_{\cal K}(g) = \Op_{\cal K}(hgh^{-1})
\ee
This result easily follows by combining the property (\ref{sinv}) with the transformation rule
(\ref{twistshift}) of the operators ${\cal W}_{{}_A}$.

It is important to note that the basis (\ref{twistbasis}) of operators is really a complete basis, in the sense
that any operator of the seemingly more general class given in (\ref{OrbOperator}), that
is not of the form (\ref{twistbasis}), vanishes identically.\footnote{Detailed construction of twisted operators as well as the proof of their gauge invariance is given in Appendix~\ref{app:twisted_observables}.} The reason is that the single site operators
${\cal W}_{{}_A}$ are all selected via the orbifold projection (\ref{projection})-(\ref{proj}) to transform
according to (\ref{twistshift}) under the finite group $\GG$. The space of orbifold operators is spanned
by the {\it orbit basis}, defined by
\be
\fld{W}_{{}_{\! A}}(g) =  {\raisebox{-1pt}{\small $1$}\over \raisebox{0pt}{\small $|\Gamma|$}} \sum_h \Rff(h)^\bB_{{}_A}\,
\fld{W}_{{}_{{}^{\! B}} ; \, h,\bar g \bar h}\, .
\ee
This definition combines the two formulas (\ref{orbitbasis1}) and (\ref{orbitbasis2}).
\smallskip

Let us now state the main conclusion of the section, that will become important later:
\smallskip

\parbox{15.4cm}{\itshape  It is possible to define a basis of twisted sector operators
(\ref{twistbasis}) such that  the single site operators ${\cal W}_{{}_A}$ that contribute in the sum
are all eigenstates of the twist matrix $\Rff(g)_{{}_A}^{\bB}$.}

\smallskip
\noindent
To prove this statement, we first note that the invariance condition
(\ref{sinv})-(\ref{conj}) allows us to restrict ourselves to a given representative $g$ in each twisted sector $[g]$.
Let ${\mathfrak S}_{k}$ denote the representation of $S_g$ that acts on the $k$-th site. Then a single element $g$ can be diagonalized,
and the eigenvalue of the corresponding spin ${\cal W}_{{}_{A_k}}$ is given by
\be
\label{g_diag}
{\mathfrak S}_{k}(g) = \sigma_k(g) \onebb,
\ee
If $g$ is an element of order $S$ in $\GG$, then $\sigma_k(g)$ is some phase factor of the form $\exp(2\pi i s_k/S)$.
This observation will be useful in the later sections, where we will study the classical and quantum properties of these twisted
sector operators.
The idea is to diagonalize the action of a given group element $g$, and then apply some of the tools used in the studies of the Abelian orbifolds.
This property will be exploited in Section~\ref{sec:BAE} where we study the Bethe Ansatz Equations for the orbifold gauge theory.\\

\medskip
\newsubsection{Quiver Gauge Theory}

In this section, we will make a comparison between the above group
theoretic description of the physical operators with the quiver representation
of the orbifold gauge theory.  The reader that is mostly interested in the integrability of the orbifold theory
may wish to skip this section at first reading.

\smallskip

The physical field content of orbifold gauge theories is made most manifest via its representation
as a quiver gauge theory. As discussed, the unbroken gauge group of the orbifold theory takes the
product form
\be
\label{sub2}
\oootimes U(NN_\lambda) \,,
\ee
where the product runs over all representations $\rrho_\lambda$ of the finite group $\GG$ and $N_\lambda = {\rm dim} V_\lambda$. Notice that, even in the case that $N\!=\!1$, that is, for the world-brane theory of a single
D3-brane near an orbifold singularity, this gauge group contains several, in general non-Abelian, factors. In the string
theoretic construction, each gauge factor is associated to a stack on $NN_\lambda$ so-called fractional D3-branes.
There is one type of fractional brane for each representation $\rho_\lambda$ of the finite group.

The vector multiplets $\Ac$ arise as the ground states of open strings attached to a given fractional brane.
Let us denote by $\Ac_\lambda$ the vector multiplet of the fractional brane associated to $\rrho_\lambda$.
Hence $\Ac_\lambda$ is an $U(NN_\lambda)$ gauge multiplet. In terms of the {\it orbit basis}  $\Ac(g)$
defined in (\ref{orbitbasis1}), the {\it quiver basis} $\Ac_\lambda$ is obtained via the Fourier-like transformation
\be
\label{quiverbasis1}
\Ac_\lambda = \sum_{g}\, \rrhobar_\lambda(g) \, \Ac(g)
\ee
Setting up the quiver terminology, we will refer to each gauge factor and its associated stack of fractional
branes, as a {\it node} of the quiver diagram. There is one quiver node for each irreducible representation of $\GG$.
\begin{figure}[t]
\begin{center}
\epsfig{figure=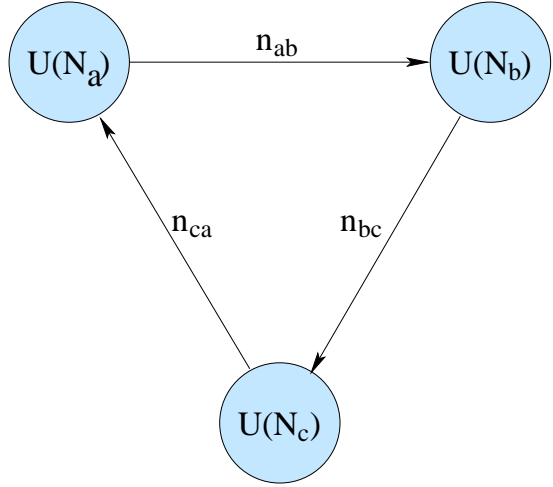,scale=0.8}
\end{center}
\caption{\small A simple example of a three node quiver. The nodes indicate the gauge group factors, and
the oriented lines indicate the number of bi-fundamental matter fields.}
\label{fig:quiver}
\end{figure}

In a quiver diagram, the nodes are connected by oriented lines: these represent the chiral matter fields. In the
string theory construction, the chiral matter fields $\Phi^\II$ arise as the ground states of open strings that
may have end-points on two different fractional branes. Correspondingly, they transform as bi-fundamental fields
under the product gauge group (\ref{sub2}). More accurately, the chiral matter fields are invariant tensors
\be
\Phi^{\lambda \overline{\mu}} \in {\rm Inv}(\Cbb^3 \otimes V_\lambda \otimes \overline{V}_\mu ).
\ee
The number $n_{\lambda\bar\mu}$ of chiral matter fields between two given nodes $\lambda$ and $\mu$
is determined by the multiplicity of $\rho_\mu$ in the decomposition of the tensor product between the
defining representation $\Rff$ and $\rho_\lambda$:\footnote{Using that multiplication of group characters reflects the representation algebra,
we can compute the multiplicities via
\be
n_{\mu\bar\lambda} =\frac{1}{|\GG|}\sum_{g} \chi_{\Rf}(g)\chi_\lambda (g) \overline{\chi}_\mu(g).
\ee
Here we used the familiar orthogonality condition on group characters $
\sum\limits_{g} \chi_\lambda(g) \overline{\chi}_\mu(g) =|\GG| \delta_{\lambda\mu}.
$}
\be
\Rff \otimes \rrho_\lambda = \omplus \, n_{\lambda\bar \mu} \, \rrho_{\mu}\,.
\label{couplings}
\ee
In the string construction, the number $n_{\lambda\bar\mu}$ is the intersection number between the two
fractional branes. The fields $\Phi^{\lambda\bar\mu}$ thus transform in the $(NN_\lambda, \overline{NN_\mu})$
bi-fundamental representation of the gauge group (\ref{sub2}).\footnote{
As a check, let us count the number of independent components of the chiral matter field $\Phi$.
For each arrow there are $N^2 N_\lambda N_\mu$ components, and
each node therefore connects to $N^2 N_\lambda \sum_{\{\mu\}} \dim V_\mu$ independent components. Since
$\Rff\otimes V_\lambda = \oplus_{\{\mu\}} V_\mu$, dimension counting gives $\sum_{\{\mu\}} \dim V_\mu = 3
\dim V_\lambda\, $.
Therefore, the total number of independent components of $\Phi$ is $3 N^2\,  \sum_\lambda N_\lambda^2 = 3 \ord{G} \,N^2\, $. This is the expected result.}
This {\it quiver basis} $\Phi^{\lambda\bar\mu}$ is related to the {\it orbit basis} $\Phi^\II(g)$ given in (\ref{orbitbasis2})
via the linear transformation
\be
\label{quiverbasis2}
\Phi^{\lambda\bar\mu}=
 \sum_{g,\II}  \, {\cal K}^{\, {}^{{}_\II}}_{\lambda\bar\mu} \, \rrhobar_\mu(g)
\, \Phi_{{}_{\! \II}}(g)
\ee
 where ${\cal K}_{\lambda\bar \mu}$ denotes one of the $n_{\lambda\bar \mu}$ basis elements that spans the space of invariant tensors in
 $\Cbb^3 \otimes V_\lambda \otimes \overline{V}_\mu$.

In the quiver basis, it is now easy to specify all possible single trace operators of the orbifold
gauge theory. For this, it is useful to introduce the notion of the {\it path algebra} of the quiver diagram.
A path is a concatenated array of arrows that connect quiver nodes connected by oriented lines.
The arrows are allowed to point back to the same node. We can multiply two paths if one ends at the
same node as where the other begins. We can then connect them head to tail to produce
a single longer path. In the quiver gauge theory, each arrow of the path represents a gauge or matter
operator ${\cal W}_{{}_A}$ of the general form (\ref{wact}), transforming in the corresponding representation
of the quiver gauge group.
Connecting two arrows amounts to taking their
gauge invariant product at the corresponding quiver node. To write gauge invariant operators,
we now simply  choose arbitrary closed paths along the quiver, pick a corresponding array of
operators, and in the end take the trace.

How does this description of gauge invariant single trace operators compare with that in terms of
twisted sector states (\ref{twistbasis}) given in the previous section?
To make this dictionary, we must first relate the quiver basis of the single site operators
${\cal W}_{{}_A}$ to the corresponding orbit basis (\ref{quiverbasis3}). This is done via the general formula
\be
\label{quiverbasis3}
\fld{W}_{\lambda\bar\mu} = \sum_{g, \aA} {\cal K}_{\lambda\bar\mu}^{{}^{\, {}_A}}\;\, \overline{\! \! \rrho}_\mu(g)\;
\fld{W}_{{}_{{}_{\! A}}}\! (g)
\ee
Here the invariant tensor ${\cal K}_{\lambda\bar \mu}$ is simply equal to the identity operator
when the index $A$ belongs to the trivial representation and $\lambda\! =\! \mu$;
i.e. in case the operator is associated to an arrow beginning and ending at the
same node. (Nevertheless, there can be the scalar lines beginning and ending at the same node
emerging from the fields having indices in a non-trivial transverse representation.)

Now let us pick some closed path ${\cal C}_\lambda$,
that starts and ends at  a given node $\lambda$ but along the way visits the following sequence of quiver nodes
\be
{\cal C}_\lambda \ \ : \ \ \lambda \ \leftarrow\ \mu \ \leftarrow\ \nu \ \leftarrow \ldots\ \leftarrow\ \sigma \ \leftarrow\ \lambda\, .
\ee
For each arrow along this path, we pick an operator of the form (\ref{quiverbasis3}) and multiply them together,
and take the trace at the $\lambda$ node
\be
{\cal O}_{{\cal C}_\lambda} = \Tr_{\! \lambda}\bigl( \fld{W}_{\lambda\bar\mu}\fld{W}_{\mu\bar\nu}\cdots
\fld{W}_{\sigma \bar\lambda} \bigr)\, .
\ee
This a manifestly gauge invariant operator of the quiver gauge theory.
The equivalence with the group algebraic description of the orbifold theory implies that this
operator must be a linear combination of twisted state operators ${\cal O}_{\cal K}(g)$ defined in
Eqn.~(\ref{twistbasis}).
A straightforward calculation, described in Appendix~\ref{app:quiver_vs_orbit}, indeed shows that
\be
\label{transition_formula}
\Oc_{{\cal C}_\lambda} = \Oc_\Kc(g)  \,,
\ee
where the invariant tensor $\Kc$ is given by
\be
\label{K_transition}
{\cal K}(g)^{{}^{A_1 A_2 \ldots A_L}}=\Tr_{\! \lambda}\bigl(\overline{\rrho}_{\lambda}(g)\,{\cal K}_{\lambda\bar\mu}^{\aAe}%\! \! \cdot
{\cal K}_{\mu \bar\nu}^{\aAt}\cdots {\cal K}_{\sigma \bar \lambda}^{\aAL}\bigr)
\ee
This expression indeed satisfies the relation (\ref{sinv}).
The class of operators associated to closed loops on the quiver diagram
span a complete basis of twisted sector operators, and vice versa.\\

\bigskip
\newsection{Field Theory Dynamics}
\label{sec:FT}

The field theoretic problem we are trying to solve on the gauge theory side is diagonalization of  the matrix of anomalous dimensions. Generally if there is a basis in the space of operators, $\{\Oc^i\}$, the renormalization involves some matrix valued $Z$-factor. It means that the renormalized operators are given as
\be
\tilde\Oc^i = Z^i_j\, \Oc^j \,.
\ee
There exists a distinguished basis in the space of operators such that the correlation functions become diagonal,
\be
\label{diag_cf}
\avg{\Oc^i(0)\, \bar\Oc^j(x)} \sim \frac{\delta^{ij}}{x^{2(L+\Delta^i)}} \,.
\ee
Here $L$ is the bare dimension of operators and $\gamma^i$ is called the anomalous dimension. It can be found as the corresponding eigenvalue of the matrix of anomalous dimensions
\be
D = \frac{dZ}{d\log \Lambda}\, Z^{-1} \,.
\ee
Here $\Lambda$ is the cutoff scale. These are exactly the eigenvectors of this matrix that make the correlation functions diagonal as in (\ref{diag_cf}). We will mainly be interested in diagonalization of the two point correlation functions of single trace operators. We restrict ourselves to the planar diagrams (the large $N$ expansion).\\

\medskip
\newsubsection{Feynman Rules}
\label{sec:feynman_rules}

As outlined above, we can think of the fields in the orbifold as a special subset of ${\cal N}\! = \! 4$ fields,
defined by the projection (\ref{projection})-(\ref{proj}), where $g$ acts as given in Eqn.~(\ref{Action}). A
natural strategy is to feed the ${\cal N}\! = \! 4$ Feynman rules through this identification.
Let us therefore label the orbifold fields by the inverse image under $\PG$. Clearly, this is a redundant
representation: via this labeling, we introduce a $|\GG|$-fold excess, since the orbifold projection identifies
%\footnote{Here we suppress the $U(N)$ indices, and only indicate the $\GG$-valued Chan-Paton indices.}
\begin{eqnarray}
\label{A_inv_cond}
\Ac_{fg,\bar{f}\bar{h}} \; = &\! \! \Ac_{g,\bar{h}}\, , \qquad  & \nonumber \\[2mm]
 \Phi^{{}^{{}_{\, \II}}}_{fg, \bar{f}\bar{h}}\;
  =  &\!\!  \Rff(f)^{{}^{{}_{\II}}}_{\JJ} \,
 \Phi_{g,\bar{h}}^{{}^{{}_{\, \JJ}}} \, . &
\end{eqnarray}
This notation, keeping the redundant group indices, proves to be very useful for the comparison to the unorbifolded theory. Namely, one now easily verifies that the ${\cal N}\!=\! 4$ Feynman rules imply the
following non-zero Wick contractions for the orbifold theory\footnote{We ignore the ghost fields. Gauge fixing is easy to do via the
Feynman gauge. Since the gauge field $\Ac$ can be treated as a group algebra valued,
the gauge fixing and Faddeev-Popov ghosts can also be treated as group algebra valued.}\,\footnote{Detailed derivation of the Feynman rules is given in Appendix~\ref{app:feynman_rules}.}
% % ===========
% % end of footnote
% % ===========
\ba
\label{quiver_quad}
\avg{A^\mu_{h, \bar{g}}\, A^\nu_{f h ,  \bar{f} \bar{g}} } &=& \frac{\delta^{\mu\nu}}{\ord{\Gamma} p^2} \,,
\\[2mm]
\label{quiver_quad_2}
\avg{\Phi^{\, \II}_{h,\bar{g}}\; \Phi^{\, \JJ}_{fg, \bar{f}\bar{ h}}} &=&  \frac{\Rff(f^{-1})^{\II\JJ}}{\ord{\Gamma} p^2} \,.
\ea
The factor of $1/ |\GG|$ compensates for the overcounting of fields. Generally, for elementary fields (or their derivatives) $\fld{W}_{{}_A}$ there takes place the following replacement in the propagator:
\bea
\avg{\fld{W}_{{}_A}\, \fld{W}_{{}_B}}_{\Nsuper 4} = G(p)\, \delta^{\aA\bB} &\to& \avg{\fld{W}_{{}_A}\, \fld{W}_{{}_B}} = G(p)\, \Rff^{\aA\bB}(f) \,.
\eea

\begin{figure}[t]
 \begin{center}
\includegraphics[width=15cm]{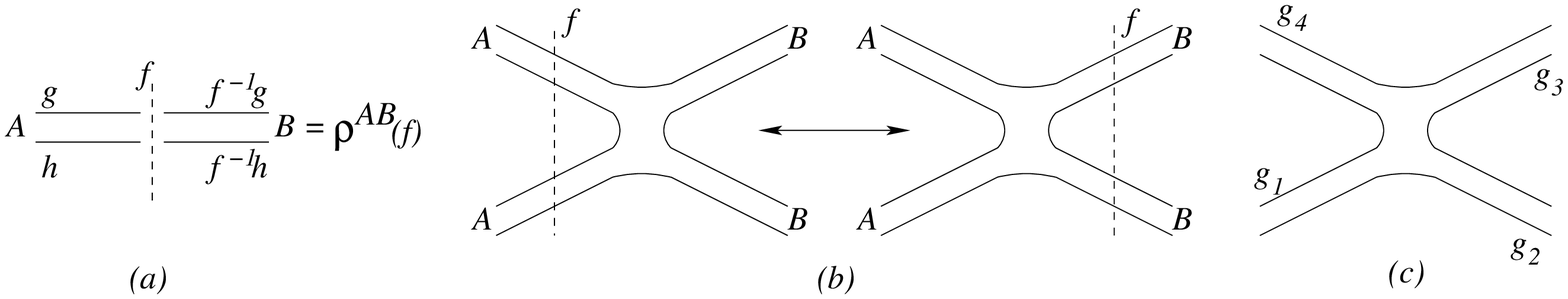}
\end{center}
\caption{\small $(a)$ When a line of the Feynman graph crosses the cut the Wick contraction
$\avg{\fld{W}_{{}_A}\, \fld{W}_{{}_B}}$ is non-diagonal, and proportional to
the matrix element $\Rff^{\aA\bB}(f)$. $(b)$ The twist lines can be deformed and moved through interaction vertices.
$(c)$ In the notation using the group valued fields untwisted vertices obey the conservation condition, similar to the conservation of momentum: for the vertex shown the product $g_1 g_2 g_3 g_4=1$.}
\label{fig:Feynman_rules}
\end{figure}

We notice the important feature that the propagator is not simply diagonal on the group valued Chan-Paton indices
$(g,h)$, but there  can be a twist by some group element $f$, that acts simultaneously on both the left and right index.
The advantage of this redundant double line notation is that the interaction vertices coincide with those of the original theory, and the only modification is the introduction of these twists along the propagators.

Equivalently, we can think of the twist as the assignment of a group element $f$ to each line of the dual graph to the Feynman diagram.
We will call these lines on the dual graph `cuts'.
When a propagator crosses a cut, the propagator
$\avg{\phi^\II\, \phi^\JJ}$ is non-diagonal: the conventional factor $\delta^{\II\JJ}$ gets replaced by the matrix element $\Rff^{\II\JJ}(f^{-1})$ with $f$ the twist along the~cut.
Vertices of the dual graph correspond to loops of the original Feynman graph. The product of the group elements that meet at
a dual vertex must multiply to the identity element in $\GG$. (Unless the amplitude involves the insertion
of some twist operator at this dual vertex, see below.)
Note that this approach is similar to that of~\cite{BJ}; instead of introducing cuts they insert projectors into each propagator.\\

\medskip
\newsubsection{Spin Chain Hamiltonian}
\label{sec:hamiltonian}

\begin{figure}[htb]
\begin{center}
\includegraphics[width=10cm]{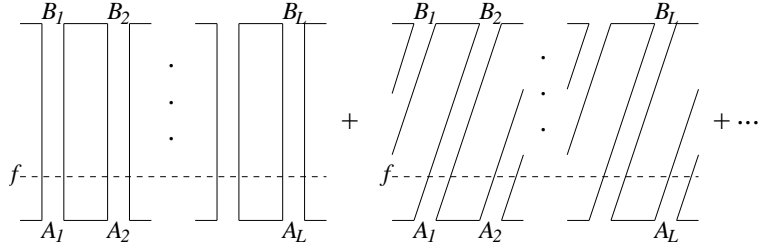}
\end{center}
\caption{\small The modification of the Feynman rules results in the left diagram being multiplied by $\sum_f \rep[]{A_1 B_1}{f}\ldots \rep[]{A_L B_L}{f} \sim \Pi^{A_1\ldots A_L}_{B_1\ldots B_L}$ --- the projector onto the singlet states w.r.t.\ $\Gamma$. This is also true for the loop diagrams.}
\label{fig:projection}
\end{figure}

Now we can proceed to our goal, diagonalization of the anomalous dimension matrix. As we said, it is convenient to represent the field theory operators as some spin chain states,
\be
\label{spin_chain_state}
\Oc^{\aAe \aAt {}^{\ldots} \aAL}[g] \equiv \Tr \bigl(\ggamma(g) \Wc_{{}_{A_1}} \Wc_{{}_{A_2}} \ldots \Wc_{{}_{A_L}} \bigr) = \ket{A_1 A_2\ldots A_L}_g \,.
\ee
Note that this basis is overcomplete --- some of the states are projected out. Another subtlety is that one can perform a cyclic permutation in the trace leading to a seemingly different spin chain representation. In the untwisted case this results in an extra requirement on the physical spin chain state (\ie, one emerging from some gauge theory operator) --- invariance w.r.t.\ the translation operator, the zero momentum condition. This particular choice of a representative is fixed by the necessity of taking into account all the diagrams shown in Fig.~\ref{fig:projection}). We will see that when a non-trivial twist field $\ggamma(g)$ is introduced, the zero momentum constraint gets slightly modified.
Using this terminology, the matrix of anomalous dimensions is represented as some spin chain Hamiltonian. We derive the relation between the Hamiltonian emerging from the unorbifolded $\Nsuper 4$ theory and that of the quotient quiver gauge theory.

Since all the terms in the action are untwisted, in the planar limit there should be no mixing between the sectors with different twists.
This can be seen very clearly if we use the notation in which the fields depend on one group index. Then for a generic operator of the form $\Wc_{\aAe}(g_1)\ldots \Wc_{\aAL}(g_L)$ with some fields $\Wc_\aA$  its twist class is determined as $[g_1\ldots g_L]$. Obviously, in the planar limit interactions cannot change the cyclic product $[g_1\ldots g_L]$ (Fig.~\ref{fig:Feynman_rules}). This way we can restrict ourselves to the operators $\Oc[g]$ with a fixed class $[g]$. A twisted sector is therefore a superselection sector: the twist $[g]$
is preserved under time-evolution defined by ${\cal H}$. However, the representation of the ${\cal H}$ as
a spin chain Hamiltonian does depend on the twist sector.

This dependence can be derived based on the particular form of the Feynman rules.
The sum over the twist factors locally decouples from the remainder of the Feynman integral. In particular,
the $\Gamma$ invariance of the interaction vertices (of the original Feynman diagram) ensures  that the cuts can be deformed and moved through the vertices, as it is indicated in Fig.~\ref{fig:Feynman_rules}.
Following this procedure one can move the cuts, and translate them along the
the worldsheet spanned by the Feynman diagram. Evidently, we can merge cuts that are along homologous
cycles on the worldsheet; the group element associated with the merged cut is the product of the
original twists.\footnote{Summation over the different configurations leading to the same overall cut results in renormalization $N\to\ord{\GG}N$ in the $1/N$ expansion.}
Proceeding in this way, we can merge all the cuts and reduce the sum over the twist factors
to a single set of twists associated to a generating set of non-contractible loops of the worldsheet spanned by the
Feynman diagram.

Note that each operator insertion corresponds to a hole (puncture) on the graph surface.That is why a planar diagram,
that describes the leading order large $N$ limit of amplitudes of some operators
of the orbifold gauge theory (\ref{spin_chain_state}), can be drawn on a cylinder (or a sphere with the two punctures).
In the untwisted sector there is only one non-contractible loop wrapping the cylinder.
Summation over this twist leads to projection onto the $\GG$-invariant states (see Fig.~\ref{fig:projection}).
Hence in this case, the amplitudes of the orbifold coincide with those of the ${\cal N}\!=\! 4$ theory, as advocated. The miraculous
integrability of the ${\cal N}\! =\! 4$ theory therefore directly carries over to the untwisted sector of the orbifold gauge theory, provided it is supersymmetric.
\begin{figure}[htb]
\begin{center}
\epsfig{figure=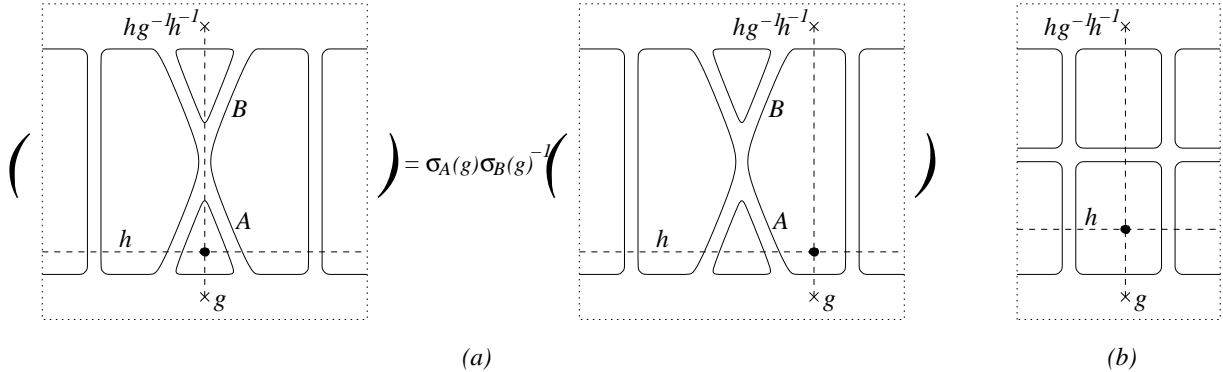,scale=0.6}
\caption{\small $(a)$ A planar diagram on a cylinder. Introduction of the twist field causes appearance of an extra vertical cut in the dual graph (dotted lines). Should this cut be located in the interaction region it can be shifted away using the interchange relation (\ref{twistshift}). Representation matrix $\Rf(g)$ being diagonalized as in (\ref{g_diag}), this shift results in a mere phase factor $\s_A(g) \s_B(g)^{-1}$. Note that the twist field gets conjugated, $g\to hgh^{-1}$; and this conjugation compensates for the action of the horizontal cut owing to the invariance condition (\ref{sinv}). Then the summation over the cut $h$ results in the projection onto the $S_g$-invariant subspace. $(b)$ Diagrams with high number of loops can contain the wrapping interactions which do not reduce to the untwisted case. The diagram shown would be multiplied by an extra factor, the character $\Tr\Rf(g)$ as a result of the horizontal loop wrapping the cylinder.}
\label{fig:cylinder_diag}
\end{center}
\end{figure}

The story with the twisted sectors is slightly more complicated. In terms of the dual graph each twist field can be represented as a tadpole ending in the corresponding puncture. A direct consequence is the fact that the standard form of the dual graph consists of one horizontal cut wrapping the cylinder and one vertical cut corresponding to the twist (Fig.~\ref{fig:cylinder_diag}). However, the extra cut can be moved away from the interaction region using the commutation relation (\ref{twistshift}). After this transformation the graph coincides with that of the $\Nc\!=\!4$ theory modulo renormalization $N\to\ord{\GG}N$ and projection onto the $S_g$-invariant states in (\ref{spin_chain_state}). Unfortunately, this equivalence extends only up to the $\ell<L$ loops. The reason for this restriction is that  the $\ell$-loop gauge theory Hamiltonian translates into a semi-local spin chain Hamiltonian that
connects $\ell+1$ adjacent spins. So when $\ell\geq L$, the Hamiltonian becomes fully delocalized,
and includes the so-called wrapping terms, non-local interactions that wrap around the full length
of the spin chain. When this is the case, the extra cut emerging from insertion of the twist field $\ggamma(g)$ can no longer be shifted away from the interaction region, and some propagators inevitably cross it (Fig.~\ref{fig:cylinder_diag}$b$).

The conclusion is that locally,  on any nearest neighbor set of spins,
the interaction terms in ${\cal H}$ all act identically
to the local interaction terms of the ${\cal N}\!= \!4$ Hamiltonian, as long as the local set of spins does not contain the
twist operator $\ggamma(g)$. If the twist generator is present in the interaction region, one could shift the twist operator
to either side, until it is outside the interaction region. In this way we derive, for example, that the nearest neighbor interaction term, when acting on two spins separated by a twist $\ggamma(g)$,  gets modified via
\be
\label{SpinTwist}
\ham_{[12]}\,\fld{W}_{{}_A}\ggamma(g)\, \fld{W}_{{}_B}\; = \;
\widetilde\ham_{{}_{AB}}^{{}^{\, CD}} \; \fld{W}_{{}_{C}}\ggamma(g) \, \fld{W}_{{}_D},
\ee
where
\be
\label{SpinTwist2}
\widetilde\ham_{{}_{AB}}^{{}^{\, CD}} \; = \;
\ham_{{}_{AB'}}^{{}^{CD'}}\; \Rff(g)_{{}_{\! B}}^{{}^{{}_{\!  B'}}}\; \overline{\Rff}(g )_{{}_{D'}}^{{}^{{}_D}}\,  .
\ee
This relation (and analogous relations for the higher order terms) expresses the $\Gamma$-invariance of the local interaction terms of ${\cal H}$ --- the twist field can be moved either to the left or to the right, which results in the same phase factor.\\

\bigskip
\newsection{Integrability: Orbifolding the Bethe Ansatz}
\label{sec:BAE}

From the standpoint of the string theory dual it is
easy to argue that, in the strong 't~Hooft coupling limit, orbifold
field theories of $\Nsuper 4$ SYM are integrable.
We shall investigate the gauge theory Bethe ansatz for the orbifold model.
First we explore the $\su_2$ subsector consisting of the two scalar fields and explain the derivation of the Bethe equations~\cite{Bet},\cite{Fey} as well as their modification for the orbifold theory.\footnote{There exists a different approach using the $\Rc$-matrices; \eg,~\cite{Fadd}. However, in our exposition we mainly restrict ourselves to the concept of BAE emerging from analysis of multiple scattering waves. The same approach was used in the study of the integrable two-dimensional field theories~\cite{ZZ}.}
%end footnote
Then we extend the complete set of Bethe equations from the $\Nc\!=\!4$ gauge theory to the quotient quiver gauge theory.\\

\medskip
\newsubsection{Bethe Equations: A Brief Introduction}
We will start with the simplest example, the periodic Heisenberg $\su_2$ spin chain of length $L$. Each of the $L$ spins has a two-dimensional space of states $\Cbb^2$ with the basis vectors $\ket{\da}$ and $\ket{\ua}$ corresponding to the spin being oriented downward or upward. (In the field theory language these correspond to the two scalar fields, $Z$ and $W$.) Thus for the whole chain the space of states is $\Cbb^{\,2^L}$. Our goal is to diagonalize the Hamiltonian
\bea
\Hc &=& \sum_{i=1}^L \Bigl(1-\Pb_{i,i+1}\Bigr) \,.
\eea
Here $\Pb_{i,i+1}$ is the interchange operator acting between the $i$-th and the $i+1$-th sites.
It is known that the $\su_2$ subsector consisting of the two scalar fields is closed in $\Nc\!=\!4$ theory, and its matrix anomalous dimension is given exactly by the Heisenberg spin chain Hamiltonian.

First, there exists a vacuum state $\ket{\da\,\da\ldots \da}$ with all spins pointing down. ($\Tr Z^L$ operator in field theory.)
The next step is to consider states with one excitation,
\bea
\ket{n} &=& \ket{\da\,\da\ldots \da\,\ua_n\,\da\ldots \da}
\eea
with the spin up being at the $n$-th position. Acting with the Hamiltonian we get
\bea
\Hc\, \ket{n} &=& -\ket{n-1}+2\ket{n}-\ket{n+1}
\eea
with the identification $\ket{0}\equiv\ket{L}$ and $\ket{L+1}\equiv\ket{1}$.
One can try to find a plane wave solution in the form
\bea
\ket{k} &=& \sum_{n=1}^L \e^{ikn}\, \ket{n} \,.
\eea
Acting with the Hamiltonian and equating the coefficients before a generic $\ket{n}$, we get for the eigenvalue $\eps(k)$
\be
\epsilon\, \e^{ikn} = \Bigl( -\e^{ik}+2-\e^{-ik} \Bigr) \e^{ikn} = 2\, (1-\cos k)\, \e^{ikn}\,,
\ee
thus $\eps(k)=1-\cos k$. In order for these equations to hold for $\ket{0}$ and $\ket{L}$ one has to impose an extra periodicity condition,
\bea
\e^{ikL} &=& 1 \,.
\eea
Physically such a solution corresponds to a standing wave.

One can proceed and introduce states with the two excitations located at positions $n_1$ and $n_2$,
\bea
\ket{n_1,n_2} &=& \ket{\da\,\da\ldots \da\,\ua_{n_1}\,\da\ldots \da\,\ua_{n_2}\,\da\ldots \da}\,.
\eea
The solution can now be found in the form of the two scattering waves,
\bea
\label{two_wave}
\ket{k_1,k_2} &=& \sum_{1\leq n_1<n_2\leq L} \Bigl(\e^{i(k_1 n_1+k_2 n_2)}+S(k_1,k_2)\,\e^{i(k_2 n_1+k_1 n_2)}\Bigr)\, \ket{n_1,n_2} \,.
\eea
Acting with the Hamiltonian and equating the coefficients before generic $\ket{n_1,n_2}$, we get the two non-interacting waves having the energy
\bea
\epsilon(k_1,k_2) &=& \epsilon(k_1) + \epsilon(k_2) \,.
\eea
Coefficients before $\ket{n,n+1}$ are responsible for the interaction, and the corresponding equation determines the scattering phase
\bea
\label{scat_phase}
S(k_1,k_2) \;=\; -\frac{\e^{i(k_1+k_2)}+1-2\e^{ik_2}}{\e^{i(k_1+k_2)}+1-2\e^{ik_1}} \;=\; \frac{\lambda_1-\lambda_2-i}{\lambda_1-\lambda_2+i} \,;
\eea
where we have introduced the rapidity $\l=\half\,\cot\frac{k}{2}$. Note that the momentum and energy in terms of rapidity are
\bea
\e^{ik(\l)} = \frac{\l+\ihalf}{\l-\ihalf} \,, && \eps(\l)= \frac{1}{\l^2+\frac{1}{4}} \,.
\eea

A remarkable feature of this system is its integrability. It manifests itself in the fact that the scattering reduces to the two-particle scattering, and this two-particle scattering is a mere exchange of quantum numbers. The solution (\ref{two_wave}) can be generalized to the states with the $l$ spins up, and it becomes
\bea
\label{n_excitations}
\ket{k_1,k_2,\ldots k_l} &=& \sum_{1\leq n_1<\ldots<n_l\leq L} a_{n_1,n_2,\ldots n_l}(k_1,k_2,\ldots k_l)\, \ket{n_1,n_2,\ldots n_l} \,.
\eea
The corresponding coefficients
\bea
\label{a_coeffs}
a_{n_1,n_2,\ldots n_l}(k_1,k_2,\ldots k_l) &=& \sum_{\sigma\in \Sc_l} S(\sigma,k)\, \exp i[k_{\sigma(1)}n_1+\cdots+k_{\sigma(l)}n_l] \,.
\eea
Here $\Sc_l$ is the group of permutations, and the phase factor $S(\s,k)$ obeys the group property
\bea
S(\s_1 \s_2,k) = S(\s_2,k)\, S(\s_1,\s_2k) \,.
\eea
For the interchange of the two neighboring excitations $\s_{i,i+1}$ the phase factor
\bea
S(\s_{i,i+1},k) &=& S(k_i,k_{i+1})
\eea
reduces to the two-particle scattering phase (\ref{scat_phase}).

Having taken this into account one can easily formulate the periodicity condition for the standing wave:
\bea
a_{n_1,n_2,\ldots n_l}(k_1,k_2,\ldots k_l) &=& a_{n_2,\ldots n_l,n_1+L}(k_2,\ldots k_l,k_1)
\eea
(note the order of momenta $k_i$). Using explicit form of the coefficients $a(k)$ in (\ref{a_coeffs}) and the properties of the phase $S(\sigma,k)$ one finds
\bea
\e^{i k_1 L} \prod_{j\neq 1} S(k_1,k_j) &=& 1 \,.
\eea
(Similar equations hold for the other momenta.) In terms of rapidities $\l_j$ these periodicity conditions read
\bea
\label{BAE_SU2}
\Bigl( \frac{\l_j+\ihalf}{\l_j-\ihalf} \Bigr)^L = \prod_{k\neq j} \frac{\lambda_j-\lambda_k+i}{\lambda_j-\lambda_k-i} \,,\quad  j=1,2,\ldots l \,.
\eea
The set of equations (\ref{BAE_SU2}) is known as the Bethe ansatz equations (BAE).

The zero momentum constraint results in the condition
\bea
\e^{i\sum_j k_j} &=& 1\,;
\eea
equivalently,
\bea
\prod_j \frac{\l_j+\ihalf}{\l_j-\ihalf} &=& 1 \,.
\eea

These results can be generalized to $\su_2$-symmetric chains with higher spins. We give them without derivation. The BAE for a spin $s$ chain read
\bea
\label{BAE_higher_spin}
\Bigl( \frac{\l_j+is}{\l_j-is} \Bigr)^L = \prod_{k\neq j} \frac{\lambda_j-\lambda_k+i}{\lambda_j-\lambda_k-i} \,,\quad  j=1,2,\ldots l \,.
\eea
The energy and momentum
\bea
\e^{iP} = \frac{\l+is}{\l-is} \,, && \eps(\l)= \frac{2s}{\l^2+s^2} \,.
\eea
\\

\medskip
\newsubsection{Bethe Equations for the Orbifold Gauge Theory: $\su_2$ Subsector}
The first step towards the construction of the BAE for the orbifold gauge theory would be to consider a $\su_2$ subsector and go through the construction in detail. As it was argued, interaction terms are unaffected by the orbifoldization procedure except for the interaction between the first and the $L$-th site. This means that the bulk solution (\ref{n_excitations}) will remain unaltered, though the periodicity condition as well as the zero momentum constraint will get modified.

As it was stated in (\ref{g_diag}), the action of the twist field $\ggamma(g)$ can be diagonalized for each given element $g$. With reference to the $\su_2$ subsector it means that the action on the fields $Z$ and $W$ can be brought to the form
\bea
g &:&
\left(\begin{array}{c}
 Z \\ W
\end{array}
\right) \;\to\;
\left(\begin{array}{cc}
 \omega^{s_Z} & 0
\\
0 & \omega^{s_W}
\end{array}
\right) \left(\begin{array}{c}
Z \\ W
\end{array}
\right) \,,
\eea
where $\omega=\sqrt[S]{1}=\e^{2\pi i/S}$; $S$ being the order of the element $g$, \ie, $g^S=1$.
The Hamiltonian $\Hc=\sum_{i=1}^L \Hc_{i,i+1}$, where $\Hc_i = 1-\Pb_{i,i+1}$ for $i=1,\ldots L-1$; while $\Hc_{L,1}$ gets modified according to (\ref{SpinTwist2}). Explicitly it means
\bea
\Hc_{L,1} \ket{Z \ldots Z}_g &=& 0 \,,
\\
\Hc_{L,1} \ket{Z \ldots W}_g &=& \ket{Z\ldots W}_g - \omega^{s_W-s_Z} \ket{W\ldots Z}_g \,,
\\
\Hc_{L,1} \ket{W \ldots Z}_g &=& \ket{W\ldots Z}_g - \omega^{s_Z-s_W} \ket{Z\ldots W}_g \,,
\\
\Hc_{L,1} \ket{W \ldots W}_g &=& 0 \,.
\eea
Note that for almost all the sites the Hamiltonian is unaltered. As it was emphasized in~\cite{BC}, one can use the same ansatz, and the only novelty is the modification of the periodicity conditions.

The simplest way to derive the new periodicity condition is to consider a plane wave solution
\bea
\ket{k}_g &=& \sum_{n=1}^L \e^{ikn} \ket{n}_g \,.
\eea
Acting with the Hamiltonian and equating the coefficients before a generic $\ket{n}_g$, we get
\bea
\eps \e^{ikn} \ket{n}_g &=& \bigl(-\e^{ik}+2-\e^{-ik}\bigr)\, \e^{ikn} \ket{n}_g \,;
\eea
giving the same eigenvalue $\eps(k)=2(1-\cos k)$. Equations for $\ket{1}_g$ and $\ket{L}_g$ have to be considered separately because of the modified action of the Hamiltonian
\bea
\Hc \ket{1}_g &=& -\omega^{s_Z-s_W}\ket{L}_g  + 2\ket{1}_g - \ket{2}_g \,,
\\
\Hc \ket{L}_g &=& -\ket{L-1}_g + 2\ket{L}_g - \omega^{s_W-s_Z}\ket{1}_g \,.
\eea
Then equating the coefficients we get the two equations
\bea
\ket{1}_g &:& \eps \;=\; -\e^{ik} +2 -\e^{ikL}\omega^{s_W-s_Z} \e^{-ik} \,,
\\
\ket{L}_g &:& \eps \;=\; -\e^{-ikL}\omega^{-(s_W-s_Z)}\e^{ik} +2 -\e^{-ik} \,.
\eea
The periodicity condition for a single wave solution then reads
\bea
\e^{ikL} \omega^{s_W-s_Z} &=& 1 \,.
\eea
Analogously, for several interacting waves the BAE read
\bea
\label{BAE_SU2_orbifold}
\e^{iP_j L} \;\equiv\; \Bigl( \frac{\l_j+\ihalf}{\l_j-\ihalf} \Bigr)^L \;=\; \omega^{s_Z-s_W} \prod_{k\neq j} \frac{\lambda_j-\lambda_k+i}{\lambda_j-\lambda_k-i} \,.
\eea

In order to formulate the ``zero momentum constraint'' one has to go back to its physical origin. In the unorbifolded theory the zero momentum condition reflects the cyclicity of the trace. However, in the twisted sectors the identification (\ref{spin_chain_state}) suggests a fixed position of the twist fields $\ggamma(g)$ in the l.h.s.\ and thus breaks the cyclic invariance. Recall that generally the cyclicity condition is to fix a spin chain representative of a given field theory operator in such a way that all the diagrams in Fig.~\ref{fig:projection} are accounted for. This consideration shows that the modified zero momentum constraint should be stated as follows: should one perform a cyclic shift of the fields under the trace (including the twist field) and use the commutation relation (\ref{twistshift}) to move the twist back, the corresponding state would remain invariant. Then the shift operator
\bea
\Uc &:& \ket{A_1\ldots A_L} \to \ket{A_2\ldots A_L A_1}
\eea
in the twisted sector should be modified according to
\bea
\Uc_g &:& \ket{A_1\ldots A_L}_g \to \s_{A_1}(g) \ket{A_2\ldots A_L A_1}_g \,.
\eea
The zero momentum condition is nothing but the $\Uc_g$-invariance.

In the $\su_2$ subsector the action of the shift operator on a state with $l$ excitations becomes
\bea
\Uc_g \ket{n_1,n_2,\ldots n_l}_g &=& \left\{
\begin{array}{lr}
\omega^{-s_Z} \ket{n_1-1,n_2-1,\ldots n_l-1}_g \,,  & n_1\neq 1 \,,
\\
\omega^{-s_W} \ket{n_2-1,\ldots n_L-1,L}_g \,, & n_1=1 \,.
\end{array}
\right.
\eea
Thus the action of the shift operator depends on whether the last spin is up or down. We will see in a moment that this difference is canceled when we take into account the form of the periodicity condition.
Thus taking the solution in the form (\ref{n_excitations}) and acting with the shift operator we have to analyze the terms with $n_1\neq 1$ and $n_1=1$ separately. For $n_1\neq 1$ we get
\bea
\nn \Uc_g &:& \sum_{1< n_1<\ldots<n_l\leq L} a_{n_1,n_2,\ldots n_l}(k_1,k_2,\ldots k_l)\, \ket{n_1,n_2,\ldots n_l}_g
\\
\nn &\to& \omega^{-s_Z} \sum_{1< n_1<\ldots<n_l\leq L} a_{n_1,n_2,\ldots n_l}(k_1,k_2,\ldots k_l)\, \ket{n_1-1,n_2-1,\ldots n_l-1}_g
\\
\nn &=& \omega^{-s_Z} \sum_{1\leq n_1<\ldots<n_l< L} a_{n_1+1,n_2+1,\ldots n_l+1}(k_1,k_2,\ldots k_l)\, \ket{n_1,n_2,\ldots n_l}_g
\\
&=& \omega^{-s_Z} \e^{iP} \sum_{1\leq n_1<\ldots<n_l< L} a_{n_1,n_2,\ldots n_l}(k_1,k_2,\ldots k_l)\, \ket{n_1,n_2,\ldots n_l}_g \,;
\eea
where we have used the fact that for $n_l<L$
\bea
a_{n_1+1,n_2+1,\ldots n_l+1}(k_1,k_2,\ldots k_l) &=& \e^{iP} a_{n_1,n_2,\ldots n_l}(k_1,k_2,\ldots k_l)\,.
\eea
Considering the terms with $n_1=1$ gives
\bea
\nn \Uc_g &:& \sum_{1< n_2<\ldots<n_l\leq L} a_{1,n_2,\ldots n_l}(k_1,k_2,\ldots k_l)\, \ket{1,n_2,\ldots n_l}_g
\\
\nn &\to& \omega^{-s_W} \sum_{1< n_2<\ldots<n_l\leq L} a_{1,n_2,\ldots n_l}(k_1,k_2,\ldots k_l)\, \ket{n_2-1,\ldots n_l-1,L}_g
\\
\nn &=& \omega^{-s_W} \sum_{1\leq n_2<\ldots<n_l< L} a_{1,n_2+1,\ldots n_l+1}(k_1,k_2,\ldots k_l)\, \ket{n_2,\ldots n_l,L}_g
\\
&=& \omega^{-s_Z} \e^{iP} \sum_{1\leq n_2<\ldots<n_l< L} a_{n_2,\ldots n_l,L}(k_2,\ldots k_l,k_1)\, \ket{n_2,\ldots n_l,L}_g \,.
\eea
In the last line we have used the periodicity condition leading to the orbifold BAE (\ref{BAE_SU2_orbifold}),
\bea
a_{1,n_2+1,\ldots n_l+1}(k_1,k_2,\ldots k_l) &=& \omega^{s_W-s_Z} a_{n_2+1,\ldots n_l+1,L+1}(k_2,\ldots k_l,k_1)
\\
\nn &=& \omega^{s_W-s_Z} \e^{iP} a_{n_2,\ldots n_l,L}(k_2,\ldots k_l,k_1) \,.
\eea
The conclusion is that the action of the shift operator on all the terms is the same, and
\bea
\Uc \ket{k_1,k_2,\ldots k_l}_g &=& \omega^{-s_Z} \e^{iP} \ket{k_1,k_2,\ldots k_l}_g \,.
\eea
Finally, the orbifold zero momentum condition in a given twisted sector is
\bea
\omega^{-s_Z} \e^{iP} &=& 1 \,.
\eea
\\

\medskip
\newsubsection{Bethe Equations: Chains with Arbitrary Symmetry Algebrae}
Let us first review integrability for
the full parent $\superN=4$ supersymmetric model.
The algebra behind the $\Nc=4$ supersymmetry is the $\su_{2,2|4}$ superalgebra.
Thus generic operators of the field theory get identified with some states of the $\su_{2,2|4}$-symmetric spin chain. Provided the spin chain is integrable, there exists a generic formulation of the BAE for an arbitrary underlying symmetry algebra.
These BAE were formulated in~\cite{Re} for orthogonal and symplectic algebrae and then generalized to an arbitrary Lie algebrae in~\cite{OW}.
Extension to the superalgebrae case was given in~\cite{Ku},\cite{MR},\cite{Sa}.

Let us introduce the notations regarding generic Lie algebrae. A semisimple Lie algebra $\gf$ has the maximal commuting subalgebra $\hf\subset\gf$ --- the Cartan subalgebra. The number $r=\dim\hf=\rk\gf$ is called the rank of the Lie algebra $\gf$. The adjoint action of $\hf$ on the complement $\hf_\perp$ can be diagonalized, so that for any $h\in\hf$
\bea
\ad{}_h E^\alpha = \alpha(h) E^\alpha \,,
\eea
$\alpha\in\hf^\ast$ being the root and $E^\alpha$ the corresponding root vector. It is known that whenever $\alpha$ is a root, $-\alpha$ is also a root; and that the roots are non-degenerate. The set of roots $\Rf=\bigcup_\alpha\Rf^\alpha$ can be split into the opposite positive and negative roots, $\Rf=\Rf_+\cup\Rf_-$ such that the sum of any two positive (negative) roots is again a positive (negative) root. There exist exactly $r$ simple roots such that any positive root can be obtained as their sum with the positive integer coefficients. Then in the Chevalley-Serre basis,
\bea
\label{Chevalley_Serre}
[H_i,H_j] = 0\,, \quad [H_i,E_j^\pm] = \pm a_{ji} E_j^\pm\,, \quad [E_i^+,E_j^-] = H_i \delta_{ij} \,.
\eea
The root vectors are normalized so that the bilinear form $B(E_i^+,E_i^-)=-1$; equivalently, it means that
$\alpha_i(H)=-B(H_i,H)$ for $H\in\hf$. The coroots $\alpha_i^\vee=2\alpha_i/(\alpha_i,\alpha_i)$, and the Cartan matrix $a_{ij}=(\alpha_i\,\alpha^\vee_j)$ are the elements of the Cartan matrix. Each triple $(H_i,E_i^+,E_i^-)$ forms an $\su_2$ subalgebra; though the raising and lowering operators from the $r$ different subalgebrae do not necessarily commute. Forming all their possible commutators we get the complete basis in $\gf$. There are the additional relations
\be
\ad{}_{E_i^\pm}^{1-a_{ji}} E_j^\pm = 0 \,.
\ee

Similarly, any irreducible representation $V$ can be decomposed into a sum of linear spaces with definite weights; and there exists a unique highest weight for which the sum of the coefficient in its expansion in simple roots is maximal. (Roots can be viewed as weights of the adjoint representation.) Given the vector of the highest weight, one can build the space of a unique irreducible representation by acting on it with the lowering operators $E_i^-$. It is convenient to use the basis of the so-called fundamental weights $\omega_i$ normalized by
\be
\label{omega_normalization}
(\omega_i,\alpha^\vee_j) = \delta_{ij} \,.
\ee
Then any highest weight can be expanded as
\be
\Omega = \sum_i V_i \omega_i\,;
\ee
where the coefficients $V_i$ are called the \emph{Dynkin labels}. Restricting the representation to one of these $\su_2$ subalgebrae $(H_i,E_i^\pm)$, one concludes that the Dynkin labels are to be integer for the finite dimensional representations. (Note the factor of two in the definition of the coroot.) There exist the $r$ distinguished representations with the highest weights $\omega_i$, $i=1,2,\ldots r$, called the fundamental representations. For the $i$-th fundamental representation the Dynkin labels $V_j=\delta_{ij}$. Since tensoring the two vectors results in addition of their weights, any irreducible representation can be extracted from the Clebsch-Gordan  series of the tensor product of some fundamental representations. In particular, the representation  with the highest weight $n\omega_i$ can be obtained as the $n$-th symmetric power of the $i$-th fundamental representation.

Now one can formulate the set of the BAE for a generic\footnote{The word ``generic'' here refers to the generic symmetry algebra; of course, the integrability condition fixes the form of the Hamiltonian.} $\gf$-symmetric spin chain. There exist the $r$ types of excitations, corresponding to the $r$ simple roots. Since there can be multiple excitations of the same type it is convenient to number the corresponding spectral parameters as $\l_{j,k}$; where $j=1,2,\ldots r$ and $k=1,2,\ldots K_j$, $K_j$ being the number of excitations of type $j$. The set of the BAE becomes
\bea
\label{BAE_generic}
\e^{iP_{j,k}L} &=& \prod_{(j^\prime,k^\prime)\neq (j,k)} S_{jj^\prime}(\l_{j,k},\l_{j^\prime,k^\prime}) \,;
\eea
where the scattering matrix
\bea
S_{jj^\prime} &=& \frac{\l_{j,k}-\l_{j',k'}+\ihalf a_{j,j'}}{\l_{j,k}-\l_{j',k'}-\ihalf a_{j,j'}} \,.
\eea
Momentum
\bea
\e^{iP_{j,k}} &=& \frac{\l_{j,k}+\ihalf V_j}{\l_{j,k}-\ihalf V_j} \,.
\eea
(Here $V_j$ are the Dynkin labels of the representation via which the algebra acts on each site --- twice the spin in the $\su_2$ case.) The total energy of the corresponding eigenstate
\bea
\eps \;=\; \sum_{j=1}^r \sum_{k=1}^{K_j} \eps_j(\l_{j,k})\,, & & \eps_j(\l_{j,k}) \;=\; \frac{V_j}{\l_{j,k}^2+\frac{1}{4}V_j^2} \,.
\eea

The superconformal symmetry algebra of the $\Nc\!=\!4$ SYM theory is the $\su_{2,2|4}$ superalgebra.
Its bosonic subalgebrae $\su_{2,2}$ and $\su_4$ are nothing but the algebra of the conformal group in four dimensions and the $\Rc$-symmetry algebra. Unlike the bosonic semisimple Lie algebrae, the Dynkin diagram of a superalgebra is not unique. For the $\su_{2,2|4}$ there exist the two distinguished choices of the root system, the so-called ``Beauty'' and the ``Beast''. Though the ``Beast'' is the most obvious system with one fermionic root, the ``Beauty'' root system proves useful in the context of $\Nc\!=\!4$ supersymmetry. Exact embedding of these root systems into $\su_{2,2|4}$ as well as the corresponding Cartan matrices is shown in Table~\ref{tab:root_systems}.
More details regarding these root systems are given in Appendix~\ref{app:root_systems}.
The whole $\Nc\!=\!4$ theory was proved to be integrable~\cite{BKS},\cite{Be},\cite{BS}. Note that the spin chains with the different representations of $\su_{2,2|4}$ correspond to different subclasses of operators.\\

\medskip
\newsubsection{Generalization to General Orbifolds}
As it was argued, there is no mixing between the different twisted sectors. Furthermore, in each given twisted sector $[g]$ one can construct all the states inserting one twist field $\ggamma(g)$, $g$ being any fixed representative of the conjugacy class $[g]$.
In conjunction with the result (\ref{g_diag}) --- the fact that one can diagonalize the action of any given element $g\in\GG$ --- the problem reduces to the Abelian case modulo some subtleties.
In particular, the $S_g$-invariance does not completely incorporate into Bethe equations; and it is to be imposed by hand --- that is why some of the Bethe eigenstates may be projected out.

Therefore, one can apply techniques similar to those used in~\cite{BR} for the study of Abelian orbifolds.
Then each given element $g\in\GG\subset \SU_4$ can be brought to the diagonal form so that in $\SU(4)$ it becomes
\bea
\Rf(g) &=& \left(\begin{array}{cccc}
\e^{-2\pi it_1/S}& & & 0 \\
& \e^{2\pi i(t_1-t_2)/S} & & \\
& & \e^{2\pi i(t_2-t_3)/S} & \\
0 & & & \e^{2\pi i t_3/S}
\end{array}\right) \,.
\eea
Here $S$ is the order of the element $g$, \ie, $g^S=1$. For supersymmetric orbifolds that we consider the group $\GG$ embeds into $\SU(3)\subset\SO(6)\simeq\SU(4)/\Zbb_2$, and in this case
\bea
\Rf(g) &=& \left(\begin{array}{ccc}
\e^{-2\pi i(t_1-t_3)/S}& & 0 \\
& \e^{2\pi i(t_1-t_2+t_3)/S} & \\
0 & & \e^{2\pi it_2/S} \\
\end{array}\right) \,.
\eea
(Even though we need only the two independent parameters in order to describe embedding $\GG\subset \SU(3)$, and here $\e^{2\pi it_3/S}=\pm 1$; it is useful to keep all the three parameters $t_1$, $t_2$ and $t_3$ in the future calculations. In particular, it may account for different embeddings $\SU(3) \subset \SU(4)$ or different choices of the vacuum state.)

The Bethe equations (\ref{BAE_SU2_orbifold}) generalize to the complete $\su_{2,2|4}$ algebra:
\bea
\label{BAE_orbifold}
\Bigl( \frac{\l_{j,k}+\ihalf V_j}{\l_{j,k}-\ihalf V_j} \Bigr)^L &=& \Rf_j(g) \prod_{(j^\prime.k^\prime)\neq (j,k)} \frac{\l_{j,k}-\l_{j',k'}+\ihalf a_{j,j'}}{\l_{j,k}-\l_{j',k'}-\ihalf a_{j,j'}} \,.
\eea
Similarly, the momentum constraint reads
\bea
\label{momentum_constraint_orbifold}
\Rf_0(g) \prod_{j=1}^7 \prod_{k=0}^{K_j} \frac{\l_{j,k}+\ihalf V_j}{\l_{j,k}-\ihalf V_j} &=& 1 \,.
\eea
The phase factors
\bea
\Rf_j(g) &=& \e^{2\pi i q_j/S} \,,
\eea
where the integers $q_j$ depend on the choice of the root system:\footnote{Conventions regarding the choice of the root system as well as the Cartan matrix are summarized in Appendix~\ref{app:root_systems}.}
\newcommand{\join}{\!\!\frac{\qquad\quad}{\qquad\quad}\!\!}
\bea
\text{``Beauty''} &:& \stackrel{-t_2}{\odot} \quad \stackrel{0}{\ominus}\join\! \stackrel{-t_1}{\otimes}\!\join\!\!\!\! \stackrel{2t_1-t_2}{\oplus}\!\!\!\!\join \!\!\!\!\!\!\!\! \stackrel{2t_2-t_1-t_3}{\oplus}\!\!\!\!\!\!\!\!\join\!\!\!\! \stackrel{2t_3-t_2}{\oplus}\!\!\!\!\join \stackrel{t_3}{\otimes}\!\join \stackrel{0}{\ominus}
\\
\text{``Beast''} &:& \stackrel{0}{\odot} \quad \stackrel{0}{\oplus}\join \stackrel{0}{\oplus}\join \stackrel{0}{\oplus}\join \stackrel{t_1}{\otimes}\join \!\!\!\!\stackrel{t_2-2t_1}{\ominus}\!\!\!\!\join \!\!\!\!\!\!\!\!\stackrel{t_1-2t_2+t_3}{\ominus}\!\!\!\!\!\!\!\!\join\!\!\!\! \stackrel{t_2-2t_3}{\ominus}
\eea
(The leftmost ``root'' corresponds to the phase $\Rf_0(g)=\e^{2\pi i q_0/S}$.)
Let us stress that this structure is the direct generalization of that in the $\su_2$ subsector: the bulk ansatz remains unaltered, while the boundary conditions get modified. Recall that in the $\su_2$ case there is a single root $\gamma_1=\a_{12}$, and the weight $q_1=s_W-s_Z\equiv s_2-s_1$.
Analogously, for an excitation associated with some simple root $\gamma=\a_{ij}$ the corresponding weight $q_\gamma=s_j-s_i$ is the difference of the two corresponding charges. The number $q_0$ is determined by the choice of the Bethe vacuum.

There is an elegant way to summarize all the Bethe equation and momentum constraint together. In order to do this one introduces the two new types of excitations to the existing seven types ($j=1,\ldots,\rk \su_{2,2|4}=7$).
The quasi-excitation of type $j=0$ corresponds to the insertion of
a new spin chain site. In order to have a length $L$ chain one is to insert exactly the $K_0=L$ excitations of type $0$.
The quasi-excitation of type $j=-1$ corresponds to the twist field.
The scattering phases of the excitations\footnote{Note that the scattering phase $S_{-1,-1}$ is not needed as we restrict ourselves to one excitation of type $-1$. Even though one may introduce several such excitations it would cause some unnecessary technical difficulties. As it was shown, insertion of a single twist field suffices to produce all the orbifold states.}
\bea
S_{j,j'} &=& \frac{\l_{j,k}-\l_{j',k'}+\ihalf a_{j,j'}}{\l_{j,k}-\l_{j',k'}-\ihalf a_{j,j'}}\,, %\quad j=1,\ldots,7 \,;
\\
S_{j,0} &=& \frac{\l_{j,k}-\ihalf V_j}{\l_{j,k}+\ihalf V_j}\,, \quad S_{j,-1} \;=\; \Rf_j(g) \,;
\\
S_{0,0} &=&1\,, \quad S_{0,-1} \;=\; \Rf_0(g) \,.
\eea
Type $0$ excitation do not have an associated spectral parameter,
while type $-1$ excitations can have different twist classes $[g]$.
Excitations of both type $0$ and $-1$ do not contribute to the total energy.

With these notations we can therefore summarize all the Bethe equations as
\be
\label{BAE_orbifold_all}
\mathop{\prod_{j'=-1}^J\prod_{k'=1}^{K_{j'}}}_{(j',k')\neq(j,k)}
S_{j,j'}(\l_{j,k},\l_{j',k'}) =1 \,.
\ee
The equations for $j=1,\ldots,7$ give the BAE (\ref{BAE_orbifold}), equation for $j=0$ gives the momentum constraint (\ref{momentum_constraint_orbifold}),\footnote{Although there are $L$ quasi-excitations of type $0$,
there is only one corresponding Bethe equation, because
all of these quasi-excitations are equivalent, and they have no
spectral parameter which might distinguish them.} and equation for $j=-1$ gives the ``zero charge condition''
\bea
\Rf_0(g)^L \prod_{j^\prime=1}^7 \Rf_{j^\prime}(g)^{K_{j^\prime}} \,.
\eea
It implies the $g$-invariance of the corresponding state in the field theory.
Again, let us stress that for a generic orbifold this condition is not sufficient, and there should be imposed a more restrictive invariance condition w.r.t.\ the full stabilizer $S_g$. That is why some of the Bethe eigenstates may be projected out in field theory.\\

\bigskip
\newsection{Applications of the BAE}
\label{sec:BAE_apps}
Now we study some application of the Bethe equations.
First we show that in the large $L$ limit one can reproduce the known results for the $\su_2$ subsector, which adhere to the closed string dual.
Next we consider particular quivers, ones with both Abelian and non-Abelian orbifold group. For these simple examples one can easily determine the anomalous dimensions of operators in the twisted sectors.
Then these operators can be recast into the quiver notation.\\

\medskip
\newsubsection{Comparison with the Closed String Sector}

It is known that the Bethe equations can be solved in the limit of the long spin chain, $L\gg 1$. Another restriction is $L\ll N$, which prevents the number of the non-planar graphs from growing. In this limit one can perform a rescaling of the Bethe roots, $\l_j\to L\l_j$~\cite{LZ}.

Then the momenta and energies of individual excitations
\bea
P_{j,k} &=& -i\log \frac{L\l_{j,k}+\ihalf V_j}{L\l_{j,k}-\ihalf V_j} \;\approx\; \frac{V_j}{L\l_{j,k}} \quad \eps_{j,k} \;\approx\; \frac{V_j}{L^2 \l_{j,k}^2} \,.
\eea
Using the same approximation, eqs.~(\ref{BAE_orbifold}) and (\ref{momentum_constraint_orbifold})
after taking the logarithm become\footnote{Here we will ignore the possibility of some states being projected out.}
\bea
\frac{V_j}{\l_{j,k}} &=& 2\pi\bigl(m_j+\frac{q_j}{S}\bigr) + \frac{1}{L} \sum_{(j^\prime,k^\prime)\neq(j,k)} \frac{a_{jj^\prime}}{\l_{j,k}-\l_{j^\prime,k^\prime}} \,,
\\
0 &=& 2\pi \bigl(m_0+\frac{q_0}{S}\bigr) + \frac{1}{L} \sum_{j=1}^7 \sum_{k=1}^{K_j} \frac{V_j}{\l_{j,k}} \,.
\eea

It is well known how to solve these equations when there are present excitations of only one type. It is convenient to define the resolvent (we drop the index $j$ of the eigenvalues since they are all of the same type)
\begin{equation}
 G(x)=\frac{1}{L} \sum_{k=1}^K \frac{1}{x-\l_k}.
\end{equation}
In terms of the resolvent the total momentum and total energy (anomalous
dimension) of the spin chain are
\begin{equation}
 P = -V_j\,G(0)\,, \quad \Delta = -\frac{\lambda V_j}{L} G'(0) \,.
\end{equation}
The Bethe equations are
\bea
\frac{V_j}{\l_{k}} &=& 2\pi\mt_j + \frac{2}{L} \sum_{l\neq k} \frac{1}{\l_{k}-\l_l} \,,
\\
0 &=& 2\pi \nt + \frac{1}{L} \sum_{k=1}^{K} \frac{V_j}{\l_{k}} \,;
\\
\label{mn_def} \mt&=&m_j+q_j/S\,, \qquad \nt\;=\;m+q_0/S \,.
\eea
The standard trick is to multiply the first equation with $\frac{1}{x-\l_k}$ and sum over $k$. After performing some useful transformations,
$$
\sum_k \frac{1}{\l_k(x-\l_k)} = \frac{1}{x} \sum_k \Bigl[ \frac{1}{\l_k}+\frac{1}{x-\l_k} \Bigr] = L\, \frac{G(0)+G(x)}{x} = -\frac{2\pi\nt}{x}+ \frac{L\,G(x)}{x} \,,
$$
and
$$
\mathop{\sum_{k,l}}_{k\neq l} \frac{2}{(x-\l_k)(\l_k-\l_l)} = \mathop{\sum_{k,l}}_{k\neq l} \frac{1}{\l_k-\l_l}\, \Bigl( \frac{1}{x-\l_k} - \frac{1}{x-\l_l} \Bigr)  = \mathop{\sum_{k,l}}_{k\neq l} \frac{1}{(x-\l_k)(x-\l_l)}
$$
$$
= \sum_{k,l} \frac{1}{(x-\l_k)(x-\l_l)} - \sum_k \frac{1}{(x-\l_k)^2} = L^2\, G(x) - L\, G^\prime(x) \,;
$$
the Bethe equations take the form
\bea
xG^2(x) + (2\pi\mt x-V_j) G(x) + 2\pi\nt V_j &=& \frac{x}{L}\, G^\prime(x) \,.
\eea
In the large $L$ limit the term in the r.h.s.\ can be dropped, and there remains a mere quadratic equation on $G(x)$. The solution is
\bea
G(x) &=& \frac{V_j-2\pi\mt x + \sgn\mt\, \sqrt{(2\pi\mt x-V_j)^2-8\pi\nt V_j x}}{2\,x} \,.
\eea
The choice of the solution is determined by the correct asymptotics at infinity,
\bea
G(x) &\sim& -\frac{\nt V_j}{\mt x} \;=\; \frac{K_j}{Lx}\,, \qquad x\gg 1 \,.
\eea
(Recall that $K_j$ is the number of excitations --- that is why for large $x$ we write $G(x)\sim \frac{K_j}{Lx}$.)
This imposes the requirement $\mt\nt<0$.
In its turn the regularity of $G$ at zero requires $\mt<0$; then $\nt>0$.
Expanding the resolvent at $x=0$ yields
\bea
\label{mn_KL}
G(x) &\sim& 2\pi\nt x - \frac{4\pi^2}{V_j}\, \nt(|\mt|-\nt)x \,, \qquad x\ll 1 \,.
\eea
Then the anomalous dimension is
\bea
 \Delta = \frac{4\pi^2\lambda}{L}\, \nt\, (|\mt|-\nt) \,.
\eea
Recalling the definitions of $\mt$ and $\nt$ (\ref{mn_def}), it can be rewritten as
\be
\Delta = \frac{4\pi^2\lambda}{L}\, \Bigl(m+\frac{q_0}{S}\Bigr) \Bigl(m^\prime-\frac{q_0+q_j}{S}\Bigr) \,,
\ee
where $m$ and $m^\prime$ are some positive integers.
In the corresponding scalar subsector formed by the two fields $(Z,W)$ the corresponding weights $q_0=-s_Z$ and $q_j=s_Z-s_W$; then it gives
\be
\Delta = \frac{4\pi^2\lambda}{L}\, \Bigl(m-\frac{s_Z}{S}\Bigr) \Bigl(m^\prime+\frac{s_W}{S}\Bigr) \,,
\ee
which matches the closed string energy (\ref{string_energy}).

Note that using (\ref{mn_KL}) one can express $\nt=K_j|\mt|/L$. Introducing the filling fraction
\bea
\a_j &\equiv& \frac{K_j}{L} \,,
\eea
one can rewrite the anomalous dimension as
\bea
\Delta = \frac{4\pi^2\lambda}{L}\, \a_j (1-\a_j) \Bigl(m_j+\frac{q_j}{S}\Bigr)^2 \,, \qquad m_j\in\Zbb \,.
\eea
It is worth mentioning that the same technique can be applied when there are the excitations of different types which are not directly coupled.
We get the anomalous dimension as
\bea
\Delta = \frac{4\pi^2\lambda}{L}\sum_j \a_j (1-\a_j) \Bigl(m_j+\frac{q_j}{S}\Bigr)^2 \,.
\eea
The condition of such decoupling of different types of excitations is that the corresponding nodes in the Dynkin diagram are not connected.
\\

\medskip
\newsubsection{An Example: Abelian $\Zbb_6$ Quiver}

\begin{figure}[htb]
\begin{center}
\begin{tabular}{ccc}
\includegraphics[scale=0.5,angle=0]{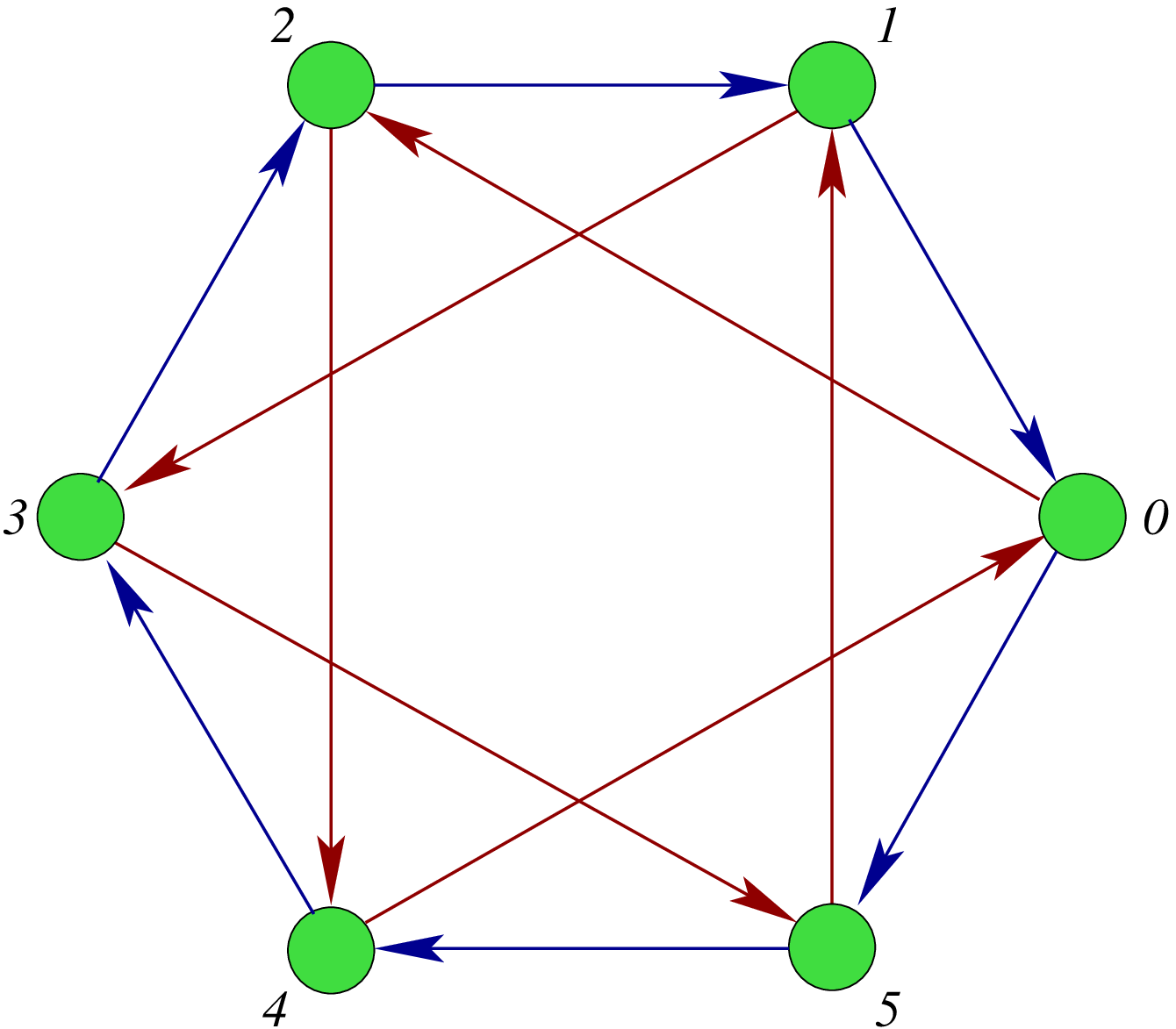} & \quad &
\includegraphics[scale=0.5,angle=0]{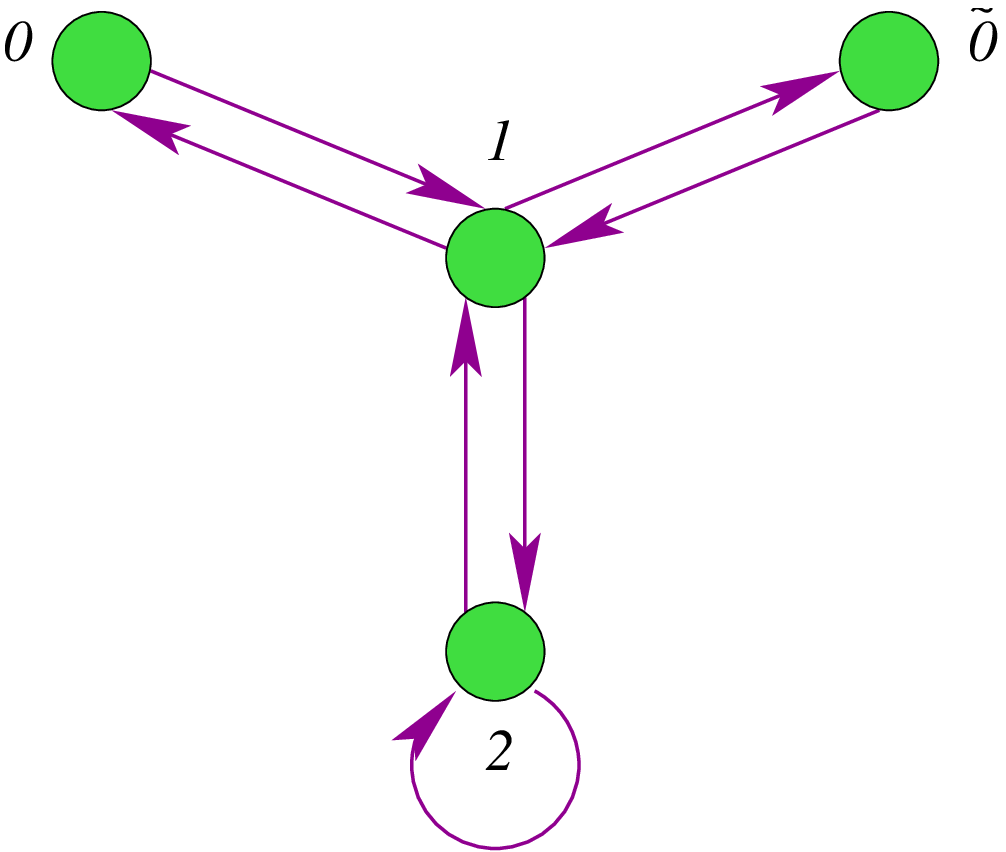} \\
$(a)$ & & $(b)$
\end{tabular}
\caption{\small $(a)$~The $\Zbb_6$ quiver. There are the six nodes corresponding to the six representations of $\Zbb_6$. We show only the scalar lines corresponding to the fields $Z$ (blue lines) transforming in $\Rf_Z\simeq \rrho_1$ and $W$ (red lines) transforming in $\Rf_W\simeq \rrho_4\simeq \rrho_{-2}$. $(b)$~The $D_5$ quiver with the two-dimensional defining representation $\Rf\simeq\rrho_1$. Note that for these two quivers we show only the lines corresponding to the $\su_2$ subsector; \ie, the two scalar fields.}
\label{fig:quivers}
\end{center}
\end{figure}

Here we consider a simple example, $\Zbb_6$ quiver (see Fig.~\ref{fig:quivers}). We restrict ourselves to the $\su_2$ subsector formed by the two scalars, $Z$ with charge $s_Z=1$ and $W$ with charge $s_W=-2$. We will study the twisted sector with the twist $\ggamma^n$, $n=0,\ldots,5$; $\ggamma$ being the generating element of $\Zbb_6$. Let us choose the length of the spin chain $L=3$; then the vacuum can be chosen as $\Tr \bigl[\ggamma ZZZ\bigr]$ --- note that it will be projected out. There also exist the excited states with one or three $W$'s, while the states with the two excitations will also be projected out.
Our goal will be to describe these Bethe vectors in terms of the quiver notation. By $\Oc_{ijki}\equiv \Tr \Phi^i_j \Phi^j_k \Phi^k_i$ we will denote the quiver gauge theory operator corresponding to the closed cycle between the three nodes $k\to j\to i$ in the quiver.

Note that the state with the three excitations is unique in each given twisted sector, and it corresponds to the field theory operator $\Tr \bigl[ \ggamma^n WWW \bigr ]$. Commuting the twist field $\ggamma^n$ with one of the fields $W$ we find that
\be
\Tr \bigl[ \ggamma^n WWW \bigr ] = \e^{2\pi i n s_W/6} \Tr \bigl[ \ggamma^n WWW \bigr ] \,;
\ee
i.e., this state is projected out in all sectors except for $n=0$ and $n=3$.
The reason for this is the extra symmetry: it is sufficient to commute the twist field with only one of the three $W$ fields. Note that
the total  charge of the three fields $W$ is zero, and normally one would expect $\Tr\bigl[\ggamma WWW\bigr]$ to be a non-trivial operator.

Using the formula (\ref{transition_formula}) we find that
\bea
\Oc_{0420} &=& \Tr \bigl[ WWW \bigr ] + \Tr \bigl[ \ggamma^3 WWW \bigr ] \,,
\\
\Oc_{1531} &=& \Tr \bigl[ WWW \bigr ] - \Tr \bigl[ \ggamma^3 WWW \bigr ] \,;
\eea
or
\bea
\Tr \bigl[ WWW \bigr ] &=& \half\Oc_{0420} + \half\Oc_{1531} \,,
\\
\Tr \bigl[ \ggamma^3 WWW \bigr ] &=& \half\Oc_{0420} - \half\Oc_{1531} \,.
\eea
Graphically the operators $\Oc_{042}$ and $\Oc_{153}$ correspond to the two closed triangles formed by the red lines.
Applying the Hamiltonian we find the anomalous dimensions
\be
\Delta_{\Tr \left[ \mathstrut{\ggamma^3} WWW \right ]}=\Delta_{\Tr \left[ \ggamma^3 WWW \right ]}=0 \,.
\ee

The states with one excitation have the form $\Tr \bigl[ \ggamma^n ZZW \bigr ]$, and there is one such state in each given twisted sector. These operators correspond to the triangles with the two blue (field $Z$) and one red (field $W$) line. There are six such triangles and there are six different operators with $n=1,\ldots,5$ --- these numbers coincide as we expect. The transition formula between these two descriptions is
\bea
\Oc_{l,\, l+1,\, l+2,\,l} &=& \sum_{n=0}^5 \e^{-2\pi i \frac{ln}{6}} \Tr \bigl[ \ggamma^n ZZW \bigr ] \,;
\eea
performing the Fourier transform yields
\bea
\Tr \bigl[ \ggamma^n ZZW \bigr ] &=& \frac{1}{6} \sum_{l=0}^5 \e^{2\pi i \frac{ln}{6}} \Oc_{l,\, l+1,\, l+2,\,l} \,.
\eea
These operators diagonalize the matrix of anomalous dimensions.
Direct application of the Hamiltonian shows that the corresponding eigenvalues are
\bea
\Delta_{\Tr \left[ \ggamma^n ZZW \right ]} &=& 4\l \sin^2\frac{\pi n}{6} \,.
\eea
This simple example illustrates the interrelation of the two descriptions in the orbifold gauge theory.
First, the quiver description gives a very clear understanding of what the physical fields and gauge invariant operators are, while in the ``orbit'' description using the twist fields some of the operators may be projected out.
On the other hand, the description using the twist fields proves to be more robust for studying the field theory dynamics (the matrix of anomalous dimensions).
In order to illustrate this let us write the part of the action responsible for the non-trivial part of the interaction Hamiltonian, $\Tr\bigl[ ZWZ^\dagger W^\dagger + WZW^\dagger Z^\dagger \bigr]$. In terms of the quiver notation
\bea
\Tr\bigl[ ZWZ^\dagger W^\dagger + WZW^\dagger Z^\dagger \bigr] &=& \sum_l \Bigl[ \Oc_{l,l+1,l-1,l-2,l} + \Oc_{l,l-2,l-1,l+1,l} \Bigr]
\\ \nn
&=& \sum_l \Tr\Bigl[ Z^l_{l+1}\, W^{l+1}_{l-1}\, Z^{l-2\,\dagger}_{l-1}\, W^{l-2\,\dagger}_l + W^{l}_{l-2}\, Z^{l-2}_{l-1}\, W^{l+1\,\dagger}_{l-1}\, Z^{l\,\dagger}_{l+1} \Bigr] \,.
\eea
Here $Z^k_l$ denotes the field corresponding to the quiver arrow going from node $l$ to node $k$. Note that the conjugation changes the direction of the corresponding arrow; \eg, $Z^{1\,\dagger}_2$ is an arrow going from node $1$ to node $2$.
Indeed, as we see, studying the matrix of anomalous dimensions using the quiver notation would have been more complicated.\\

\newsubsection{An Example: non-Abelian $D_5$ Quiver}
Next we consider a simple orbifold with a non-Abelian discrete group $D_5$ (the facts about the dihedral group $D_S$ as well as its representation ring are given in Appendix~\ref{app:dihedral_group}.) The corresponding quiver is shown in Fig.~\ref{fig:quivers}. Again, we study the $\su_2$ sector, and the scalar field $\Phi^\II$ transforms in the two-dimensional representation $\Rf\simeq\rrho_1$. From the quiver representation it is clear that there are the four different operators of length $L=2$; namely, those are
\be
\Oc_{010}\,, \quad \Oc_{\tilde 0 1\tilde 0}\,, \quad \Oc_{121}\,, \quad \Oc_{222} \,.
\ee
On the other hand, there are the four different twist classes, $\{[e]$, $[r]$, $[r^2]$, $[\s]\}$. Applying the definitions of the operators (\ref{gen_operator}), we see that in each twist class there is exactly one non-trivial operator; thus there are the total of four operators of length two:
\be
\Oc_e=\Tr\bigl[ZW\bigr]\,, \;\; \Oc_r=\Tr\bigl[\ggamma(r)ZW\bigr]\,, \;\; \Oc_{r^2}=\Tr\bigl[\ggamma(r^2)ZW\bigr]\,, \;\; \Oc_\s=\Tr\bigl[\ggamma(\s)ZZ\bigr]\,.
\ee
Here $Z$ and $W$ denote the first and second components of the field $\Phi^\II$.
Note that the product of the two fields $ZZ$ has transforms non-trivially under the action of $r$; nevertheless, in the sector with twist $[\s]$ the operator $\Oc_\s=\Tr\bigl[\ggamma(\s)ZZ\bigr]$ is non-trivial as $r\!\not\in\!S_\s$.
The absence of mixing between the different twist classes ensures that the set of operators $\{\Oc_e$, $\Oc_r$, $\Oc_{r^2}$, $\Oc_\s\}$ diagonalize the matrix of anomalous dimensions.
Acting with the Hamiltonian we find the corresponding anomalous dimensions as
\be
\Delta_{\Oc_e}=0 \,, \quad \Delta_{\Oc_r}=4\l\sin^2\frac{\pi}{5}=\frac{5-\sqrt{5}}{2}\,\l\,, \quad \Delta_{\Oc_{r^2}}=4\l\sin^2\frac{2\pi}{5}=\frac{5+\sqrt{5}}{2}\,\l\,, \quad \Delta_{\Oc_\s}=0\,.
\ee

The same eigenvalues can be obtained solving the Bethe equations.  The three operators $\Oc_e$, $\Oc_r$ and $\Oc_{r^2}$ are the states with one excitation. Diagonalizing the twist field as
\be
\gamma(g) = \left(
\begin{array}{cc}
\e^{i\a} & 0 \\
0 & \e^{-i\a}
\end{array}
\right) \,, \qquad g=e,r,r^2 \,;
\ee
we find that the Bethe equation and the momentum constraint reduce to
\be
\frac{\l+\ihalf}{\l-\ihalf} = \e^{i\a}\,,\qquad \eps=\frac{1}{\l^2+\frac{1}{4}} \,.
\ee
This gives
\be
\l=\half\cot\frac{\a}{2} \,, \qquad \eps=4\sinh^2\frac{\a}{2} \,.
\ee
For the twist element $g=e$, $r$, $r^2$ we have $\a=0$, $2\pi/5$, $4\pi/5$ correspondingly. This reproduces the correct result.
The twist field $\ggamma(g)$ is non-diagonal.
After the diagonalization of $\ggamma(\s)$ operator $\Oc_\s$ maps to the vacuum state, and that is why $\Delta_{\Oc_\s}=0$.

The next step is to find the dictionary between the two notations. In order to do this one can start with the quiver notation and rewrite the corresponding operators using the transition rules (\ref{loop_operator}) and (\ref{inv_tensor_loop}) from Appendix~\ref{app:twisted_observables}.
The two operators $\Oc_{010}$ and $\Oc_{\tilde 01\tilde 0}$ correspond to the closed paths $\rrho_0\leftarrow \rrho_1\leftarrow \rrho_0$ and $\rrho_{\tilde 0}\leftarrow \rrho_1\leftarrow \rrho_{\tilde 0}$. Since the representations $\rrho_0$ and $\rrho_{\tilde 0}$ are one-dimensional, the corresponding invariant tensors
\be
\Kc_{AB}{}_1^1 = \Kc_{AB}
\ee
(the indices $A$, $B$ belong to the defining representation $\Rf\simeq \rrho_1$.)
The non-zero components are
\bea
\Kc_{12} = \Kc_{21} = \frac{1}{\sqrt{2}}
\eea
(note that the normalization respects the unitarity condition.)
Then
\be
\Kc_{AB}(g) = \Kc_{AB} \repconj[]{\l}{g}\,, \qquad \l=0,\tilde{0}\,.
\ee
This gives
\bea
\nn \Oc_{010} &=& \sqrt{2} \Tr \Bigl[ ZW + (1+\omega) \ggamma(r) ZW + (1+\omega^2) \ggamma(r^2) ZW + 5\ggamma(\s)ZZ \Bigr]
\\
&=& \sqrt{2} \Bigl[ \Oc_e + (1+\omega) \Oc_r + (1+\omega^2) \Oc_{r^2} + 5\Oc_\s \Bigr] \,;
\\
\nn \Oc_{\tilde 01\tilde 0} &=& \sqrt{2} \Tr \Bigl[ ZW + (1+\omega) \ggamma(r) ZW + (1+\omega^2) \ggamma(r^2) ZW - 5\ggamma(\s)ZZ \Bigr]
\\
&=& \sqrt{2} \Bigl[ \Oc_e + (1+\omega) \Oc_r + (1+\omega^2) \Oc_{r^2} - 5\Oc_\s \Bigr] \,.
\eea
(We have used the permutation relation (\ref{perm_rel}).)

Next, $\Oc_{121}$ corresponds to the product of the two tensors,
\be
\label{tensor_O12}
\Kc_{AB}{}^k_l = \sum_{p\in\rho_2} \Kc^p_{Al}\, \Kc^k_{Bp} \,, \qquad k,l\in\rrho_1 \,.
\ee
The non-trivial coefficients corresponding to the decomposition $\Rf\otimes \rrho_1\to \rrho_2$ are $\Kc^1_{11}=\Kc^2_{22}=1$, while those corresponding to the decomposition $\Rf\otimes \rrho_2\to \rrho_1$ are $\Kc^1_{21}=\Kc^2_{12}=1$. This gives the corresponding invariant tensor in (\ref{tensor_O12}):
\be
\Kc_{12}\,{}^1_1 = \Kc_{21}\,{}^2_2 = 1\,.
\ee
Therefore, one identifies
\bea
\Oc_{121} &=& 2\bigl[ \Oc_e + (\omega^2+\omega^4)\Oc_r + (\omega^3+\omega^4)\Oc_{r^2} \bigr] \,.
\eea

Similarly, for the operator $\Oc_{222}$ we need to find the decomposition $\Rf\otimes\rrho_2\to\rrho_2$. The non-trivial coefficients are $\Kc^2_{11}= \Kc^1_{22}=1$. Consequently,
\be
\Kc_{12}\,{}^1_1 = \Kc_{21}\,{}^2_2 = 1
\ee
and
\bea
\Oc_{222} &=& 2\bigl[ \Oc_e + 2\omega^3\Oc_r + (1+\omega)\Oc_{r^2} \bigr] \,.
\eea

These formulae give the transition between the two bases in the operator space.
Again, the conclusion is that generally operators corresponding to the closed paths in the quiver are \emph{not} the eigenvectors of the matrix of anomalous dimensions. In other words, an operator corresponding to a closed loop in the quiver is typically a mix of operators with different conformal dimensions; neither does it belong to a given twisted sector.\\

\bigskip
\newsection{Concluding Remarks}
As we have seen, methods of studying the Abelian orbifold gauge theories can be extended to arbitrary non-Abelian setups with minor modifications. The key argument is that the diagonalization of the twist field in each given twisted sector allows one to apply the techniques used for the Abelian case. Indeed, in a given twisted sector $[g]$ the BAE reduce to those in the Abelian theory; though some of the states may still be projected out. As a general rule, which states survive the projection is determined by the invariant tensors of the stabilizer subgroup $S_g$; although there can be present extra symmetries projecting out some of the conceivably non-trivial states.
A useful feature is that there is no mixing between the different twisted sectors; and this superselection rule simplifies diagonalization of the matrix of anomalous dimensions.
On the other hand, one can use the quiver gauge theory notation. In this language the problems with some states being projected out do not appear, but the matrix of anomalous dimensions becomes more complicated.

It seems that the higher loop techniques described in~\cite{BR}, \cite{BR2} apply to the non-Abelian case as well. This would open a possibility of applying the existing powerful integrability techniques to the quiver gauge theories with reduced supersymmetry. This will be done elsewhere.
\\

\bigskip
\vglue .0cm \pagebreak[3]
\noindent{\large \bf Acknowledgements}\nopagebreak[4]\par\vskip .3cm

First of all, I would like to thank my advisor H.~Verlinde for the formulation of the problem and extensive help during the work on the project.
I would also like to thank N.~Beisert for his comments on different stages of the project.  The work was supported by the National Science Foundation under grant PHY-0243680. Any opinions, findings, and conclusions or recommendations expressed in this
material are those of the author and do not necessarily reflect the views of the National Science Foundation.\\

\appendix

\renewcommand{\newsection}[1]{
\refstepcounter{section} %\setcounter{equation}{0}
\setcounter{subsection}{0} \addcontentsline{toc}{section}{\protect
\numberline{\Alph{section}}{{\rm #1}}} \vglue .0cm \pagebreak[3]
\noindent{\large \bf  \thesection. #1}\nopagebreak[4]\par\vskip .3cm}
\renewcommand{\newsubsection}[1]{
\refstepcounter{subsection}
\addcontentsline{toc}{subsection}{\protect
\numberline{\Alph{section}.\arabic{subsection}}{ #1}} \vglue .0cm
\pagebreak[3] \noindent{\it \thesubsection. #1}\nopagebreak[4]\par\vskip .3cm}

%\newpage
\bigskip
\newsection{Group Theory Facts}
\label{app:group_definitions}
\newsubsection{Basic Notations and Definitions}
For a discrete group $\Gamma$ the group algebra $\cg$ consists of the formal sums of the form
$$
\cg \ni \xq = \sum_{g\in \Gamma} x_g\, g \,,
$$
with $x_g\in \Cbb$ being some numbers.
% Therefore each element $\xq\in\cg$ can be considered as a function $\xq:\Gamma\to\Cbb$,
% $x(g)=x_g$. Algebra multiplication in $\cg$ is induced by that in $\Gamma$:
% $$
% \xq \yq = \sum_{g_1,\, g_2} x_{g_1} y_{g_2} \, g_1 g_2 \,.
% $$
% In terms of functions on the group this corresponds to the
% convolution,
% $$
% (x\circ y) (g) = \sum_{h\in \Gamma} x(gh^{-1})\, y(h) \,.
% $$
% The group algebra $\cg$ is a $\star$-algebra with the conjugation
% operation given by
% $$
% \star \xq = \sum_g \bar{x}_g\, g^{-1} \,.
% $$
% There also exists a linear functional $\avg{\ldots}: \cg \to \Cbb$
% given by
% $$
% \avg \xq = \frac{1}{|\Gamma|}\, \Tr_{V_\reg} \xq = x_e\,.
% $$
For matrix elements of unitary irreducible representations $\{\Rff_\lambda\}$ of a discrete group $\Gamma$ over $\Cbb$ there hold the following orthogonality relations,
\be
\label{orth_rel} \frac{1}{|\Gamma|}\, \sum_{g\in \Gamma} \rrho^\lambda_{ij}(g)\, \overline{\rrho^\mu_{kl}(g)} = \frac{1}{\dim V_\lambda}\, \delta^{\lambda\mu}\, \delta_{ik}\, \delta_{jl} \,.
\ee
Thus any function on the group can be expanded in Fourier series
\be
\label{Fourier} x(g) = \sum_\lambda \sum_{m,n=1}^{\dim V_\lambda} \frac{1}{S_\lambda}\, X^\lambda_{mn}\, \repconj[\lambda]{mn}{g} \,; \text{ where } S_\lambda=\frac{\ord{\Gamma}}{\dim V_\lambda} \,.
\ee
Then the orthogonality relation (\ref{orth_rel}), implies the inverse Fourier transform
\be
X^\lambda_{mn} = \sum_{g\in \Gamma} x(g)\, \rep[\lambda]{mn}{g} \,.
\ee
In particular, the decomposition (\ref{Fourier}) provides the famous Peter-Weyl decomposition of the regular representation $\Rff_\reg$,
\be
V_\reg \simeq \bigoplus_\lambda \Cbb^{N_\lambda}\otimes V_\lambda\,, \quad N_\lambda=\dim V_\lambda \,.
\ee

Tensor product of the two unitary decomposable representations can be decomposed as a direct sum of irreducible representations,
$$
V_\lambda \otimes V_\mu = \bigoplus_\nu V_\nu \,.
$$
(Some representations can enter the sum with multiplicities.) It terms of the basis vectors this means
\be
\e_{\lambda l} \otimes \e_{\mu m} = \sum_{\nu n} \K{\nu n}{\lambda l,\mu m}\, \e_{\nu n} \,;
\ee
and when the representations in the l.h.s.\ are irreducible the numbers $\K{\nu n}{\lambda l,\mu m}=\clebsch{\nu n}{\lambda l}{\mu m}$ are called the Clebsch-Gordan coefficients. Since all the representations are unitary, Clebsch-Gordan coefficients must also be unitary; \ie,
\be
\e_{\nu n} = \sum_{\lambda l,\, \mu m} \invclebsch{\nu n}{\lambda l}{\mu m}\, \e_{\lambda l} \otimes \e_{\mu m} \,.
\ee
A more general expansion is
\be
\e_{\lambda_1 l_1} \otimes \cdots \otimes \e_{\lambda_k l_k} = \sum_{\mu m} \K{\mu m}{\lambda_1 l_1,\ldots \lambda_k l_k}\, \e_{\mu m} \,.
\ee
(The same representation $\mu$ can enter the sum more than once.) In particular, acting on both sides with a group element $h$ one derives the following useful expansion,
\be
\label{rho_product_expansion} \rep[\lambda_1]{l_1 m_1}{h}\ldots \rep[\lambda_k]{l_k m_k}{h} = \sum_{\mu,\, pq} \rep[\mu]{pq}{h}\, \Kinv{\mu p}{\lambda_1 l_1,\ldots \lambda_k l_k}\, \K{\mu q}{\lambda_1 m_1,\ldots \lambda_k m_k} \,.
\ee
Summation over the group yields
\be
\label{ave_rho_product} \sum_h \rep[\lambda_1]{l_1 m_1}{h}\ldots \rep[\lambda_k]{l_k m_k}{h} = \ord{\Gamma} \sum_a \Kinv{(a)}{\lambda_1 l_1,\ldots \lambda_k l_k}\, \K{(a)}{\lambda_1 m_1,\ldots \lambda_k m_k} \,,
\ee
where the invariant tensors
$$
\K{}{\lambda_1 l_1,\ldots \lambda_k l_k} = \K{\triv}{\lambda_1 l_1,\ldots \lambda_k l_k}
$$
are simply the coefficients projecting onto the trivial representation (there is an extra index $a$ numbering them if the trivial representation enters the product $V_{\lambda_1}\otimes \cdots \otimes V_{\lambda_k}$ more than once). They are orthogonal in the sense that
\be
\label{K_orth_rel} \sum_{\{l_i\}} \Kinv{(a)}{\lambda_1 l_1,\ldots \lambda_k l_k}\, \K{(b)}{\lambda_1 l_1,\ldots \lambda_k l_k} = \delta_{ab} \,.
\ee
\\

\medskip
\newsubsection{Root System of the $\su_{2,2|4}$ Superalgebra}
\label{app:root_systems}
Here we summarize the relevant facts regarding the superconformal algebra $\su_{2,2|4}$ and its complex form $\alg{sl}_{2,2|4}(\Cbb)$.
The corresponding supergroup $\SU(2,2|4)$ is the group of transformations of the superspace $\Cbb^{2,2|4}$ preserving the corresponding quadratic form $\eta=\diag(\onebb_2,-\onebb_2|\onebb_4)$, with an extra condition on the superdeterminant; $\sdet A=1$, $A\in\SU(2,2|4)$.
The two bosonic subalgebrae\footnote{Note that in general the bosonic part of $\su_{2,2|N}$ is $\su_{2,2}\oplus \su_N \oplus \alg{u}_1$. An artifact of the $N=4$ case is that the $\alg{u}_1$ part is represented by the unit matrix and thus completely decouples.} $\su_{2,2}$ and $\su_4$ correspond to the conformal algebra and the $\Rc$-symmetry algebra in field theory.
\renewcommand{\join}{\!\!\frac{\qquad}{\qquad}\!\!}
\begin{table}[htb]
\begin{center}
\caption{\small Root systems and Cartan matrices for the $\su_{2,2|4}$ superalgebra. For superalgebrae the root system is not fixed uniquely. The ``beast'' system is the most straightforward choice; while the ``beauty'' system is especially useful from the perspective of the supersymmetric theories. With this choice of root system the Bethe vacua are the half-BPS states.}
\label{tab:root_systems}
\begin{tabular}{|c|c|c|}
\hline
 &  \emph{Beauty} & \emph{Beast}
\\ \hline
Cartan Matrix & $\scriptsize \left(\begin{array}{c|c|ccc|c|c}
 -2 & +1 & & & & & \\ \hline
 +1 & & -1 & & & &\\ \hline
 & -1 & +2 & -1 & & & \\
 & & -1 & +2 & -1 & & \\
 & & & -1 & +2 & -1 & \\ \hline
 & & & & -1 & & +1 \\ \hline
 & & & & & +1 & -2
\end{array} \right)$
& $\scriptsize\left(\begin{array}{ccc|c|ccc}
 +2 & -1 & & & & & \\
 -1 & +2 & -1 & & & &\\
 & -1 & +2 & -1 & & & \\ \hline
 & & -1 & & -1 & & \\ \hline
 & & & -1 & -2 & -1 & \\
 & & & & -1 & -2 & +1 \\
 & & & & & +1 & -2
\end{array} \right)$
\\  \hline
Dynkin diagram & $\tiny\ominus\,\join \otimes\join \oplus\join \oplus\join \oplus\join \otimes\join\, \ominus $ & $\tiny\oplus\,\join \oplus\join \oplus\join \otimes\join \ominus\join \ominus\join\, \ominus $
\\ \hline
\end{tabular}
\end{center}
\end{table}
As opposed to bosonic semisimple algebrae, the root system and Cartan matrix of the superalgebrae are not fixed uniquely; and there is some freedom in the choice of the fermionic roots.
The most simple choice would be to mimic the bosonic case. Then the simple roots are $\a_{12}$, $\a_{23}$, \ldots $\a_{78}$.
(The root vector for the root $\a_{ij}$ is the matrix with the only non-zero entry at the intersection of the $i$-th row and $j$-th column.)
In other words, the root vectors for the simple roots are represented as the following matrices:
$$
\scriptsize\left(\begin{array}{cccc|cccc}
 0 &\ast & & & & & & \\
 &0 & \ast & & & & & \\
 & & 0 & \ast & & & & \\
 & & & 0 & \ast & & & \\ \hline
 & & & & 0 & \ast & & \\
 & & & & & 0 & \ast & \\
 & & & & & & 0 & \ast \\
 & & & & & & & 0 \\
\end{array}\right)
$$
This choice corresponds to the three positive and three negative bosonic roots and one fermionic root. Following~\cite{BS} we will call it the \emph{beast} root system. However, as it was emphasized in~\cite{DP}, there exists a more convenient choice from the point of view of the supersymmetry, the so-called \emph{beauty} root system.

In order to construct it one can choose the quadratic form $\eta$ in $\Cbb^{2,2|4}$ in a non-standard way,
\be
\eta = \left(\begin{array}{cc|c}
0 & \onebb_2 & \\
\onebb_2 & 0 & \\
\hline
 & & \onebb_4
\end{array} \right) \,.
\ee
Then the $\su_{2,2|4}$ algebra is defined by the following constraints,
\be
X^\dagger \eta + \eta X =0\,, \qquad \STr X=0 \,; \qquad\qquad X\in \su_{2,2|4} \,.
\ee
The general solution is
\be
X=\left(\begin{array}{cc|c}
A & B & P \\
C & -A^\dagger & Q \\
\hline
-Q^\dagger & -P^\dagger & F
\end{array} \right) \,;
\ee
where $B=-B^\dagger$, $C=-C^\dagger$, $F=-F^\dagger$ and $\Tr(A-A^\dagger)=\Tr F=0$. (Recall that the diagonal $\alg{u}_1$ part decouples, and thus we can impose the tracelessness condition on $A$ and $F$ separately.)

Now the $\su_{2,2}$ part gets identified with the algebra of the conformal group as follows. The element $\s=\left(\begin{array}{cc} \onebb_2 & 0 \\ 0 & -\onebb_2\end{array}\right)$ is the generator of dilatations. It commutes with the $\alg{sl}_2(\Cbb)$ subalgebra
$$
\alg{m} = \left\{ \left(\begin{array}{cc}
A & 0 \\ 0 & -A^\dagger
\end{array}\right)\,, \quad \Tr A=0\right\} \,,
$$
which is nothing but the algebra of the Lorentz group. Finally, matrices of the form $\left(\begin{array}{cc} 0 & B \\ 0 & 0 \end{array}\right)$ and $\left(\begin{array}{cc} 0 & 0 \\ C & 0 \end{array}\right)$ (with $B$ and $C$ anti-hermitian) represent the special conformal transformations and translations, as on can figure out from their commutation relations with $\s$.

The Cartan subalgebra of $\alg{sl}_{2,2|4}\simeq \alg{su}_{2,2|4}\otimes \Cbb$ is the direct sum $\alg{h}\simeq \alg{h}_{\alg{sl}_{2,2}} \oplus \alg{h}_{\alg{sl}_{4}}$.Both Cartan subalgebrae $\alg{h}_{\alg{sl}_{2,2}}$ and $\alg{h}_{\alg{sl}_{4}}$ can be chosen as diagonal matrices; thus $\dim \alg{h}=6$. The set of simple roots can be chosen as $\gamma_1=\a_{12}$, $\gamma_2=\a_{34}$, $\gamma_3=\a_{25}$, $\gamma_4=\a_{48}$, $\gamma_5=\a_{56}$, $\gamma_6=\a_{67}$, $\gamma_7=\a_{78}$. Thus the root vectors for the simple roots are represented as
$$
\scriptsize\left(\begin{array}{cccc|cccc}
 0 &\ast & & & & & & \\
 &0 & & & \ast & & & \\
 & & 0 & \ast & & & & \\
 & & & 0 & & & & \ast \\ \hline
 & & & & 0 & \ast & & \\
 & & & & & 0 & \ast & \\
 & & & & & & 0 & \ast \\
 & & & & & & & 0 \\
\end{array}\right)
$$

Note that it is impossible to choose a simple root system consisting of exactly six roots, and the minimal system contains seven roots~\cite{DP}. A consequence of this redundancy is the relation
\be
\gamma_1 +2\gamma_3 +\gamma_5 = \gamma_2 +2\gamma_4 +\gamma_7 \,.
\ee
With this choice there are three positive, two negative and two fermionic roots. Cartan matrices and Dynkin diagrams for both choices of the root system are shown in Table~\ref{tab:root_systems}.
\\

\medskip
\newsubsection{Representation Ring of the Dihedral Group}
\label{app:dihedral_group}
The dihedral group $D_S$ is generated by the two elements, $r$ and $\s$, with the additional relations
\bea
r^S=\s^2=1\,, & \qquad & r\s=\s r^{-1} \,.
\eea
The order of the group $\ord{D_S}=2S$.
We will restrict ourselves to the odd $S=2n+1$. Then there are the $n+2$ conjugacy classes, $\Oc_1=\{e\}$, $\Oc_2=\{r,r^{2n}\}$,\ldots, $\Oc_{n+1}=\{r^n,r^{n+1}\}$, $\Oc_{n+2}=\{\s,\s r,\,\ldots,\s r^{2n}\}$. Thus there exist the $n+2$ irreducible representations. Among them there are the $n$ two-dimensional representations $\rrho_m$:
\be
\rrho_m(r) = \left(\begin{array}{cc}
\omega^m & 0 \\
0 & \omega^{-m}
\end{array}\right) \,, \qquad
\rrho_m(\s) = \left(\begin{array}{cc}
0 & 1 \\
1 & 0
\end{array}\right) \,, \qquad
m=1,2,\,\ldots n\,.
\ee
Here $\omega=\e^{2\pi i/S}$
There are also the two one-dimensional representations $\rrho_0$ and $\rrho_{\tilde 0}$:
\bea
\rrho_0(r)=1\,, \quad \rrho_0(\s)=1\,; &\qquad& \rrho_{\tilde 0}(r)=1\,, \quad \rrho_{\tilde 0}(\s)=-1\,.
\eea
The table of characters as well as the representation ring and the stabilizer subgroups of each element of the group $D_{2n+1}$ are shown in Table~\ref{tab:dihedral_group}.
\begin{table}[htb]
\caption{\small Table of characters and representation ring (multiplication table) of the dihedral group $D_{S=2n+1}$. Stabilizer subgroups $S_g$ for a representative of each conjugacy class of the group $D_5$.}
\label{tab:dihedral_group}
\begin{center}
\begin{tabular}{|c|c|c|c|}
\hline
 & $[e]$ & $[r^m]$ & $[\s]$ \\
\hline
$\chi_0$ & $1$ & $1$ & $1$ \\
\hline
$\chi_{\tilde 0}$ & $1$ & $1$ & $-1$ \\
\hline
$\chi_l$ & 2 & $2\cos\bigl(2\pi\frac{lm}{S}\bigr)$ & 0 \\
\hline
\end{tabular}
\qquad
\begin{tabular}{|c||c|c|c|}
\hline
$\otimes$ & $\rrho_0$ & $\rrho_{\tilde 0}$ & $\rrho_k$
\\ \hline \hline
$\rrho_0$ & $\rrho_0$ & $\rrho_{\tilde 0}$ & $\rrho_k$
\\ \hline
$\rrho_{\tilde 0}$ & $\rrho_{\tilde 0}$ & $\rrho_0$ & $\rrho_k$
\\ \hline
$\rrho_l$ & $\rrho_l$ & $\rrho_l$ & $\left\{\begin{array}{lr}
 \rrho_{k+l} \oplus \rrho_{k-l}\,, & k\neq l \\
 \rrho_{2l} \oplus \rrho_0 \oplus \rrho_{\tilde 0}\,, & k=l
\end{array}\right.$
\\ \hline
\end{tabular}
\\
\begin{tabular}{|c||c|c|c|c|}
\hline
$g$ & $e$ & $r$ & $r^2$ & $[\s]$ \\
\hline \hline
$S_g$ & $D_5$ & $\{e,r,\, \ldots,r^4\}\simeq \Zbb_5$ & $\{e,r,\, \ldots,r^4\}\simeq \Zbb_5$ & $\{e,\s\}$ \\
\hline
\end{tabular}
\end{center}
\end{table}
These data are needed for us to work with the corresponding orbifold theory.

\bigskip
\newsection{Quiver vs Orbit Description}
\label{app:quiver_vs_orbit}
Here we show that the two descriptions of the quiver gauge theory field content are equivalent and develop explicit transition formulae between them. The ``orbit'' description is based on explicit parameterization of the orbits of discrete group acting in the field space. It proves to be the most useful for the study of the orbifold theory in comparison with the original $\Nc=4$ one. The ``quiver'' description in its turn emerges from the algebraic analysis of the transformation properties of the fields and decomposition of representations. It leads to the description of the field content in terms of the quiver. This section is rather technical.\\

\medskip
\newsubsection{Basis in the Field Space}
As we saw in section~\ref{sec:gauge_theory}, Chan-Paton indices of the fields transform in the regular representation of the orbifold group $\GG$. Namely, the fields belong to
\bea
\label{field_rep}
V^\perp\otimes V_\reg^{\oplus N}\otimes \bar{V}_\reg^{\oplus N} \simeq V^\perp\otimes V_\reg\otimes \bar{V}_\reg \otimes \Cbb^N \otimes \Cbb^{\ast N}
\eea
$V^\perp$ being the representation corresponding to the transverse indices.
That is why it is important to introduce a basis in $V_\reg\simeq\cg$. We are going to use the two different bases in the group algebra $\cg$,
\be
\{ \e_g = g \}
\ee
and
\be
\{ \E^\lambda_{mn} = \frac{1}{\sqrt{S_\lambda}} \sum_g \repconj[\lambda]{nm}{g}\, g\} \,, \quad S_\lambda=\frac{\ord{\Gamma}}{\dim V_\lambda}\,.
\ee
The group acts on them according to
\be
h \,:\, \e_g \to \e_{hg}
\ee
and
\be
h \,:\, \E^\lambda_{mn} \to \sum_k \repconj[\lambda]{nk}{h^{-1}}\,
\E^\lambda_{mk} = \sum_k \E^\lambda_{mk}\, \rep[\lambda]{kn}{h} \,.
\ee
The relation between these bases is
\ba
\e_g &=& \ds \sum_\lambda \frac{1}{\sqrt{S_\lambda}}\, \sum_{mn}
\rep[\lambda]{nm}{g}\, \E^\lambda_{mn} \,,
\\
\E^\lambda_{mn} &=& \ds \frac{1}{\sqrt{S_\lambda}}\, \sum_g \repconj[\lambda]{nm}{g}\, \e_g \,.
\ea
Both bases are orthonormal. The dual vector space is introduced as the space of linear functions on $\cg$. Its basis vectors can also be defined in different ways, $\e^\ast_g(\e_h)=\delta_{gh}$ or $\E^{\lambda\ast}_{kl}(\E^\mu_{mn})=\delta^{\lambda\mu}\, \delta_{km}\, \delta_{ln}$. Considerations of elementary linear algebra give the following relation between the dual bases,
\ba
\E^{\lambda\ast}_{mn} &=& \ds \frac{1}{\sqrt{S_\lambda}}\, \sum_g
\rep[\lambda]{nm}{g}\, \e_g^\ast\,,
\\
\e^\ast_g  &=& \ds \sum_\lambda \frac{1}{\sqrt{S_\lambda}}\, \sum_{mn} \repconj[\lambda]{nm}{g}\,
\E^{\lambda\ast}_{mn} \,.
\ea

Next we construct the two bases in the field space. These basis vectors are to label the invariant configurations in the field space. Thus for a generic field we need to find the invariant configurations
\bea
\Big(V^\perp\otimes V_\reg^{\oplus N}\otimes \bar{V}_\reg^{\oplus N}\Big)^G \simeq \Big(V^\perp\otimes V_\reg\otimes \bar{V}_\reg\Big)^G \otimes \Cbb^N \otimes \Cbb^{\ast N}
\eea
(see (\ref{field_rep}).) In what follows we are going to drop the trivial $\Cbb^N \otimes \Cbb^{\ast N}$ factor

Let us start with the gauge field. It has no transverse indices, and we need to find the invariant subspace $\Big(\End(V_\reg)\Big)^G$. The ``orbit'' basis
\be
\label{orbit_basis}
\tb_g = \sum_h \e_h \otimes \e^\ast_{hg}
\ee
has a natural interpretation in terms of invariant combinations of strings stretching between image branes. Similarly, the product $\tb_g \circ \tb_h = \tb_{gh}$ has a natural interpretation in terms of gluing the ends of open strings. Using orthonormality of the basis $\{\e_g\}$, one can identify the basis and the dual basis via $\e_g^\ast(x)=\avg{\e_g,x}$. This allows one to perform hermitian conjugation, and it gives $\tb_g^\dagger=\tb_{g^{-1}}$.

In order to build the ``quiver'' basis we note that $\E^\lambda_{mn}$ do \emph{not} transform in the first index (recall that each representation $\Rff_\lambda$ enters $\Rff_\reg$ with multiplicity equal to $N_\lambda=\dim V_\lambda$ {---} and this is the first index of $\E^\l_{mn}$ that numbers these copies). Therefore, the combination
\be
\label{quiver_basis} \Tb^\lambda_{mn} = \sum_k \E^\lambda_{mk} \otimes
\E^{\lambda\ast}_{nk}
\ee
is $G$-invariant. The multiplication rule is $\Tb^\lambda_{mn} \circ \Tb^\mu_{kl} = \delta^{\lambda\mu}\, \delta_{kn}\, \Tb^\lambda_{ml}$. Hermitian conjugate $\Tb^{\lambda\dagger}_{mn}=\Tb^\lambda_{nm}$. Here we recognize the matrix algebra $ \bigoplus_\lambda \gl(v_\lambda) \simeq \cg$.

Thus, in these two calculations we get the same answer; \ie, the algebrae of $\tb_g$ and $\Tb^\lambda_{mn}$ are both isomorphic to the group algebra. However, it is interesting to find an explicit relation between these two bases. Now it is easy to find
$$
\tb_g = \sum_h \e_h \otimes \e^\ast_{hg} = \sum_h
\sum_{\lambda,\,mn} \sum_{\mu,\,kl} \frac{\dim V_\lambda}{|G|}\,
\rep[\lambda]{nm}{h}\, \repconj[\mu]{lk}{hg}\, \E^\lambda_{mn}
\otimes \E^{\mu\ast}_{kl} \,.
$$
Taking into account the group property and orthogonality relation,
one evaluates
$$
\frac{\dim V_\lambda}{|G|}\, \sum_h \rep[\lambda]{nm}{h}\,
\repconj[\mu]{lk}{hg} = \delta^{\lambda\mu}\, \delta_{nl}\,
\repconj[\lambda]{nl}{g} \,.
$$
Finally we get
\be
\label{t_via_T} \tb_g = \sum_{\lambda} \sum_{klm}
\repconj[\lambda]{mk}{g}\, \E^\lambda_{ml} \otimes
\E^{\lambda\ast}_{kl} = \sum_{\lambda} \sum_{km}
\repconj[\lambda]{mk}{g}\, \Tb^\lambda_{mk} \,.
\ee
Hence the two bases are simply related by a discrete Fourier
transform.

Now we can do a similar calculation for the scalar and spinor fields which do have transverse indices with non-trivial transformation rules. This is a generic case, and we have to find the invariant subspace $\Big(V^\perp\otimes V_\reg\otimes \bar{V}_\reg\Big)^G$. Denote the basis of the transverse representation $V^\perp$ as $\{\f_A\equiv \e_{\alpha,A}\}$. Then the ``orbit'' basis has the form
\be
\label{orbit_basis_nontriv} \tb_{A,g} = \sum_h (h\act\f_A) \otimes
\e_h \otimes \e^\ast_{hg} \,.
\ee
In terms of components the action is
\bea
h \act \f_A = \sum_B \rep[\alpha]{BA}{h}\, \f_B \,.
\eea
Since the representation $\Rff_\alpha$ is real, hermitian conjugation acts according to
$$
\tb^\dagger_{A,g}=\sum_B \rep{BA}{g^{-1}} \tb_{B,g^{-1}} \,.
$$
To find the ``quiver'' basis we will need to find the invariant tensors in the product
$$
\Big(V^\perp\otimes V_\reg\otimes \bar{V}_\reg\Big)^G \simeq \bigoplus_{\lambda,\mu} \Big(V_\alpha\otimes V_\lambda\otimes V_\mu^\ast\Big)^G \otimes \Cbb^{N_\l} \otimes \Cbb^{\ast N_\mu} \,.
$$
In other words, we are to find the invariant operators from $V_\mu$ to $V_\alpha\otimes V_\lambda$,
$$
\Big(V_\alpha\otimes V_\lambda\otimes V_\mu^\ast\Big)^G \simeq \Big(\Hom(V_\mu,V_\alpha\otimes V_\lambda)\Big)^G \,.
$$
The way to do it is to decompose the product of representations $\Rff_\alpha$ and $\Rff_\lambda$ into a direct sum of irreducible representations. In particular, in terms of basis vectors
$$
\e_{\alpha A} \otimes \e_{\lambda l} = \sum_{\mu m} \K{\mu m}{\alpha A,\lambda l}\, \e_{\mu m} \,.
$$
Then the unitarity implies that
$$
\e_{\mu m} = \sum_{A,l} \Kinv{\mu m}{\alpha A,\lambda l}\, \e_{\alpha A} \otimes \e_{\lambda l} \,.
$$
Therefore, the invariant configuration is
$$
\sum_m \e_{\mu m} \otimes \e_{\mu m}^\ast = \sum_{A,l,m} \Kinv{\mu m}{\alpha A,\lambda l}\, \e_{\alpha A}
\otimes \e_{\lambda l} \otimes \e_{\mu m}^\ast \,.
$$
(The field components of the invariant configuration are given by the invariant tensor, $\Phi^{Al}_{m} \sim \Kinv{\mu m}{\alpha A,\lambda l}$.) This gives
\be
\label{quiver_basis_nontriv} \Tb^{\lambda\mu}_{lm} = \sum_{A,i,j} \Kinv{\mu j}{\alpha I,\lambda i}\, \f_A \otimes
\E^\lambda_{li} \otimes \E^{\mu\ast}_{mj} \,.
\ee
Note that here the indices $l$ and $m$ are the indices of the gauge groups in the corresponding nodes $\Rf_\l$ and $\Rf_\mu$. These indices appear owing to the fact that each representation $\Rf_\l$ enters the decomposition of the regular representation $\Rf_\reg$ with multiplicity $N_\l$. If the defining representation $\Rff_\alpha$ is trivial, $C^{\mu j}_{\lambda i} = \delta^\mu_\lambda\, \delta^j_i$, and we  recover the previous expression for the gauge field.

A simple exercise would be to calculate the number of independent components of the scalar field $\phi$ after the reduction. To do this  note that for every fixed node $\Rf_\lambda$ there are some arrows going to some nodes $\{\Rf_\mu\}$. (There can be multiple arrows going to the same node.) For each arrow there are $N^2 N_\lambda N_\mu$ components since there are $N N_\lambda$ copies of $\Rf_\lambda$ and $N N_\mu$ copies of $\Rf_\mu$ in $\Cbb^N \otimes \Rf_\reg$. Therefore, to each node $\Rf_\lambda$ there correspond $N^2 N_\lambda \sum_{\{\mu\}} \dim V_\mu$ independent components. Since $V_\alpha \otimes V_\lambda \simeq \bigoplus_{\{\mu\}} V_\mu$, the dimension counting gives $\sum_{\{\mu\}} \dim V_\mu = (\dim V_\lambda)\, (\dim V_\alpha)$. Therefore, the total number of independent components of $\phi$ is $N^2\, \dim V_\alpha \sum_\lambda N_\lambda^2 = \ord{G} \,N^2\, \dim V_\alpha$. This fact is not surprising  since the invariant configurations of the fields are orbits in the field space, and one can parameterize them in terms of the group algebra valued fields. This way the dimension counting confirms equivalence of the two descriptions. Furthermore, it is possible to find an exact transition formula between them. Indeed, substituting for the basis vectors $\e_g$ via $\E^\lambda_{lm}$, we get for $\tb_{A,g}$
$$
\tb_{A,g} = \sum_B \sum_h \sum_{\lambda,\,mn} \sum_{\mu,\,kl} \frac{\dim V_\lambda}{|G|}\, \rep[\alpha]{BA}{h}\, \rep[\lambda]{nm}{h}\, \repconj[\mu]{lk}{hg}\, f_B \otimes \E^\lambda_{mn} \otimes \E^{\mu\ast}_{kl} \,.
$$
Using expansion
$$
\rep[\alpha]{BA}{h}\, \rep[\lambda]{nm}{h} = \sum_{\nu, pq} \Kinv{\nu p}{B,\lambda n}\, \K{\nu q}{A,\lambda m}\,
\rep[\nu]{pq}{h} \,,
$$
and orthogonality relation, one can evaluate
$$
\sum_h \frac{\dim V_\lambda}{|G|}\, \rep[\alpha]{BA}{h}\, \rep[\lambda]{nm}{h}\, \repconj[\mu]{lk}{hg} = \sum_i \frac{\dim V_\lambda}{\dim V_\mu}\, \Kinv{\mu l}{B,\lambda n}\, \K{\mu i}{A,\lambda m}\, \repconj[\mu]{ik}{g} \,.
$$
Finally, after the use of (\ref{quiver_basis_nontriv}) this gives
\be
\label{t_via_T_nontriv}
\tb_{A,g} = \sum_{\lambda, l} \sum_{\mu, km} \frac{\dim V_\lambda}{\dim V_\mu}\, \K{\mu k}{A,\lambda l}\,
\repconj[\mu]{km}{g}\, \Tb^{\lambda\mu}_{lm} \,.
\ee
The presence of the factor $\K{\mu\cdot}{\alpha\cdot,\lambda\cdot}$ restricts the sum over $(\lambda,\mu)$ only to those pairs which are connected by a line in the quiver. Again, when the defining representation
$\Rff_\alpha$ is trivial, the correct formula (\ref{t_via_T}) is recovered.\\

\medskip
\newsubsection{Relation Between the Field Components}
It is very convenient to write the field operators in the quiver notation (then the gauge invariance is manifest), while the Feynman rules most closely resemble those for the unorbifolded theory in the notation with fields depending on the group index. Here we proceed to develop the connection between the field components and the algebra of fields in the two notations. Writing down the gauge field as $A=\sum_g A_g\, \tb_g = \sum_\lambda \sum_{lm} A^\lambda_{lm}\, \Tb^\lambda_{lm}$ and using (\ref{t_via_T}), we find the relation
\be
\label{a_via_A}
A^\lambda_{lm} = \sum_g \repconj[\lambda]{lm}{g}\, A_g \,.
\ee
Similarly, for the scalar fields we get
\be
\label{f_via_F}
\phi^{\lambda\mu}_{lm} = \sum_g \sum_k \frac{N_\lambda}{N_\mu}\, \K{\mu k}{A,\lambda l}\, \repconj[\mu]{km}{g}\, \phi^A_g \,.
\ee
Then to the matrix multiplication of the gauge fields at the same node there corresponds the convolution product in terms of the group algebra:
$$
\sum_l A^\lambda_{kl}\, A^\lambda_{lm} = \sum_{g,h} \sum_l \repconj[\lambda]{kl}{g}\, \repconj[\lambda]{lm}{h}\, A_g\, A_h = \sum_g \repconj[\lambda]{km}{gh}\, A_g\, A_h = \sum_g \repconj[\lambda]{km}{g}\, (A\circ A)(g) \,.
$$
The story with the scalar fields is slightly more complicated. The product of the two fields $\phi$ and $\psi$ with the transverse indices transforming in the  representations $\Rff_\alpha$ and $\Rff_\beta$ can be viewed as a field with a transverse index in the product representation $\Rff_\alpha \otimes \Rff_\beta$, though the convolution rule is slightly modified. Particularly,
$$
\sum_m \phi^{\lambda\mu}_{lm}\, \psi^{\mu\nu}_{mn} = \frac{N_\lambda}{N_\nu}\, \sum_{g,h} \K{\mu p}{\alpha A,\lambda l}\, \K{\nu q}{\beta B,\mu m}\, \repconj[\mu]{pm}{g} \, \repconj[\nu]{qn}{h}\, \phi^A_g\, \psi^B_h \,.
$$
Since the decomposition coefficients $K^\cdot_{\cdot,\cdot}$ are invariant tensors by construction, we have
$$
 \repconj[\mu]{pm}{g}\, \K{\nu q}{\beta B,\mu m} = \repconj[\beta]{BC}{g^{-1}}\, \rep[\nu]{qr}{g^{-1}}\, \K{\nu r}{\beta C,\mu p} \,.
$$
Substituting this and summing over the index $q$, we get
\be
\label{f_via_F_gen} \sum_m \phi^{\lambda\mu}_{lm}\, \psi^{\mu\nu}_{mn} = \frac{N_\lambda}{N_\nu}\, \sum_{g,h} \K{\nu r}{\alpha A, \beta B, \lambda l}\, \repconj[\nu]{nr}{h^{-1}g^{-1}}\, \rep[\beta]{BC}{g}\, \phi^A_g\, \psi^C_h \,.
\ee
Here $\K{\nu r}{\alpha A, \beta B, \lambda l}=\sum_p \K{\mu p}{\alpha A,\lambda l}\, \K{\nu r}{\beta B,\mu p}$ is (one of) the invariant tensors corresponding to the decomposition $\Rff_\alpha\otimes \Rff_\beta\otimes \Rff_\lambda \to \Rff_\nu$. Let us stress that generally there can exist a possibility of decomposing $\Rff_\alpha\otimes \Rff_\lambda \to \Rff_\mu$ and/or $\Rff_\beta\otimes \Rff_\mu\to\Rff_\nu$ in multiple ways. In this case we do \emph{not} have to sum over these different tensors. Graphically this corresponds to the case when there are multiple arrows between the same nodes in quiver, and then one is free to choose between them. Another observation is that (\ref{f_via_F_gen}) has the same structure as (\ref{f_via_F}), the product $\phi\circ\psi$ having the defining representation $\Rff_\alpha \otimes \Rff_\beta$. The convolution rule is slightly modified,
\be
\label{convolution}
(\phi\circ\psi)^{AB}_g = \sum_h \phi^A_h\, \rep{BC}{h}\, \psi^C_{h^{-1}g} \,.
\ee
Both formulae (\ref{f_via_F_gen}) and (\ref{convolution}) are also valid for the gauge fields which have no transverse indices (trivial representation). In this case some matrix elements and decomposition tensors become degenerate. Recalling the reduction formula (\ref{inv_phi}) one can see that the multiplication rule  (\ref{convolution}) naturally corresponds to the standard matrix multiplication in the $\SU(\ord{\Gamma} N)$ theory. This means that
$$
\sum_f \phi^A_{hf}\, \psi^B_{fg} = \rep[\alpha]{AA^\prime}{h}\, \rep[\beta]{BB^\prime}{h}\, (\phi\circ\psi)^{A^\prime B^\prime}_{h^{-1}g} \,.
$$
This way of multiplication is induced from the original theory, and that is why it respects the gauge transformations. Another nice feature of the formula (\ref{f_via_F_gen}) is that when there exist several arrows going between different nodes the choice of a given arrow affects only the choice of the invariant tensors and does not affect the convolution product (\ref{convolution}). It means that all the operators corresponding to the different paths (not necessarily closed) in the quiver formed by $L$ consequent scalar lines $\lambda_1\to \l_2\to \cdots \to \l_{L+1}$ are contained in the product $\phi^{A_1}\ldots \phi^{A_L}$.

We can summarize these results as follows. An operator formed in the quiver notation as the product $\phi^{\l \nu_1}\, \phi^{\nu_1\nu_2}\ldots\, \phi^{\nu_{L-1}\mu}$ can be recast as
\bea
\label{path_operator}
\Big(\phi\circ\cdots\circ\phi\Big)^{\lambda\mu}_{lm} &=& \sum_g \sum_k \frac{N_\l}{N_\mu}\, \K{}{\aAe\ldots \aAL}{}^{{}^{\mu k}}_{{}^{\l l}}\, \repconj[\mu]{km}{g}\, \Big(\phi\circ\cdots\circ\phi\Big)^{\aAe\ldots\aAL}_g \,;
\eea
where the invariant tensor $\K{}{}$ is the one corresponding to the decomposition $\Rf_\l\otimes \Rf_{\nu_1}\otimes\cdots \otimes\Rf_{\nu_{L-1}}\to \Rf_\mu$. The product of fields in the r.h.s.\ is calculated according to (\ref{convolution}). In its turn it is related to the product of the fields of the original $\Nc=4$ theory as
\bea
\label{field_product}
\Big(\phi^{\aAe}\ldots\,\phi^{\aAL}\Big)_{h,hg} &=& \sum_{\bBe,\ldots \bBL} \rep{\aAe\bBe}{h}\ldots\, \rep{\aAL\bBL}{h}\, \Big(\phi\circ\cdots\circ\phi\Big)^{\bBe\ldots\bBL}_g \,.
\eea
These formulae will be of crucial importance for constructing the gauge invariant observables.\\

\medskip
\newsubsection{Construction of Observables: Untwisted Sector}
\label{app:untwisted_observables}
Let us first study the untwisted observables which are obtained by a reduction {---} for instance, $\Tr \phi^L$. They can be rewritten in terms of the group valued field $\phiq$ in the following way:
\be
\Oc^{A_1\ldots A_k} = \Tr \phi^{A_1} \ldots \phi^{A_k} = \sum_{h,i} \phi^{A_1}_{i_1 h_1,i_2 h_2}\, \phi^{A_2}_{i_2 h_2, i_3 h_3}\, \ldots \phi^{A_n}_{i_n h_n,i_1 h_1}\,.
\ee
Using (\ref{inv_phi}), it reduces to
$$
\Oc^{A_1\ldots A_k} = \sum_{h,\, i,\, B} \rep[]{A_1 B_1}{h_1} \ldots \rep[]{A_n B_n}{h_n}\, \phiq^{B_1}_{i_1 i_2}(h_1^{-1}h_2) \ldots \phiq^{B_n}_{i_n i_1}(h_n^{-1}h_1) \,.
$$
Introducing the new variables $h$, $p_i$, $i=1, \ldots, n-1$ according to $h=h_1$, $h_{i+1} = h p_i$, we get
$$
\Oc^{A_1\ldots A_k} = \sum_{h\com p\com B} \rep[]{A_1 B_1}{h}\, \rep[]{A_2 B_2}{hp_1}\, \ldots \rep[]{A_n B_n}{hp_{n-1}}\, \times
$$
$$
\times \Tr \phiq^{B_1}(p_1)\, \phiq^{B_2}(p_1^{-1} p_2)\, \ldots \phiq^{B_{n-1}}(p_{n-2}^{-1} p_{n-1})\, \phiq^{B_n}(p_{n-1}^{-1})
\,.
$$
This rewrites as
\bea
\nn \Oc^{A_1\ldots A_k} &=& \sum_{p\com B\com C} D_{A_1 B_1\com A_2 C_2\com \ldots A_n C_n}\, \rep[]{C_2 B_2}{p_1}\, \ldots \rep[]{C_n B_n}{p_{n-1}}\, \times
\\
\nn & & \times \Tr \phiq^{B_1}(p_1)\, \phiq^{B_2}(p_1^{-1} p_2)\, \ldots \phiq^{B_{n-1}}(p_{n-2}^{-1} p_{n-1})\,
\phiq^{B_n}(p_{n-1}^{-1}) \,,
\eea
where
$$
D_{A_1 B_1\com \ldots \com A_n B_n} = \sum_h \rep[]{A_1 B_1}{h}\, \rep[]{A_2 B_2}{h}\, \ldots \rep[]{A_n B_n}{h} \,.
$$
Using (\ref{rho_product_expansion}), it can be evaluated as
$$
D_{A_1 B_1\com \ldots \com A_n B_n} = \sum_h \sum_{\mu\com pq} \rep[\mu]{pq}{h}\, \Kinv{\mu p}{A_1,\ldots A_n}\, \K{\mu q}{B_1,\ldots B_n} = |G|\, \Kinv{\triv}{A_1,\ldots A_n}\, \K{\triv}{B_1,\ldots B_n} \,.
$$
(In case there exist several different invariant tensors, one should sum over them.) Then
\bea
\label{reduction_op}
\nn \Oc^{A_1\ldots A_k} &=& \sum_a \sum_{p\com B\com C} \Kinv{(a)}{A_1,\ldots A_n}\, \K{(a)}{B_1 C_2,\ldots C_n}\, \rep[]{C_2 B_2}{p_1}\, \ldots \rep[]{C_n B_n}{p_{n-1}} \times
\\
& & \times |\Gamma|\, \Tr\phiq^{B_1}(p_1)\, \phiq^{B_2}(p_1^{-1} p_2)\, \ldots \phiq^{B_{n-1}}(p_{n-2}^{-1} p_{n-1})\,
\phiq^{B_n}(p_{n-1}^{-1}) \,.
\eea
The presence of invariant tensors implies that only the index combinations which are invariant w.r.t.\ the orbifold group are able to survive the projection (\ie, the reduction procedure extracts the invariant part of the operator only). Multiplication in the r.h.s.\ of (\ref{reduction_op}) corresponds exactly to the convolution rule (\ref{convolution}). Then orthogonality relation (\ref{K_orth_rel}) implies that the operators
\be
\Oc^{(a)}= \frac{1}{\ord\Gamma}\, \K{(a)}{A_1,\ldots A_k}\, \Oc^{A_1\ldots A_k} = \K{(a)}{A_1,\ldots A_k}\, (\phi\circ\cdots\circ\phi)^{A_1,\ldots A_k}(e)
\ee
form a basis in the space of untwisted observables obtained by a reduction from the original theory. In terms of the fields of the parent theory
\be
\Oc^{(a)} = \K{(a)}{A_1,\ldots A_L}\, \Tr \bigl(\phi^{\aAe}\ldots\phi^{\aAL}) \,.
\ee
\\

\medskip
\newsubsection{Construction of Observables: Twisted Sectors}
\label{app:twisted_observables}
It turns out that the twisted operators are constructed in a similar way. In order to construct generic observables it is convenient to use the quiver notation. Taking a closed loop in the quiver and using (\ref{path_operator}) one can write the corresponding operator as
\bea
\label{loop_operator}
\Tr_{\Rff_{\lambda_1}} \phi^{\lambda_1 \lambda_2}\, \phi^{\lambda_2 \lambda_3} \ldots \phi^{\lambda_L \lambda_1} &=& \sum_g \sum_{k,l} \K{}{{}_{A_1,\ldots A_L}}{}^{{}^{\l_1k}}_{{}_{\l_1l}}\, \repconj[\l]{kl}{g} \Big(\phi\circ\ldots\circ\phi\Big)^{A_1,\ldots A_L}_g \,;
\eea
where the invariant tensor
\bea
\label{inv_tensor_loop}
\K{}{{}_{A_1,\ldots A_L}}{}^{{}^{\l_1k}}_{{}_{\l_1l}} &=& \sum_{l_2,\ldots,\, l_L} \K{{}^{{}_{\l_2 l_2}}}{{}_{A_1\, \l_1 l}} \K{{}^{{}_{\l_3 l_3}}}{{}_{A_2\, \l_2 l_2}} \ldots \K{{}^{{}_{\l_L l_L}}}{{}_{A_{L-1}\, \l_{L-1} l_{L-1}}} \K{{}^{{}_{\l_1 k}}}{{}_{A_L\, \l_L l_L}} \,.
\eea
corresponds to the closed path $\l_1\to\l_L\to \cdots\to \l_2\to \l_1$.
Note that the l.h.s.\ is explicitly symmetric w.r.t.\ the cyclic permutations of the fields under the trace. It would be interesting to illustrate that the r.h.s.\ possesses the same symmetry. For this small digression we consider the simplest example, the Abelian $\Zbb_S$ orbifold. There is one generating element $g$ in the $\Zbb_S$ group, and $g^S=1$. There are the $S$ one-dimensional representations $\{\Rff_n\}$, $\rrho_n(g)=\epsilon^n$, $\epsilon=\e^{2\pi i/S}$, $n=0,\ldots S-1$. The complex scalar $\phi^\aA$ decomposes into three representations; \ie, $g\,:\, \phi^\aA\to \epsilon^{n_A} \phi^\aA$. The quiver has $S$ nodes corresponding to the $S$ representations of $\Zbb_S$. In this setup the invariant tensor is simply $\K{}{\aAe\dddots \aAL}=\delta(n_{{}_{A_1}}+\cdots+n_{{}_{A_L}})$, and it makes sure that the total charge of the product is zero. Choose a closed loop in the quiver going between the nodes $n\to n+n_{{}_{A_1}}\to n+n_{{}_{A_1}}+n_{{}_{A_2}}\to \cdots \to n+\sum_i n_{{}_{A_i}}=n$. (Thus the closedness of the loop translates into the condition that the operator has zero charge). The character $\chi_n(g^m)=\epsilon^{mn}$. Denote $n_1=n$, $n_i=n+\sum_{j=1}^{i-1} n_{{}_{A_j}}$, $i=2,\ldots L$. Then (\ref{loop_operator}) becomes
\bea
\Oc_{n_1\ldots n_L} = \sum_m \epsilon^{-mn_1} \Tr \phi^{\aA}\ldots \phi^{\aAL}\, t_{-m} = \sum_m \sum_{l_1+\cdots+l_L\equiv m} \epsilon^{-mn_1} \Tr \phi^{\aAe}_{l_1}\circ \cdots\circ \phi^{\aAL}_{l_L} \,.
\eea
Na\"ively the formula is not symmetric since it has explicit dependence on $n=n_1$ in the character. However, using the permutation rule (\ref{perm_rel}) we get $\Tr \phi^{\aAe}_{l_1} \ldots \phi^{\aAL}_{l_L}=\epsilon^{mn_{I_L}} \Tr \phi^{\aAL}_{l_L} \phi^{\aA1}_{l_1} \ldots \phi^{{}^{A_{L-1}}}_{l_{L-1}}$. This yields
$$
\Oc_{n_1\ldots n_L} = \sum_m \sum_{l_1+\cdots+l_L\equiv m} \epsilon^{-m(n_1-n_{A_L})} \Tr \phi^{\aAL}_{l_L} \phi^{\aAe}_{l_1} \ldots \phi^{{}^{A_{L-1}}}_{l_{L-1}}
$$
$$
= \sum_m \sum_{l_1+\cdots+l_L\equiv m} \epsilon^{-mn_L} \Tr \phi^{\aAL}_{l_L} \phi^{\aAe}_{l_1} \ldots \phi^{{}^{A_{L-1}}}_{l_{L-1}} = \Oc_{n_L n_1\ldots n_{L-1}} \,.
$$
The form of the formula (\ref{loop_operator}) is indeed invariant w.r.t.\ the cyclic permutations of the fields.

There also exists a different way to construct gauge invariant operators. Namely, let us start with the ansatz
\bea
\Oc[\Kc] &=& \sum_g \sum_{\aAe,\ldots\, \aAL} \K{}{\aAe\ldots\, \aAL}(g)\, \Big(\phi\circ\ldots\circ\phi\Big)^{A_1,\ldots A_k}_g \,.
\eea
Generally such an expression corresponds to a sum of operators corresponding to some paths in the quiver, not necessarily closed. That is why the gauge invariance condition has to be imposed separately. In order to do this we can consider a generic group algebra valued generator of gauge transformations $\chi=\sum_g \chi_g g$ and require that $\delta_\chi \Oc[\Kc]=0$. Recall that the gauge transformation rule of the adjoint fields is $\delta_\chi \phi^A=\comm{\chi}{\phi^A}=\big(\chi\circ\phi-\phi\circ\chi\big)^A$, where the convolution product is understood in the sense of (\ref{convolution}). This gives
\bea
\nn \Big(\delta_\chi\big(\phi\circ\ldots\circ\phi\big)\Big)^{A_1\ldots A_k}_g &=& \sum_h \rep{A_1B_1}{h}\ldots\, \rep{A_LB_L}{h}\, \chi_h\, \big(\phi\circ\ldots\circ\phi\big)^{B_1\ldots B_k}_{h^{-1}g} -
\\
&& -\sum_h \big(\phi\circ\ldots\circ\phi\big)^{A_1\ldots A_k}_{h}\, \chi_{h^{-1}g} \,.
\eea
Then
\bea
\nn \delta_\chi \Oc[\Kc] &=& \sum_{g,h} \sum_{B_1\ldots B_L} \Big[ \K{}{A_1\ldots A_L}(hg) \rep{A_1B_1}{h}\ldots\rep{A_LB_L}{h} - \K{}{B_1\ldots B_L}(gh) \Big]
\\
&&
\times \chi_h \big(\phi\circ\ldots\circ\phi\big)^{B_1\ldots B_k}_g \,.
\eea
This condition rewrites as
\bea
\label{gauge_inv_cond}
\K{}{\bBe\cdots\bBL}(h^{-1}gh) &=& \sum_{\aAe\cdots\aAL} \K{}{\aAe\cdots\aAL}(g) \rep{\aAe\bBe}{h}\ldots \rep{\aAL\bBL}{h} \,.
\eea
A straightforward consequence of this result is that $\Kc[g]$ has to be an invariant tensor w.r.t.\ the stabilizer subgroup $S_g$. Note that in (\ref{loop_operator}) we had
\bea
\K{}{A_1\ldots A_L}(g) &=& \sum_{k,l} \K{}{A_1,\ldots A_L}{}^{\l k}_{\l l}\, \repconj[\l]{kl}{g} \,,
\eea
and it obviously satisfies (\ref{gauge_inv_cond}). On the other side, tensor $\Kc(g)$ can be expanded in Fourier series as a function on the group,
\bea
\K{}{A_1\ldots A_L}(g) &=& \sum_\l \sum_{k,l} \tilde{\Kc}^{(\l)}_{A_1,\ldots A_L}{}^{\l k}_{\l l}\, \repconj[\l]{kl}{g} \,;
\eea
and then the condition (\ref{gauge_inv_cond}) translates into the requirement that the coefficients $\tilde{\Kc}^{(\l)}_{A_1,\ldots A_L}{}^{\l k}_{\l l}$ are invariant tensors. These considerations provide a dictionary between the two notations in the quiver gauge theory.

It is very important that the gauge invariance condition (\ref{gauge_inv_cond}) relates the values of the tensor $\Kc(g)$ within the same conjugacy class, and there is no relation between the values of $\Kc$ on the different conjugacy classes. That is why one can build a gauge invariant operator with $\Kc(g)\neq 0$ only on a given conjugacy class $[g]$. Such operators are said to belong to the \emph{twisted sector} with the twist $[g]$ (determined only up to a conjugation). One can choose a reference element $g$ in the conjugacy class $[hgh^{-1}]$ and set $\K{}{A_1\ldots A_L}(g) = \K{}{A_1\ldots A_L}$, $\K{}{A_1\ldots A_L}$ being some $S_g$-invariant tensor. Then (\ref{gauge_inv_cond}) determines the values of $\Kc(h^{-1}gh)$ on all the elements of the conjugacy class. The corresponding operator
\bea
\Oc[\Kc] &=& \sum_{g,h} \sum_{\aAe\cdots\aAL} \sum_{\bBe\cdots\bBL} \K{}{\aAe\cdots\aAL} \rep{\aAe\bBe}{h}\ldots \rep{\aAL\bBL}{h} \big(\phi\circ\cdots\phi\big)^{\bBe\cdots\bBL}_{h^{-1}gh} \,.
\eea
It can be rewritten in terms of the fields of the original $\Nc=4$ theory in a very simple way,
\bea
\label{gen_operator}
\Oc[\Kc] &=& \sum_{g,h} \sum_{\aAe\cdots\aAL} \K{}{\aAe\cdots\aAL} \Tr \big[\ggamma(g)\phi^{\aAe}\ldots\phi^{\aAL}\big] \,.
\eea
The twist field $\ggamma(g)$ acts on the dynamical fields as follows,
\bea
\big(\phi^A\, \ggamma(g)\big)_{h_1,h_2} &=& \phi^A_{h_1,gh_2} \,,
\\
\big(\ggamma(g)\, \phi^A\big)_{h_1,h_2} &=& \phi^A_{g^{-1}h_1,h_2} \,.
\eea
Invariance condition imposed by the orbifold projection on the fields implies that
\be
\label{perm_rel}
\bigl(\ggamma(g)\, \phi^A\bigr)  = \rep{AB}{g^{-1}}\, \bigl(\phi^B\, \ggamma(g)\bigr) \,.
\ee
\\

\bigskip
\newsection{Derivation of the Feynman Rules}
\label{app:feynman_rules}
As an example we go through the derivation for the scalar field  $\Phi^\II$; the other fields are treated in a similar way. Then the $\GG$-invariance condition (\ref{A_inv_cond}) for the scalar field is solved by
\be
\label{inv_phi}
\Phi^{{}^{{}_{\II}}}_{i,h;\, j,g} = \sum_{\JJ} \Rf(h)^{{}^{{}_{\II}}}_{\JJ}\, \phi^{{}^{{}_{\JJ}}}_{ij}(h^{-1}g) \,.
\ee
We can parameterize the invariant configurations of the scalar field $\Phi^{{}^{{}_{\II}}}_{ig,jh}$ in terms of this group algebra valued object $\phi^{{}^{{}_{\II}}}_{ij}(g)$, and this group algebra valued field $\phi$ is to be integrated over in the path integral.
Using the parameterization (\ref{inv_phi}) and the orthogonality of the defining representation $\Rf:\Gamma\to \SO(6)$, we get the kinetic term for the scalar field in the form
\be
\Lc_{\phiq\phiq} = \half \sum_\II \Tr \pr_\mu
\Phi^{{}^{{}_{\II}}}\, \pr^\mu \Phi^{{}^{{}_{\II}}} = \half |\Gamma| \sum_g \sum_{\II,\JJ} \Rf(g)^{{}^{{}_{\II}}}_{\JJ}\, \pr_\mu
\Tr \phi^{{}^{{}_{\II}}}(g)\, \pr^\mu \phi^{{}^{{}_{\JJ}}}(g^{-1}) \,.
\ee
Then for the quadratic propagator the only modification compared to the original theory is ``conservation of the group index'' and renormalization:
\bea
\label{free_quiver}
\avg{\phiq^{{}^{{}_{\II}}}_{ij,g}\, \phiq^{{}^{{}_{\JJ}}}_{kl,h}} &=& |\Gamma|^{-1}
\frac{\Rf(g)^{{}^{{}_{\II}}}_{\JJ}}{p^2}\, \delta_{gh,e}\, \delta_{il}\, \delta_{jk}\,,
\eea
In terms of the original $\Nc\!=\!4$ fields (we omit the obvious Latin part of the Chan-Paton indices)
\bea
\label{quiver_quad}
\avg{\Phi^{{}^{{}_{\II}}}_{h,g}\, \Phi^{{}^{{}_{\JJ}}}_{f^{-1}g\com f^{-1}h}} &=& \frac{\Rf(f)^{{}^{{}_{\II}}}_{\JJ}}{\ord{\Gamma} p^2} \,.
\eea

\end{document}